%% file: main.tex
\documentclass[11pt, leqno, a4paper, twoside]{amsart}

\input{Praeamble.tex}

\title[Cerebral Aneurysm Thrombus Formation: Modeling and Occlusion Assessment]{Device-Induced Thrombus Formation in Cerebral Aneurysms: Linking Patient-Specific Clot Modeling and Functional Occlusion to Virtual Angiographic Assessment}


\author[F. Holzberger, S. Hume, M. Muhr, M. Ngoepe, B. Wohlmuth]{Fabian Holzberger$^{1}$, Struan Hume$^{2}$, Markus Muhr$^{1}$, Malebogo Ngoepe$^{2}$, Barbara Wohlmuth$^{1}$}

\email{\href{mailto:holf@cit.tum.de}{holf@cit.tum.de}}
\email{\href{mailto:struan.hume@uct.ac.za}{struan.hume@uct.ac.za}}
\email{\href{mailto:muhr@cit.tum.de}{muhr@cit.tum.de}}
\email{\href{mailto:malebogo.ngoepe@uct.ac.za}{malebogo.ngoepe@uct.ac.za}}
\email{\href{mailto:wohlmuth@cit.tum.de}{wohlmuth@cit.tum.de}}

\begin{document}
	
\maketitle
\vspace*{-4mm}
\begin{center}
{\footnotesize
	$^{1}$TU Munich, School of Computation, Information and Technology, Department Mathematics, Chair for Numerical Mathematics, Boltzmannstr. 3, 85748, Garching, Germany\\
	$^{2}$University of Cape Town, CERECAM and Department of Mechanical Engineering, South Africa
}
\end{center}


\begin{abstract}
	Endovascular treatment of cerebral aneurysms aims to achieve functional occlusion and isolation of the aneurysm sac from bloodflow. In clinical practice, treatment success is assessed primarily through digital subtraction angiography (DSA), which visualizes contrast-agent inflow and washout but does not directly resolve thrombus formation driving early occlusion.
	
	We present a computational framework that couples acute fibrin thrombus formation with virtual angiography, enabling early thrombus growth to be interpreted through clinically familiar DSA-like imaging. Three common treatment strategies: endovascular coiling, flow diversion, and stent-assisted coiling, are modeled under pulsatile hemodynamics and linked to simulated contrast transport.
	
	Across three representative aneurysm morphologies, the simulations demonstrate that while devices reduce inflow, residual contrast access and trapping may persist, with early thrombus formation contributing substantially to perfusion suppression and altered washout patterns. These effects are clearly reflected in the virtual angiographic imaging. The importance of vortical structures in device-induced thrombosis is highligthed in one of the cases.
	
	By seeking to align modelling and simulation tools with clinically-relevant metrics, with a particular focus on occlusion outcome, this work presents a good starting point for bridging the gap between these two paradigms.
\end{abstract}
\vspace{8mm}


\input{Introduction}

\input{EndovascularDevices}

\input{ThrombusModel}

\input{OcclusionQuality}

\input{NumericalExperiments}

\input{Discussion}

\input{Conclusion}

\section*{Funding}
Fabian Holzberger, Markus Muhr and Barbara Wohlmuth gratefully acknowledge the financial support provided by the German Science Foundation (DFG) under project number 465242983 within the priority programme SPP 2311: Robust coupling of continuum-biomechanical in silico models to establish active biological system models for later use in clinical applications - Co-design of modeling, numerics and usability (WO 671/20-1 and WO 671/20-2)

\section*{Acronyms}
\input{acronyms}

\bibliographystyle{plain}   
\bibliography{references}   

\end{document}

%% file: Praeamble.tex
\usepackage[left=2.2cm, right=2.2cm, bottom=3cm]{geometry}
\usepackage{graphicx}

\usepackage{xcolor}
\definecolor{orange}{RGB}{1,0.55,0.2}

\usepackage{dsfont}

\newcommand{\mat}[1]{\textbf{#1}}
\renewcommand{\vec}[1]{\boldsymbol{#1}}

\usepackage{hyperref}
\hypersetup{
    colorlinks=true,
    linkcolor=blue,
    filecolor=blue,      
    urlcolor=blue,
    citecolor=blue
    }

\usepackage[printonlyused]{acronym}

\usepackage{lipsum}
\usepackage{subcaption}
\usepackage{float}
\usepackage{multirow} 
\usepackage{placeins}
\usepackage{multicol}
\usepackage{booktabs}

\usepackage{siunitx}
\DeclareSIUnit\Molar{M}

%% file: Introduction.tex
\section{Introduction}\label{sec:Introduction}
Cerebral aneurysms are balloon-shaped sacs that develop on blood vessels of the brain. The weakened vessel wall is at risk of rupture, with subsequent morbidity or mortality. Guidelines and decisions on when and how to treat cerebral aneurysms vary across the world \cite{Thompson2015, Steiner2013, Jeong2014}. The last three decades have brought about a shift from predominantly surgery-based treatment approaches to endovascular treatments \cite{Guglielmi1991,Lee2022}.  Despite advances in endovascular treatment techniques and management guidelines, the availability of resources in different contexts, including access to specialist care, strongly informs treatment approaches \cite{Lee2022, Ferreira2023, Mkhize2023}. Nevertheless, endovascular techniques have gained popularity for several clinical reasons, including greater ease of customizing solutions per patient and reduction of iatrogenic risk.\\

Endovascular treatment has several effects on the aneurysm, including alteration of local flow patterns and occlusion of the sac. Occlusion is achieved by the presence of an endovascular device and/or thrombus, with variable influence on aneurysm evolution \cite{Frosen2012}. Across endovascular strategies, a central therapeutic aim is full occlusion, where the aneurysm sac is isolated completely from circulation from the parent-vessel. Complete occlusion of the sac is a desirable outcome as it prevents any further flow from entering the aneurysm. This outcome can take a long time to develop, with some clinical studies reporting complete occlusion after six or twelve months of device placement \cite{Turjman2014, Bender2018, AlSaiegh2022}. In some cases, partial occlusion has been shown to exacerbate vascular degradation of an already weakened aneurysm wall and accelerates the time to rupture \cite{Hokari2015, Frosen2012}. Occlusion depends on a number of variables, including aneurysm geometry, clotting profile, deposition patterns and choice of endovascular device \cite{Epshtein2020, Czaja2018}. These interdependencies create a persistent clinical challenge in assessing, within a clinically-relevant timeframe, whether a selected endovascular configuration is likely to produce durable functional occlusion. \\

The patient-specific nature of cerebral aneurysms and their treatment has motivated the development of various image-based and  continuum-based computational models  \cite{Ngoepe2018, Zhang2022, Brindise2019, Varble2017, Goubergrits2011}. Continuum flow-based models have demonstrated how the placement of specific devices results in altered haemodynamic outcomes \cite{macraild2024accelerated}. Models reporting thrombotic outcomes typically present the effects of one type of endovascular treatment only \cite{Cebral2024, Sarrami-Foroushani2021, Sarrami-Foroushani2019, Ou2017, Ngoepe2016}. These models have also been used to analyse different quantitative metrics, ranging from wall shear stress to biochemical species concentration, and variable outcomes have been reported for treated aneurysms \cite{Baek2009}.  As the range of treatment options expand, patient-specific modelling approaches that give insight into how device placement influences occlusion could support clinical interventional planning. \\

Clinically, treatment success is often assessed through digital subtraction angiography (DSA), which visualizes contrast inflow and washout, and is commonly interpreted as a proxy for occlusion. In parallel, computational fluid dynamics (CFD)-based “virtual angiography” has been used to compare simulated hemodynamics to angiographic observations. However, much of the computational work that exists does not explicitly account for thrombus evolution, and contrast washout is most often treated as a surrogate for stasis rather than being tested against a mechanistic occlusion process. In the absence of endovascular device placement, mechanistic fibrin clot evolution in aneurysms is strongly influenced by changing vorticial structures, with smaller vortex modes supporting clot formation across a range of aneurysm morphologies \cite{Ngwenya2024}. We hypothesise that a similar mechanism is likely to drive fibrin clot formation in the presence of devices. Here we show the impact of the applied treatment devices. \\

\begin{figure}[h]
    \centering
    \includegraphics[width=0.3\linewidth]{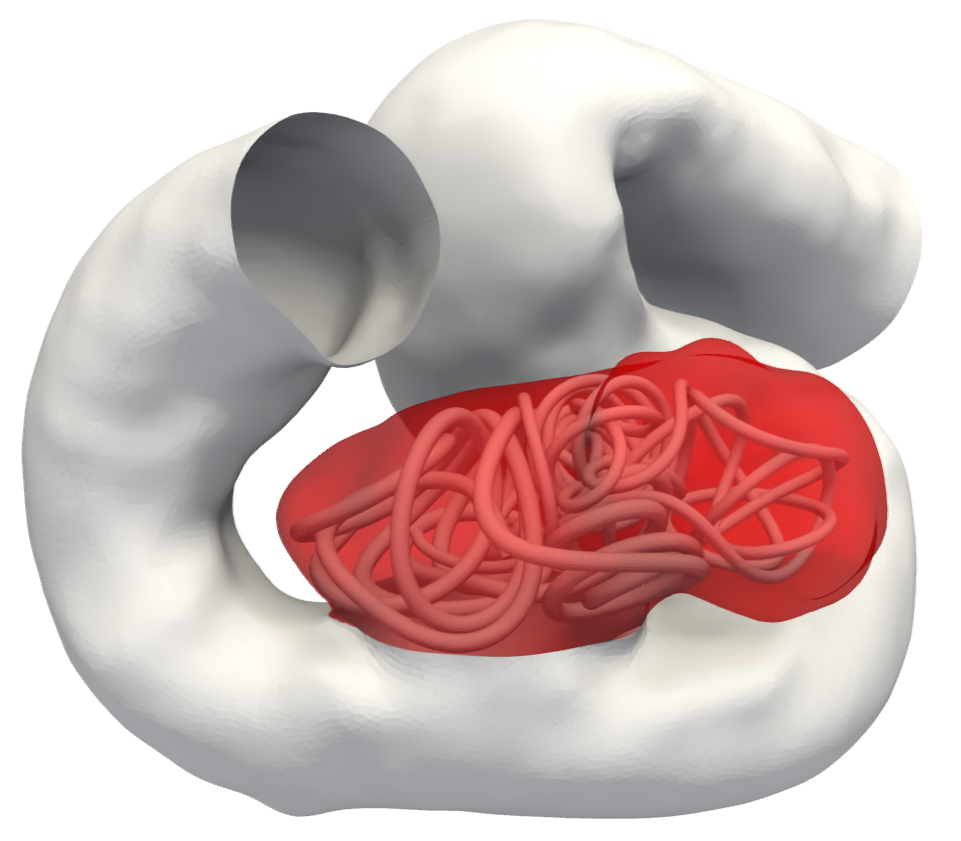}\includegraphics[width=0.3\linewidth]{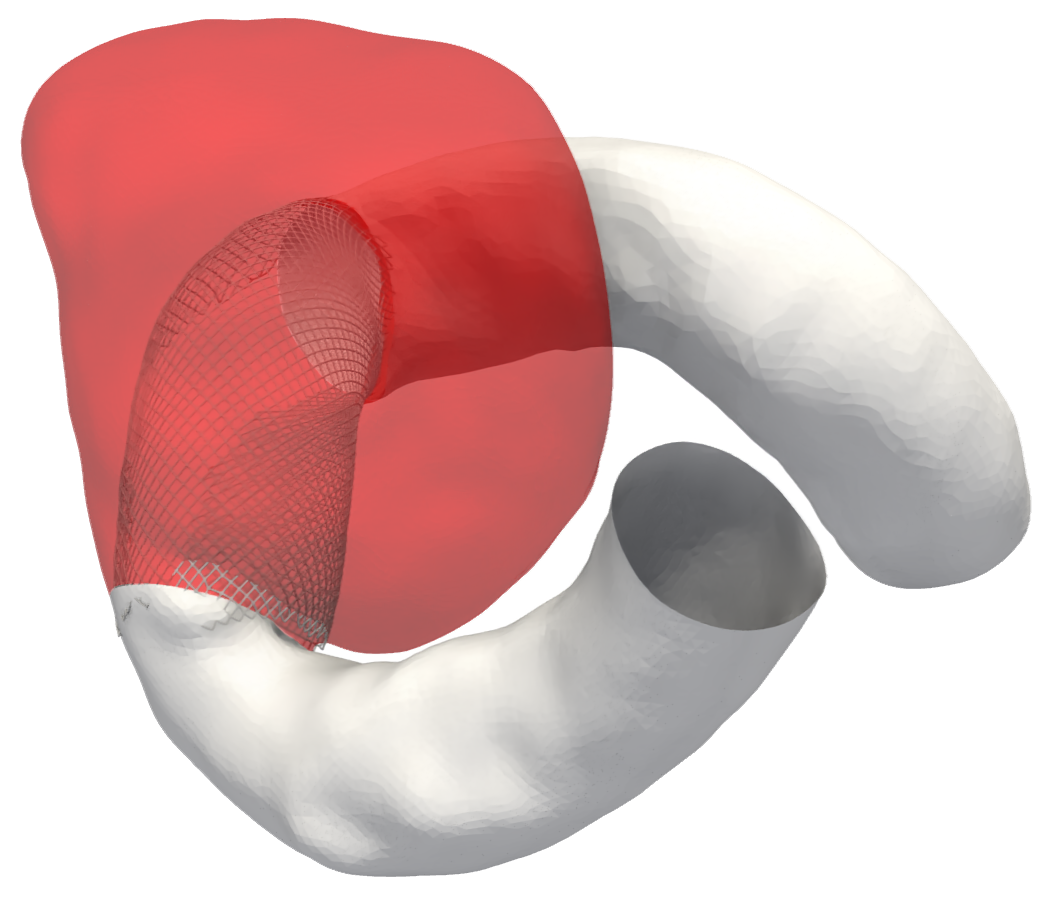}\includegraphics[width=0.4\linewidth]{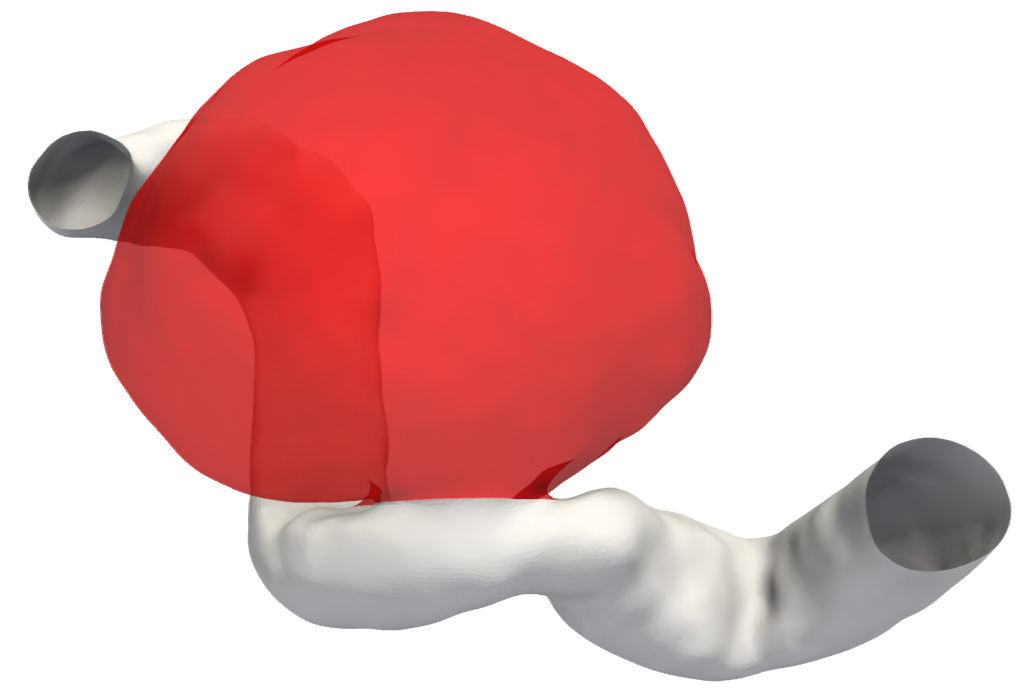}
    \caption{\footnotesize The three exemplary cases of different cerebral aneurysm morphologies treated in this work. \textbf{(Left) Case 1:} Small, saccular aneurysm where both, stenting and/or coiling could be employed. An exemplary coil is inserted for visualization of the method. \textbf{(Middle) Case 2:} Large fusiform aneurysm and a prime example for stent-assisted coiling. \textbf{(Right) Case 3:} Very large saccular aneurysm, suitable for (material intensive) coiling or stenting as an alternative.}
    \label{fig:OurThreeCases}
\end{figure}

To connect mechanistic fibrin clot formation with an angiography-like observable, in the presence of endovascular devices, we combine a fibrin clot CFD-based model with a residual-contrast transport model \cite{1546120, wu2007real}. Our thrombosis model and virtual DSA model, linked as a two-step framework for the pre- and post-thrombus state, allows us to test whether clinically familiar washout patterns provide a reliable proxy for mechanistically predicted thrombus-based occlusion. Based on our aforementioned hypothesis, we then use this framework to examine the extent to which vorticial structures inform fibrin clot patterns in the presence of endovascular devices. We do so by applying our framework to three exemplary cerebral aneurysm cases with distinct morphologies and device requirements, depicted in Fig.\,\ref{fig:OurThreeCases}. Across cases, three common endovascular strategies (endovascular coiling, flow diversion, and stent-assisted coiling) are considered to probe generality across device classes. These cases stand as prototypes for classes of statistically relevant aneurysm-types regarding their (volumetric) size, neck-width and shape (saccular vs. fusiform).

\begin{itemize}
    \item \textbf{Case 1:} A relatively small, saccular, nearly tube-shaped, and wide-necked aneurysm with a volume of approximately \SI{71}{mm^3}, an ostium diameter of around \SI{4}{mm} and an approximate length (longest dimension) of around \SI{8}{mm}. 
    \item \textbf{Case 2:} A large fusiform aneurysm with a volume of approximately \SI{412}{mm^3}, affected vessel length of \SI{6.5}{mm} and approximate diameter of \SI{10}{mm}.
    \item \textbf{Case 3:} With a volume of around \SI{2193}{mm^3} a volumetrically very large and spherically shaped saccular aneurysm, however with a narrow neck and an ostium diameter of around \SI{6}{mm}.
\end{itemize}

%% file: EndovascularDevices.tex
\section{Endovascular devices: Models and data} \label{sec:EndovascularDevices}

Endovascular techniques encompass a wide variety of treatment approaches and methods. The common feature across techniques is catheter-based delivery and deployment of an endovascular device to the aneurysm region, differentiating them from open surgical approaches such as clipping. Typical treatment devices include endovascular coils, stents, flow-diverters, Woven EndoBridge (WEB) or contouring devices, each with its own (dis)advantages and typical use cases. In some instances, combinations of these devices are deployed and additional devices, like endovascular balloons, can be employed temporarily to aid placement. An overview of different \textit{mechanical and geometric models} used for simulating and analysing endovascular treatment options is presented in \cite{frank2024numerical}. In this work, coiling, flow-diversion and stent-assisted coiling as a combination of the two techniques are modelled, with mathematical details given in subsequent sections. 

\subsection{Endovascular coiling}\label{subsec:Coiling}
\begin{figure}[h]
    \centering
    \includegraphics[width=0.295\linewidth]{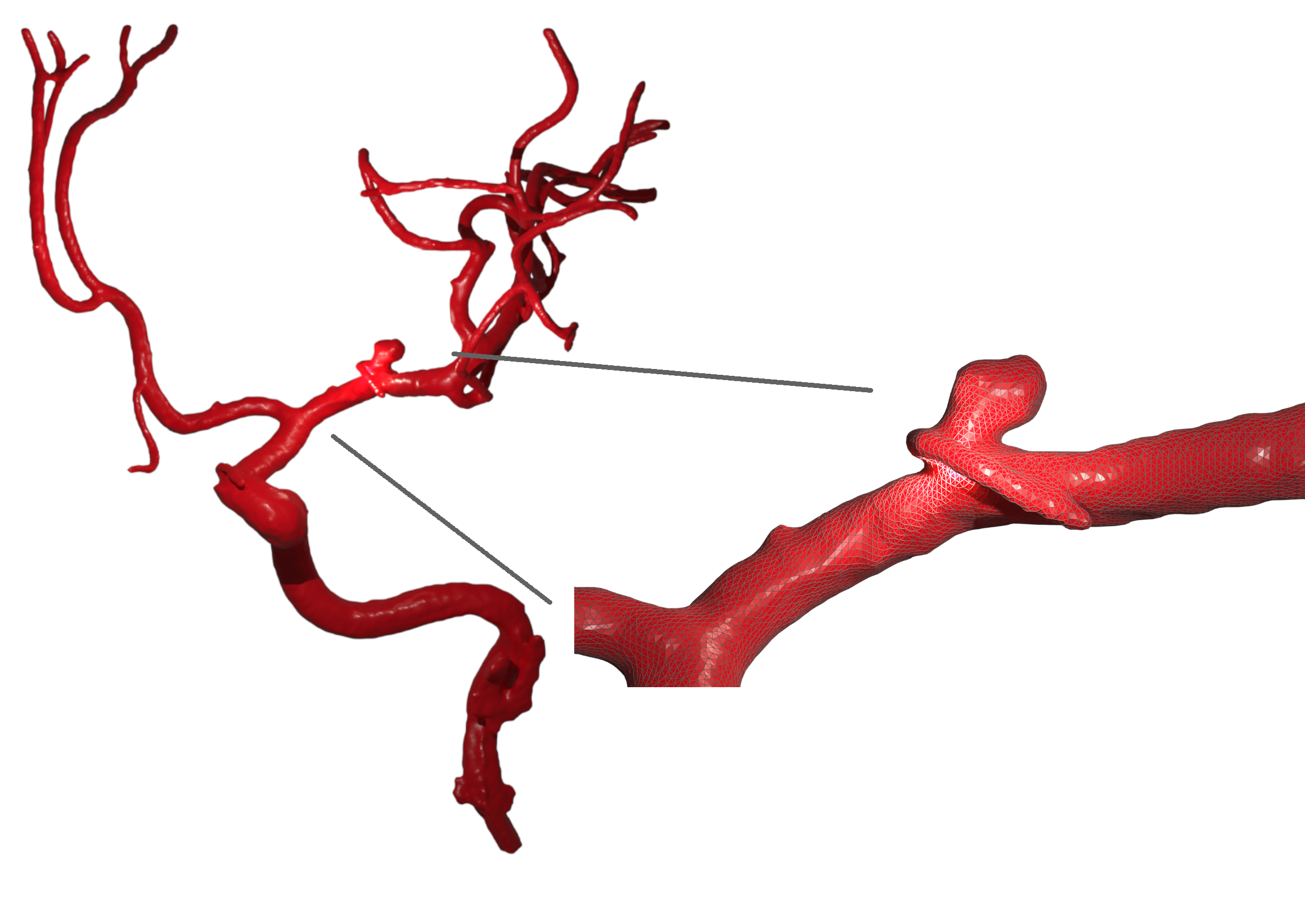}\hfill
    \includegraphics[width=0.225\linewidth]{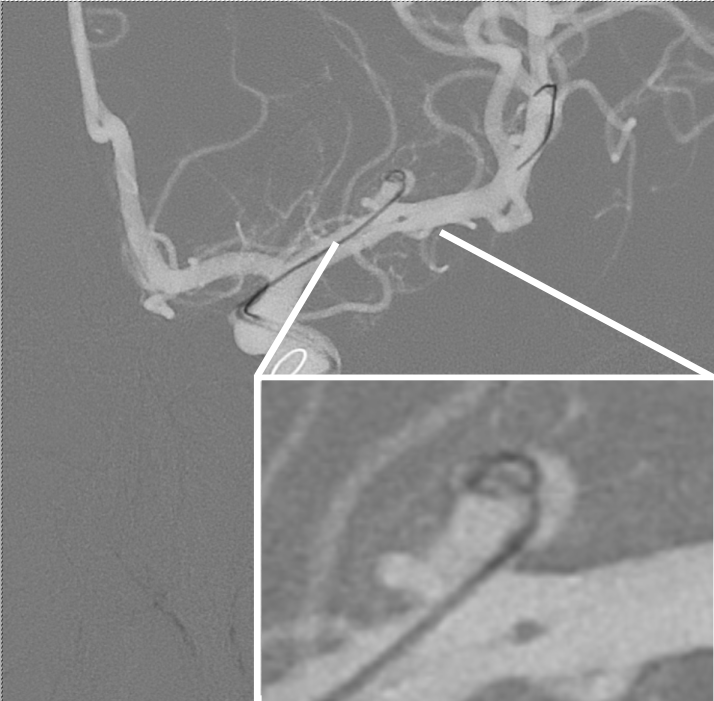}\hfill
    \includegraphics[width=0.225\linewidth]{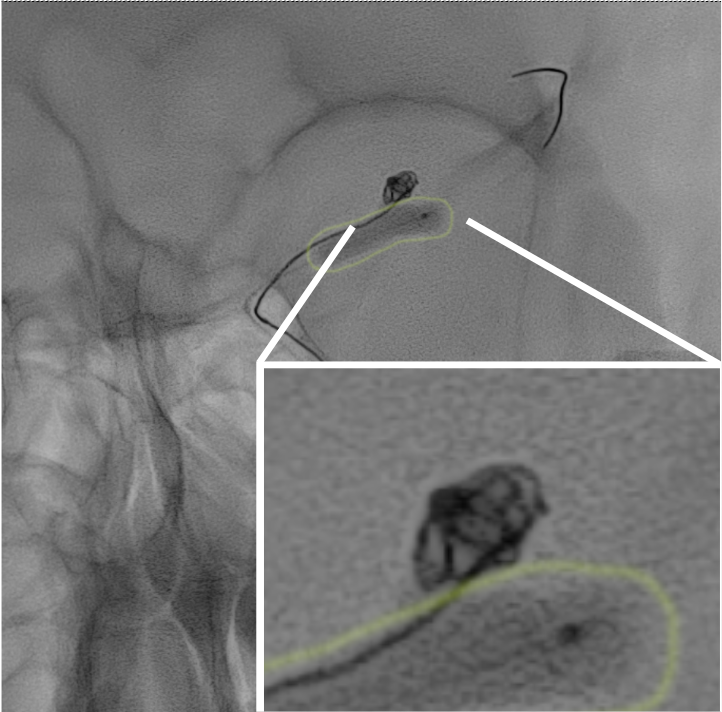}\hfill
    \includegraphics[width=0.24\linewidth]{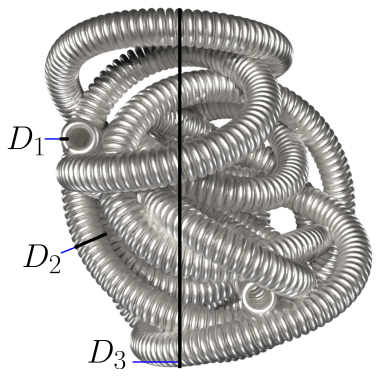}
    \caption{\footnotesize 
        \textbf{(left)} Vascular-tree geometry including an aneurysm, reconstructed from CT-images, 
        \textbf{(middle)} Angiographic images from during the coil insertion procedure on that geometry. First, at the beginning, with only the guiding wire visible, and second, towards the end of the procedure, with a surgical balloon (outlined in green) used to stabilize the coil in place,  
        \textbf{(right)} Schematic sketch of a coil's micro-wire showing its three characteristic radii $D_1<D_2<D_3$.}
    \label{fig:MedicalImageOfCoilInsertion}
\end{figure}

Endovascular coils comprise one or several soft micro-wires that are deployed into the aneurysmal sac using a micro-catheter, as illustrated in Fig.\,\ref{fig:MedicalImageOfCoilInsertion} . The coils curl up and occlude the aneurysm, achieving flow- and wall shear stress-reduction. 
Mechanically, endovascular coiling wires are complex devices. A closer look reveals an imprinted microstructure that supports natural bending and torsion behaviour. When deployed from the microcatheter into the aneurysm, the wire assumes a pre-shaped configuration, driving it towards the desired shape with a given spatial diameter. The variety of shape-design choices allows the manufacturing of coils for different purposes. From a treatment perspective, stiffer framing coils are usually placed first in the empty aneurysm to form a stabilizing "basket" aligned to the aneurysm walls.

The size of the \textit{framing} coil, in terms of its encasing volume, is typically measured by its $D_3$-radius (see Fig.\,\ref{fig:MedicalImageOfCoilInsertion}), allowing a surgeon to choose the correct coil based on prior angiographic measurements of the aneurysm's dimensions. Following the \textit{framing}-coil, softer \textit{filling}-coils are placed in the ``basket'' to fill out the aneurysm volumetrically and suppress further blood-circulation. \\

\paragraph{Mechanical coil model}
To obtain realistically shaped coil-geometries \textit{in-silico} for subsequent thrombus-formation and occlusion-quality simulations, a mechanical simulation of the coil-placement procedure is utilized. We specifically refer to the numerical model developed in \cite{holzberger2024comprehensive} and already applied to further patientspecific datasets in \cite{schwarting2024numerical}, where we summarize the most important parts for the reader's convenience.\\
In this model, a coiling wire is resolved as a one-dimensional sequence of $N-1$ discrete elastic and interconnected rods represented by edges $\Vec{e}_j=\Vec{x}_{j+1}-\Vec{x}_j$, $j=1, \dots, N-1$, where for all indices $j$ the points $\Vec{x}_j\in\mathds{R}^3$ are the coordinates of the connecting vertices between two subsequent rods. The instantaneous length of the $j$-th rod is hence given by $\|\Vec{e}_j\|$, where $\|\cdot\|$ always refers to the Euclidean norm. The whole configuration is then subjected to deformations by bending and torsion, e.g., by trying to adopt the imprinted natural shape once released from the micro-catheter. Therefore the deformation is calculated by means of minimizing the following strain energy functional:

\begin{equation}\label{eq:EnergyFunctional}
E_{\textup{tot}}=\sum_{j=0}^{N-2}\frac{1}{2}\alpha\left(\frac{\|\Vec{e}_j\|}{\|\overline{\Vec{e}}_j\|}-1\right)^2\|\overline{\Vec{e}_j}\|+\sum_{i=1}^{N-2}\frac{(\vec{\kappa}_i-\overline{\vec{\kappa}}_i)\mat{B}(\vec{\kappa}_i-\overline{\Vec{\kappa}}_i)^{\top}}{\|\overline{\Vec{e}}_{i-1}\|+\|\overline{\Vec{e}}_i\|}+\sum_{j=1}^{N-2}\frac{\beta(\tau_j-\overline{\tau}_j)^2}{{\|\overline{\Vec{e}}_{j-1}\|+\|\overline{\Vec{e}}_j\|}}
\end{equation}~\\

Here, all over-barred quantities $\overline{\cdot}$ symbolize values in the imprinted, natural ``goal'' shape, while quantities without an over-bar are the  currently attained ones. Specifically $\vec{\kappa}_i\in\mathds{R}^2$ stands for the discrete curvature-component-vector at vertex $i$, while $\tau_j=\phi_j-\phi_{j-1}$ symbolizes difference of torsion\,/\,twist(-angle) $\phi$ over the respective rod. $\mat{B}$ and $\beta$ are the respective bending and torsion parameters of the specific coiling wire depending on its material as well as microscopic helix-structure geometry, i.e. its $D_1$ and $D_2$ radii, depicted in Fig.\,\ref{fig:MedicalImageOfCoilInsertion}, and helix-pitch, see \cite[Sec.\,2.5]{holzberger2024comprehensive} for details. $\alpha$ hereby is a penalty parameter that aims to enforce an inextensibility constraint on the individual rods that is reflected by a fiber core within real coils. Minimizing the energy-functional hence drives the coil shape towards the one given by the natural pre-shape. Additional influence comes from the contact of the coiling wire's rods with the aneurysm surface and other loops of the wire itself, and in the case of multi-coil insertion, other placed coils. These additional external forces are modeled by a Coulomb stick-slip friction contact model, which allows for sliding contact while treating the aneurysm wall as an impenetrable obstacle within the insertion process. In total, these contacts exert an external force $\vec{F}_{\textup{ext}}$ onto the coil structure. For details on the contact model used, we again refer to the respective publication \cite[Sec.\,2.6,2.7]{holzberger2024comprehensive}.\\

\begin{figure}[h]
    \begin{center}
            \includegraphics[trim=22cm 0cm 10cm 0cm, clip, width=0.3\linewidth]{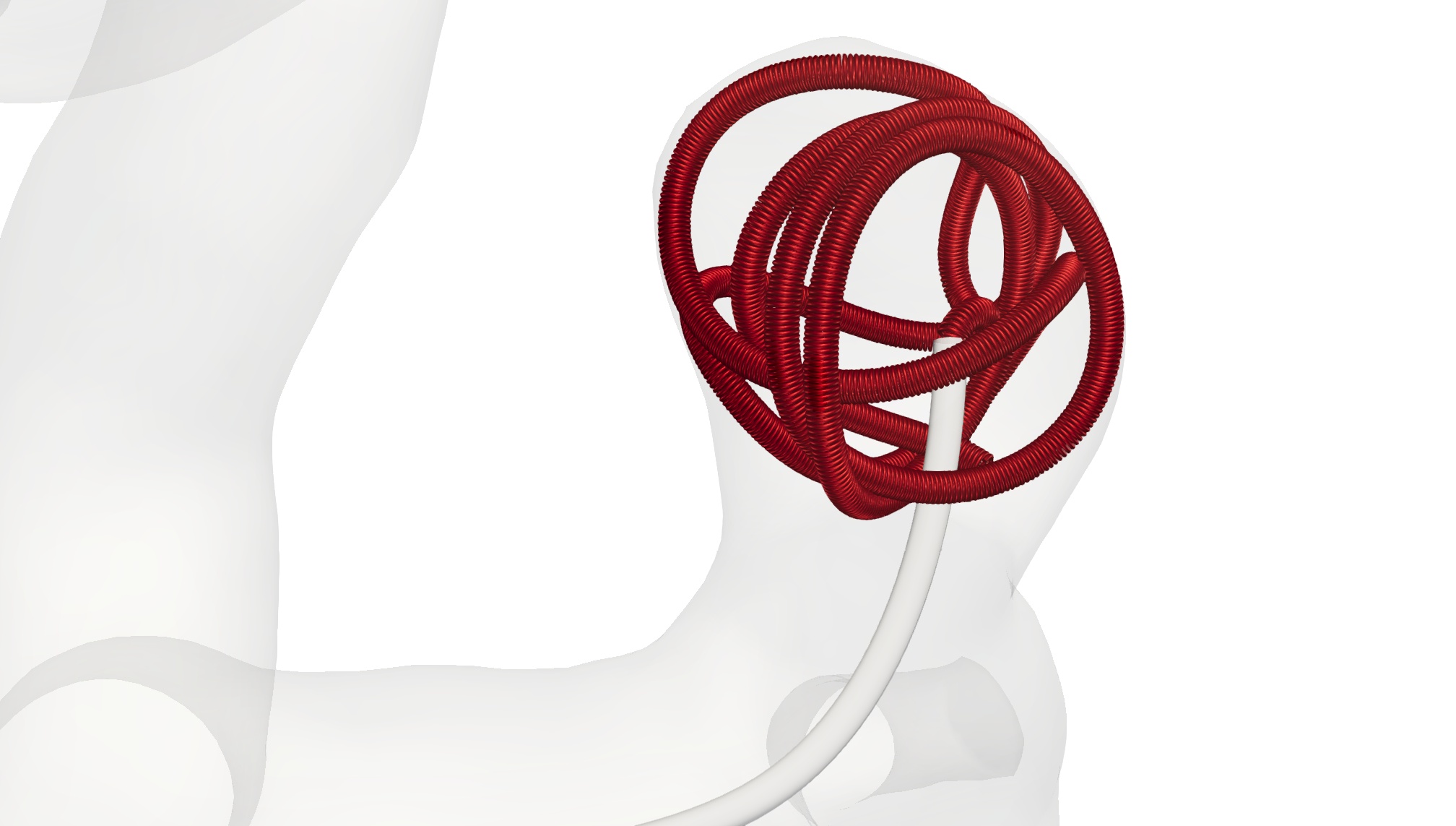} \hspace{1cm} \includegraphics[trim=22cm 0cm 10cm 0cm, clip, width=0.3\linewidth]{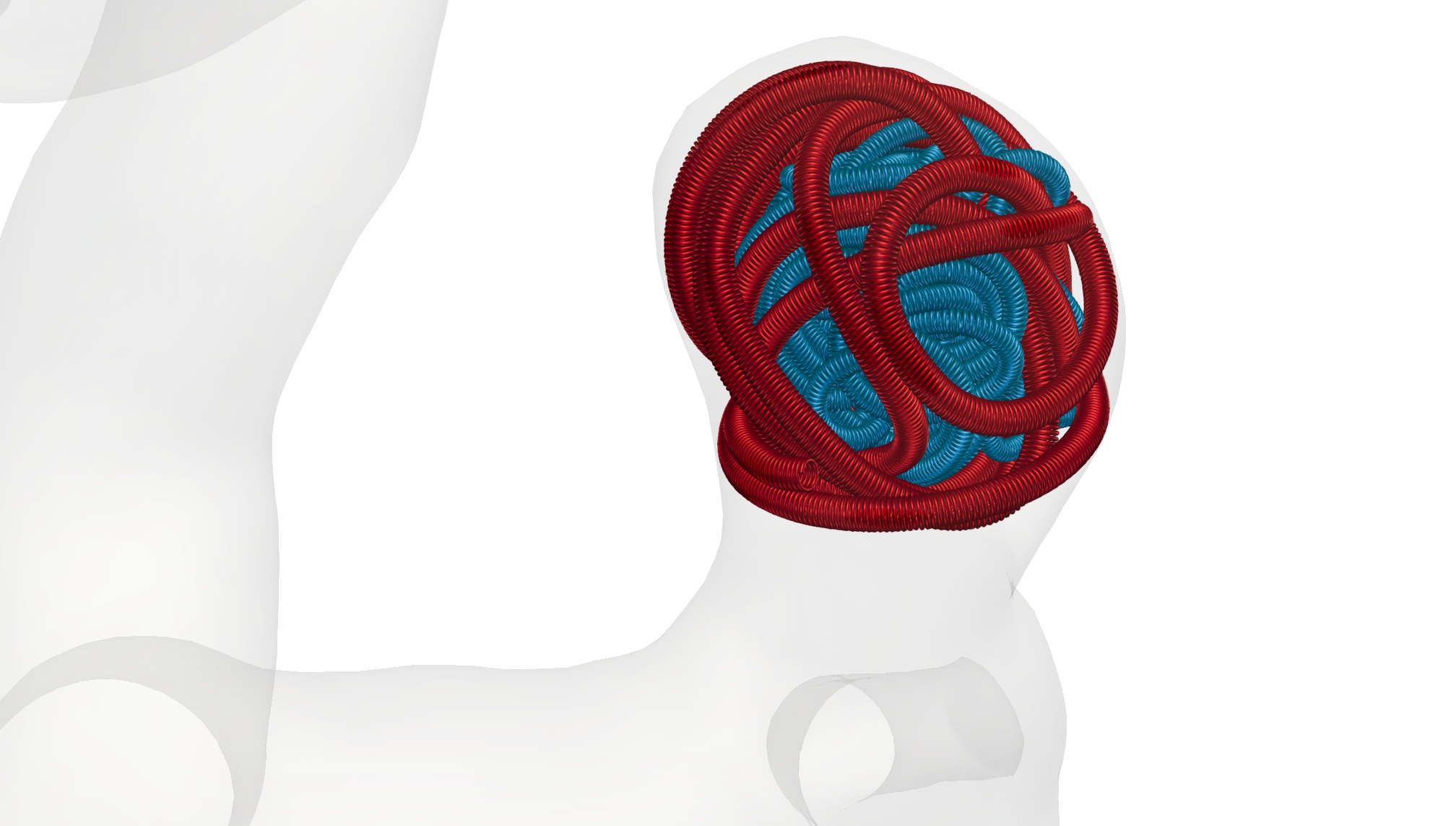}
    \end{center}
    \caption{\footnotesize Simulated insertion of two coils  from \cite{holzberger2024comprehensive}. \textbf{(Left)} shows the insertion of an initial framing coil (red) through a micro-catheter, \textbf{(right)} is a subsequently inserted filling coil (blue) with the micro-catheter being removed again.}
    \label{fig:CoilingSequenceSimulated}
\end{figure}

The system of ordinary differential equations obtained by differentiating the above energy-functional \eqref{eq:EnergyFunctional} has the rod's spatial vertex-locations as well as the individual twist-angles as degrees of freedom. It is solved numerically via the symplectic Euler method where details can again be found in \cite[Sec.\,2.3]{holzberger2024comprehensive}. Fig.\,\ref{fig:CoilingSequenceSimulated} shows a simulated coiling sequence of multiple such inserted coils. \\

\paragraph{Simulation output data}
Even though the coil model discussed above resolves the coil as a one-dimensional structure, its spatial extent by means of its $D_2$ radius is reflected with the coil-coil as well as coil-surface contact model where friction sets in once that radial distance is undercut. Hence, two discrete rods are never closer to each other than $D_2$, or closer to the aneurysm surface than $D_2/2$, and are hence not physically penetrating. In order to obtain a three-dimensional (object-) representation of the final coil shape for further (numerical) analysis, e.g., to be placed as an actual, fully resolved flow-obstacle in a hemodynamic simulation, a cylindrical surface (triangular) mesh of diameter $D_2$ is constructed around the connected one-dimensional rod-centerline curve. The mesh is then exported in \texttt{*.stl} or \texttt{*.obj} format as typical data formats for use in other software packages. 

\paragraph{Application scenarios}
Even with the introduction of newer endovascular devices, coiling has remained the primary mode of treatment since its introduction as a clinical standard. It is well-suited to narrow-necked (Case 3) aneurysms because the coils can unfold freely in the sac while remaining contained, with minimal risk of parent vessel obstruction. In very wide-necked fusiform aneurysms (see Case 2), where coil retention is harder to achieve and risk of migration into the parent vessel is higher, stenting and flow-diverting techniques are a common alternative, sometimes in conjunction with coiling.

\subsection{Flow diversion}\label{subsec:FlowDiversion}
By contrast to coils, which achieve occlusion by volumetric packing, a \textit{\ac{FD}} interrupts flow at the ostium using a very fine grid structure. Delivery and deployment of \acp{FD} is similar to coiling, where a micro-catheter is advanced to the parent vessel of the aneurysm to place the device. During deployment, the \ac{FD} unfolds into a tube-like structure and anchors against the vessel wall. Medical images depicting flow-diverter treatment and devices are reported in \cite[Fig.\,1,2,3]{chiu2020future}. \acp{FD} are candidates for treatment of, e.g., fusiform aneurysms as Case 2 of the present study, where pure coiling would not be feasible, due to the risk of occlusion of the parenting vessel. As with coiling devices, having a three-dimensional model of a \ac{FD} enables virtual placement in a reconstructed geometry for subsequent haemodynamic flow simulations. These can be used to study the impact of the device on flow behavior, rupture-risk and treatment-relevant quantities of interest inside the aneurysm.\\

\begin{figure}[h]
    \centering
    \includegraphics[width=0.3\linewidth]{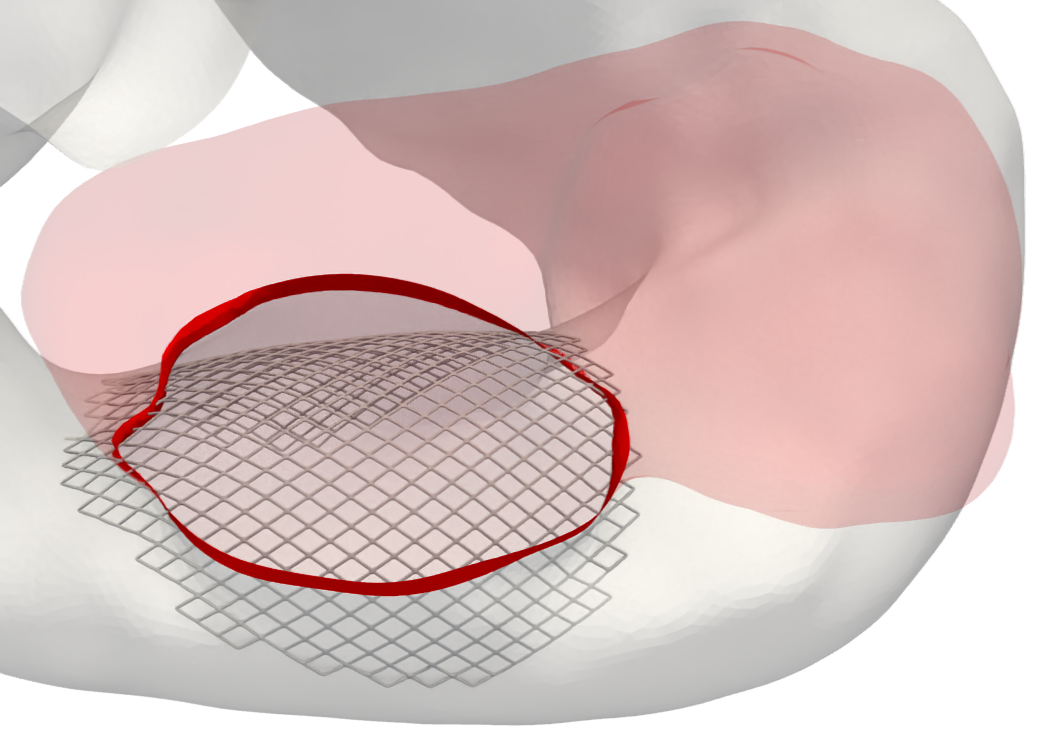}\includegraphics[width=0.3\linewidth]{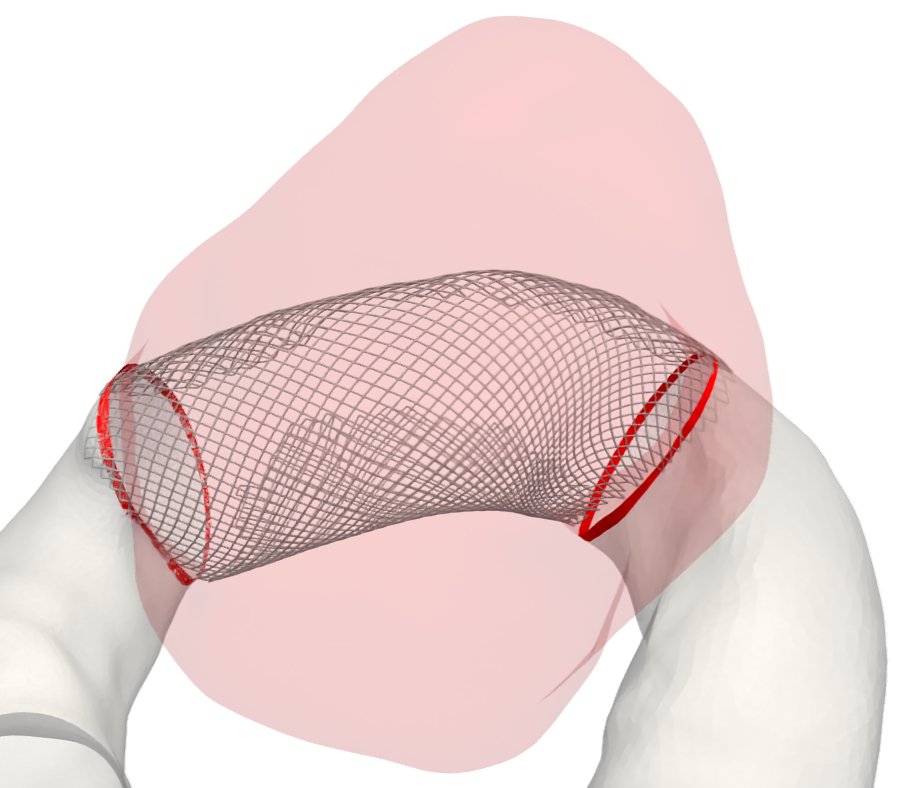}\includegraphics[width=0.3\linewidth]{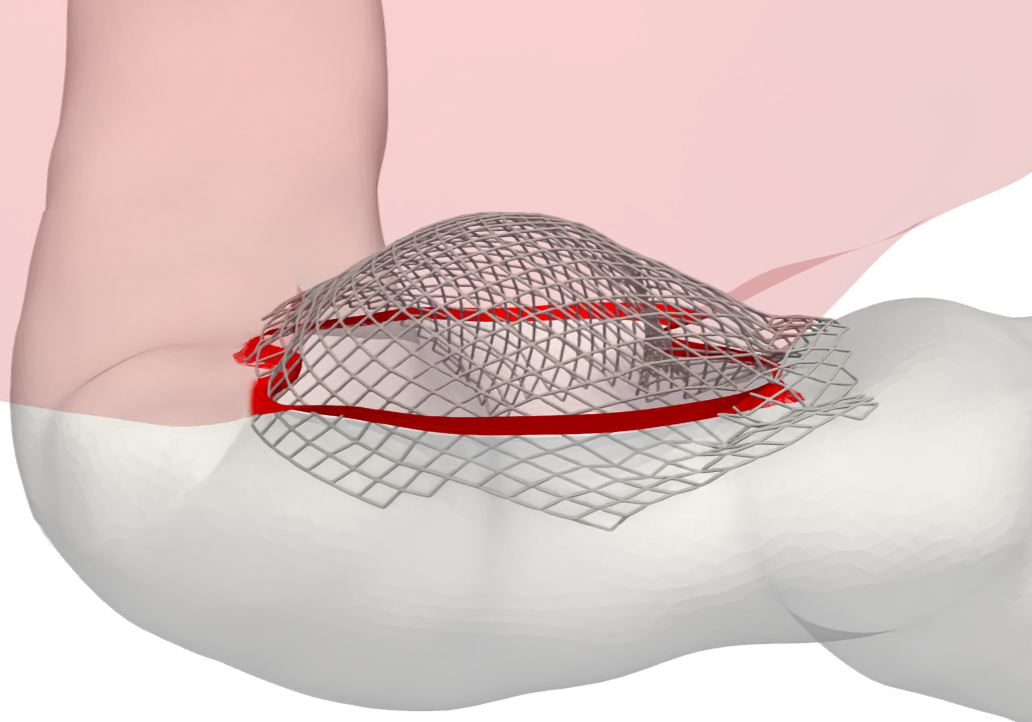}
    \caption{\footnotesize 3d-reconstructed flow diverter geometries within their respective vessel geometries (transparent): \textbf{(Left)} Case 1, \textbf{(Middle)} Case 2 and \textbf{(Right)} Case 3. The aneurysms are shown in transparent red, while the red loops mark the ostium of the respective aneurysm or, in case 2, the beginning and end of the fusiform dilation of the parenting vessel, i.e., the region of interest.}
    \label{fig:FlowDiverterImages}
\end{figure}

\paragraph{Flow diverter model} 
Mathematical modelling of \acp{FD}, or stent-grafts in general, is an extensive research area. It ranges from intricate mechanical models simulating the unfolding process that takes place during deployment to more simplistic geometric models that take into account the centerline and radial information of the respective vessel, geometrically aligning the \ac{FD} according to the vessel's geometry \cite{frank2024numerical}. For this work, we obtained virtually derived flow-diverter models and corresponding aneurysm geometries used in the Flow Diverter 2016 challenge dataset for our three cases presented in Fig.\,\ref{fig:OurThreeCases} \cite{berg2018virtualstenting}.  The flow-diverter geometries are available as \texttt{*.stl} files with resolutions of around 1.5 million vertices and 3 million faces each, which is roughly 1-2 orders of magnitude greater than the final coil models and vessel (surface) geometries depicted in Fig.\,\ref{fig:OurThreeCases}. The three flow diverter models are shown within their respective case geometries in Fig.\,\ref{fig:FlowDiverterImages}.

\subsection{Stent-assisted coiling}\label{subsec:StentassistedCoiling}
To simulate the treatment of large, recurring or complicated aneurysm cases, such as the large fusiform aneurysm seen in Case 2 (c.f. Fig.\,\ref{fig:OurThreeCases}, middle), we also model the treatment approach that combines coiling with stentgrafts, denoted as \textit{stent-assisted coiling}. In such cases, a \ac{FD} is placed in the parent vessel, while in addition the volume of the aneurysm is filled with (several) coils to achieve volumetric occlusion. The coil enhances the occlusion quality of the \ac{FD} and the \ac{FD} acts as a supporting scaffold that holds the coils in place, while preventing blockage of the parent vessel. This renders coiling feasible in Case 2 and contributes to the coil's positional stability for example also in Case 1 by avoiding protrusion into the parent vessel. Since our goal is to obtain realistic three-dimensional representations of the \ac{FD} and coils to be employed in subsequent thrombus-formation and occlusion analyses, the mechanical coiling model described in Sec.\,\ref{subsec:Coiling} is combined with the \ac{FD} model from Sec.\,\ref{subsec:FlowDiversion}.\\

\paragraph{Surrogate balloon}
The simplest  approach to incorporating stent-assisted coiling in the model would be to consider the respective stent-mesh (see Fig.\,\ref{fig:FlowDiverterImages}) as an additional contact obstacle in the coil-deployment simulation. This would extend the current framework, which already accounts for collisions with the aneurysm dome and other (parts of the) coils. While theoretically possible, the vertex- and face-counts of the stent meshes far exceed those of the vessel walls and coils, thus leading to an inordinate increase in simulation time. Furthermore, penetration of the \ac{FD} is feasible for some coil diameters and \ac{FD} ring spacings, leading to unstable or completely failed simulation runs. To alleviate these challenges, we take inspiration from clinical practice again, where coil migration from the aneurysm sac is a real risk that is mitigated by balloon-assisted device placement.\\

\textbf{- For coil deployments}, an inflatable balloon is advanced to the parent vessel together with the micro-catheter bearing the coiling wire. The balloon is inflated alongside the aneurysm, holding the coiling catheter in place and avoiding wire misguidance during the procedure. After deployment is complete, the balloon is deflated and retracted from the vessel completely. For reference, Fig.\,\ref{fig:MedicalImageOfCoilInsertion}, second from right, depicts a surgical balloon in action towards the end of a real procedure.\\[5mm]
\enlargethispage*{7mm}
\textbf{- For stent placement}, balloon-supported techniques also exist. The \textit{folded} stent is mounted on a deflated balloon catheter and is also inflated once at the correct position alongside the aneurysm. Inflating the balloon catheter unfolds the stent, anchoring it so that the balloon can be deflated and extracted with the catheter.\\

\begin{figure}[h]
    \centering
    \includegraphics[trim=0cm 5.5cm 5.2cm 0cm, clip, width=0.325\linewidth]{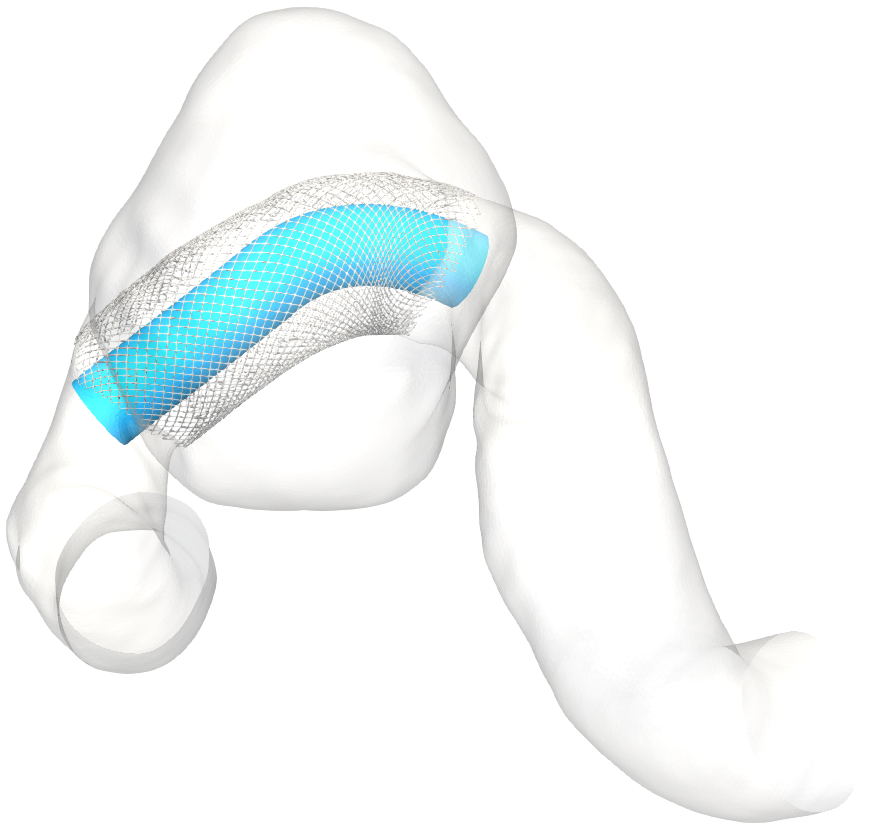}\includegraphics[trim=0cm 5.5cm 5.2cm 0cm, clip, width=0.325\linewidth]{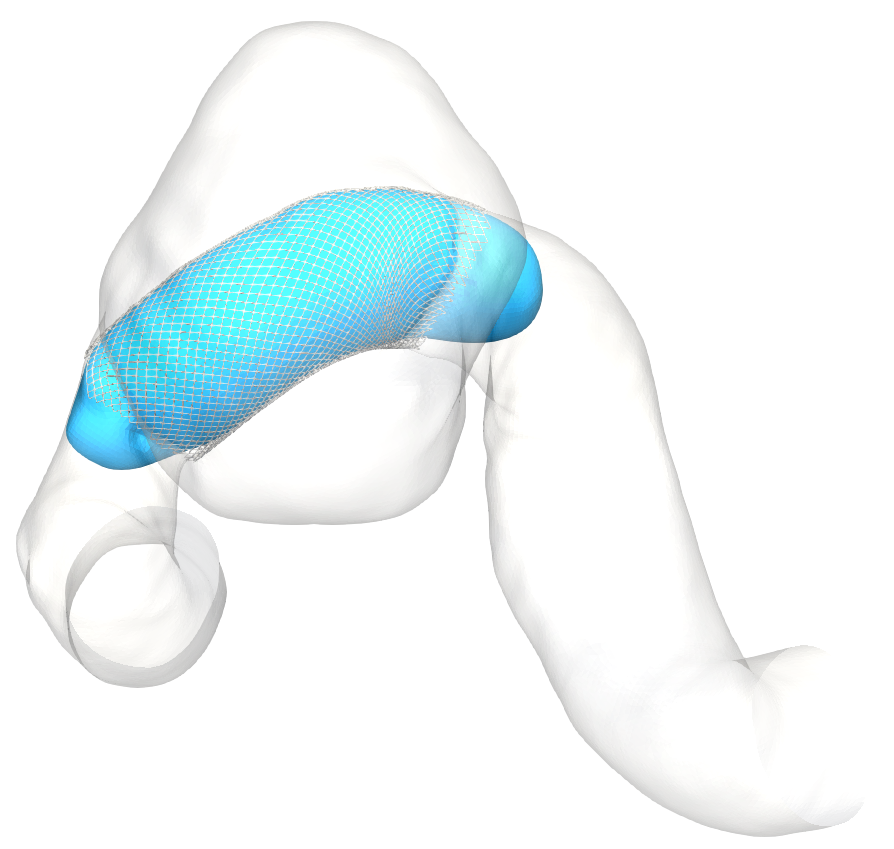}\hspace*{1cm}\includegraphics[trim=0cm 0cm 0cm 0cm, clip, width=0.25\linewidth]{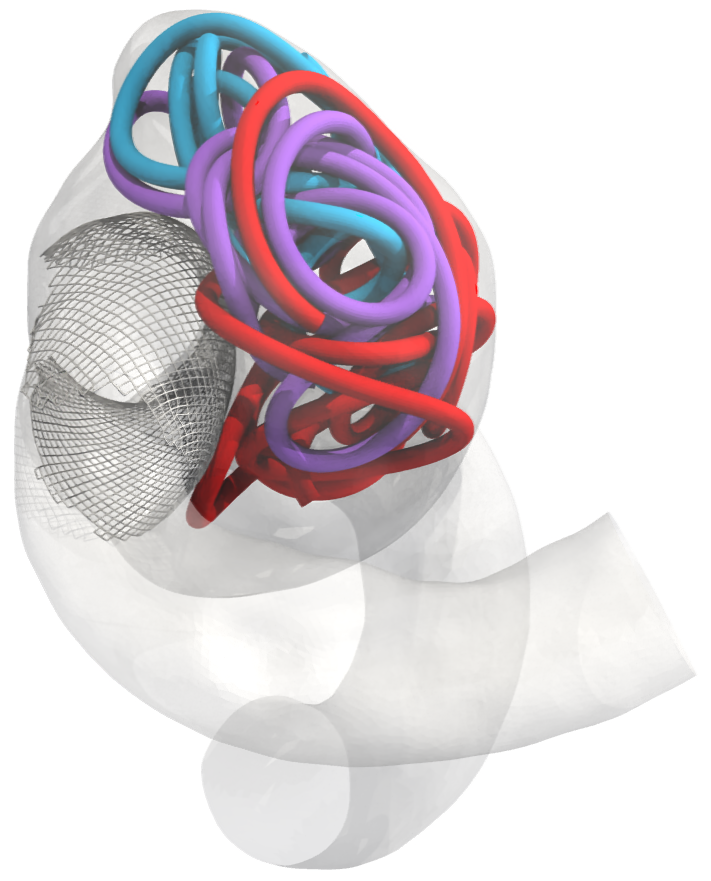}
    \caption{\footnotesize Stent-assisted coiling via balloon for Case 2. \textbf{(Left)} The initial configuration of the surrogate balloon before inflation. \textbf{(Middle)} The final shape of the inflated balloon aligning with the \ac{FD} mesh. \textbf{(Right)} the result of the subsequent stent-assisted coiling with three different coils (colors only for better distinction). The surrogate balloon is then \textit{removed} for the fluid- and thrombus growth simulations. Note that all three images show the same geometry, the last image is just shown in a rotated perspective to better see the parenting vessel cavity.}
    \label{fig:StentAssistedCoilingProcedure}
\end{figure}

In the current study, the \ac{FD} models from Fig.\,\ref{fig:FlowDiverterImages} are not ``foldable'', but are static and in their final configuration already. This allows for an inverse approach, where the balloon is selected to volumetrically fill out the final configuration of the \ac{FD}, then to be used as a coarse obstacle surrogate for the otherwise too finely resolved \ac{FD}. This is conducted using \textit{blender}\footnote{blender software webpage: \url{https://www.blender.org/}.} \cite{blenderItself}, a visualisation and rendering software that has cloth simulation\footnote{blender documentation on cloth simulation \url{https://docs.blender.org/manual/en/latest/physics/cloth/index.html}} \cite{provot1997collision, provot1995deformation} capabilities in its physics engine. Fig.\,\ref{fig:StentAssistedCoilingProcedure} shows an initial and final snapshot of the internal pressure-based surrogate balloon inflation exemplary for Case 2, starting from an approximate center-line cylinder of the \ac{FD}. Over time, the balloon achieves better alignment with the shape of the \ac{FD} and the parent vessel. The last image shows the final balloon configuration consisting of 4.354 vertices as compared to roughly 1.5 million vertices in the original \ac{FD}-mesh, making it sufficiently coarse for use as an obstacle in the coiling simulations. The mesh quality of the surrogate balloon produced by \textit{blender} during the inflation process can be sub-optimal, e.g., with distorted sliver elements. The mesh quality can be improved by applying a Laplace-Taubin smoother \cite{taubin1995signal} to the final balloon mesh as implemented in \textit{MeshLab}\footnote{MeshLab software webpage: \url{https://www.meshlab.net/.}} \cite{meshlab}. The last image in Fig.\,\ref{fig:StentAssistedCoilingProcedure} shows three subsequently inserted coils, which respect the cavity enclosed by the \ac{FD} after resubstitution of the balloon by the original \ac{FD} again.

%% file: ThrombusModel.tex
\section{Direct thrombosis model}\label{sec:BloodClotting}

A CFD-based, fibrin thrombosis model that accounts for the short timespan immediately after device placement is presented in this section. The model accounts for pulsatile flow, transport and reaction of biochemical species, and the impact of the growing clot on the flow field, as detailed in \cite{Hume2022, Jimoh-Taiwo2022, Ngoepe2016}. Although platelets are not directly modelled, the clot region assumes different porosity and permeability values based on experimentally-measured values \cite{diamond1993inner, diamond1999engineering}. In turn, the reduced flow in the clotted region affects the transport of clot proteins and their subsequent interaction. Further details, including considerations of validation and verification, are provided in the subsections that follow. 


\subsection{Navier--Stokes equations for a porous media}\label{subsec:Navier-Stokes}

Fluid flow is governed by Navier-Stokes equations for an isotropic porous medium [Equation 8.18–8.19 in \cite{ansys2025fluent}]
\begin{align} 
\frac{\partial(\phi\rho) }{\partial t}+ \nabla \cdot(\phi \rho  \mathbf{u})&= 0, & \text{ in } \Omega \times [0,T] \label{eq:Mass}\\
\frac{\partial( \phi \rho  \mathbf{u})}{\partial t} + \nabla\cdot(\phi \rho  \mathbf{u}\otimes \mathbf{u})   &=  - \phi\nabla p + \nabla \cdot\bigg(\phi\mu \bigg[\nabla \mathbf{u} + (\nabla\mathbf{u})^{\top}- \frac{2}{3}(\nabla\cdot \mathbf{u})\mathbf{I}\bigg]\bigg) - \frac{\mu}{\alpha} \mathbf{u} & \text{ in } \Omega \times [0,T] \label{eq:Momentum}
\end{align}

For these equations, $\mathbf{u}$ is the fluid velocity vector, $\rho$ is the fluid density, $\mu$ is the dynamic fluid viscosity, $t$ is time, $p$ is pressure and $\phi$ the volume fraction of the fluid in a porous region. Human blood is modeled as a Newtonian fluid with density $\rho=\SI{1e3}{\kilogram\per\meter\cubed}$ and viscosity 
$\mu=\SI{0.004}{\kilogram\per\meter\per\second}$. A porosity-based model for clot development is constituted in part via the addition of a Darcy Law, which includes a viscous resistance term $\mu/\alpha$ where \(\alpha\) is the porous medium's permeability being assumed to be isotropic. At arterial walls, a free-slip condition in $\mathbf{u}$ is adopted to account for unresolved near-wall slip-like effects observed in PIV velocity distributions for a fibrinogen-based fluid at a macrovascular scale in previous research \cite{hume2024computational}. Although cardiovascular CFD commonly employs no-slip wall conditions, recent in vivo observations of non-zero tangential velocities near arterial walls provide independent support for the broader premise that near-wall blood flow behavior may deviate from the classical no-slip approximation \cite{jarolimova2025}. The (spatially) averaged velocity, given by a time dependent Dirichlet boundary condition $u_\text{Inlet}(t)$, is applied at the inlet while at the outlet, a time dependent average pressure boundary condition $p_\text{Outlet}(t)$ is applied.

\begin{figure}[h]
    \centering
    \begin{subfigure}[b]{0.49\linewidth}
        \centering
        \includegraphics[width=\linewidth]{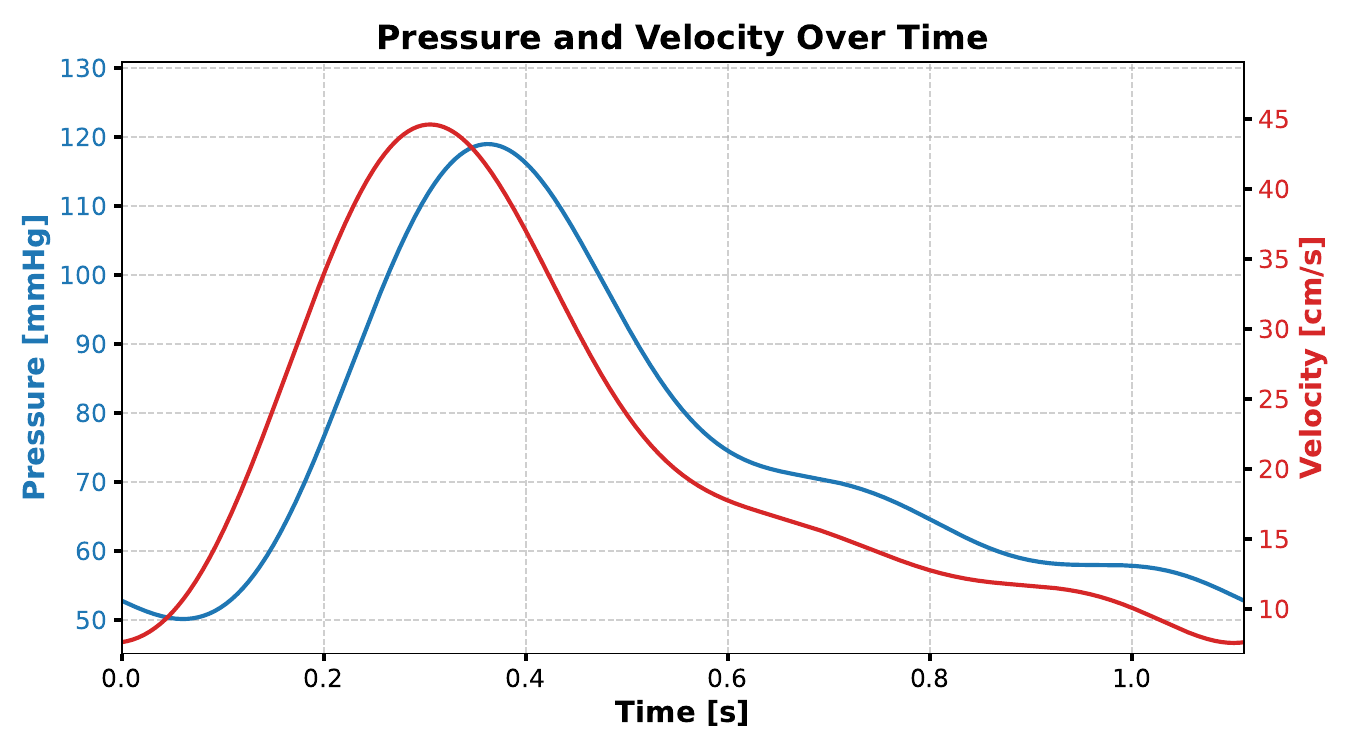}
        \caption{}
        \label{fig:fourier_plot}
    \end{subfigure}
    \hfill
    \begin{subfigure}[b]{0.49\linewidth}
        \centering
        \renewcommand{\arraystretch}{1.2}
        \begin{minipage}[b]{\linewidth}
            \centering
            \begin{tabular}{ c c c c c }
                \toprule
                Quantity & $k$ & $A_k$ & $f_k$ & $\phi_k$ \\
                \midrule
                \multirow{3}{*}{Pressure} 
                & 1 & $-30.44$ & $0.89$  & $-13.19$ \\
                & 2 & $2.96$   & $-1.93$ & $0.50$ \\
                & 3 & $1.39$   & $15.86$ & $-2.01$ \\
                \midrule
                \multirow{3}{*}{Velocity} 
                & 1 & $-15.49$ & $0.89$  & $-12.84$ \\
                & 2 & $9.42$   & $-1.78$ & $-1.61$ \\
                & 3 & $0.27$   & $18.44$ & $-1.44$ \\
                \bottomrule
            \end{tabular}
            \vspace{0.5em}
        \end{minipage}
        \caption{}
        \label{tab:fourier_coeffs_inline}
    \end{subfigure}
    \caption{
        Truncated Fourier approximation of the inlet velocity and outlet pressure profiles. 
        Pressure offset: $75.07$, Velocity offset: $19.59$, Period: $T = \frac{10}{9} \approx 1.111$ s (one heart-beat).
    }
    \label{fig:fourier_summary}
\end{figure}

The pressure and velocity profiles $p_\text{Outlet}(t)$, $u_\text{Inlet}(t)$  are approximated using truncated Fourier series of the form
\begin{equation} \label{eq:fourier_general}
f(t) = \sum_{k=1}^{3} A_k \sin\left( 2\pi f_k t + \phi_k \right) + \text{offset},
\end{equation}
where the signals can be seen in figure \ref{fig:fourier_plot} and the coefficients are given in the table in figure \ref{tab:fourier_coeffs_inline}. Details about the signals in equation \ref{eq:fourier_general} are provided in \cite{Ferns2010} and \cite{Hume2022}.

\captionsetup[subfigure]{labelformat=parens, labelsep=space, font=large}


\subsection{Biochemistry}\label{subsec:Biochemistry}
    The transport equation accounts for convection and diffusion of biochemical proteins that contribute to the clotting process, and is detailed in the transport model for a passive scalar in the presence of a porous medium [Equation 8.17 in \cite{ansys2025fluent}],
\begin{equation} \label{eq:Transport}
\frac{\partial (\phi\mathbf{c})}{\partial t} + \nabla\cdot(\phi\mathbf{u} \otimes \mathbf{c}) = \frac{1}{\rho}\nabla\cdot (\phi\mathbf{\Gamma} \nabla \mathbf{c}) + \mathbf{s}
\end{equation}
where $\mathbf{c}:=(c_\text{IIa}, c_\text{Ia}, c_\text{I})^{\top}$ are species concentrations, $\mathbf{\Gamma}=\operatorname{diag}(\Gamma_\text{IIa}, \Gamma_\text{Ia}, \Gamma_\text{I})$ is the diagonal matrix of effective diffusive conductivity coefficients of the respective species and $\mathbf{s} =(s_\text{IIa}, s_\text{Ia}, s_\text{I})^{\top}$ is the source term vector of the species. Three species, namely thrombin $c_\text{IIa}$, fibrinogen $c_\text{I}$ and fibrin $c_\text{Ia}$, are represented as scalar quantities. The coefficients in $\mathbf{\Gamma}$ for each scalar are shown  table \ref{tab:ScalarTable} and are based on the diffusion of each respective protein in vivo. For the boundary conditions, we apply a zero normal gradient boundary condition at the walls and outlet for each species. At the inlet we apply Dirichlet boundary conditions where fibrin and thrombin are set to zero and fibrinogen enters with a constant value of \SI{7000}{\nano\Molar}. We choose the initial conditions for each species such that they match up their corresponding inlet condition. 
\begin{table}[h]
\centering
\sisetup{per-mode = fraction}

\begin{tabular}{l c}
\toprule
Scalar & Effective Diffusive Conductivity 
\;($10^{-8}$\,\si{\kilogram\per\metre\per\second}) \\
\midrule
Thrombin (IIa)  & 6.79 \\
Fibrinogen (I)  & 3.25 \\
Fibrin (Ia)     & 2.59 \\
\bottomrule
\end{tabular}

\caption{Effective Diffusive Conductivity of scalar species.}
\label{tab:ScalarTable}
\end{table}

The biochemical reactions between fibrinogen (I) and thrombin (IIa) enable the formation of fibrin (Ia). This process can be modeled by a reduced Michaelis-Menten formulation in equation \eqref{eq:MichaelisMenten} \cite{Jimoh-Taiwo2022, Kremers2017}, 
\begin{equation} \label{eq:MichaelisMenten}
\text{Thrombin (IIa)} + \text{Fibrinogen (I)} \rightarrow \text{Fibrin (Ia)} \hspace*{5mm}\text{with} \hspace*{5mm} -s_\text{I} = s_\text{Ia} = \frac{k_\text{cat} c_\text{IIa}c_\text{I}}{K_m + c_\text{I}} \hspace*{5mm} \text{\&}\hspace*{5mm} s_{\text{IIa}}=0
\end{equation}
Where $k_\text{cat} c_\text{IIa}$ is the maximum reaction rate achieved by the system, $c_\text{I}$ is the concentration of the substrate (here fibrinogen), $k_\text{cat}=\SI{3540}{\per\minute}$  is the catalytic constant, and $K_\text{m}=\SI{3160}{\nano\Molar}$ is the Michaelis constant. 

\subsection{Porosity-based clotting model}\label{subsec:Porosity-Based Clotting Model}

    The porosity function alters porosity on a linear scale from 1 (no porosity) to 0.75 within computational cells where the fibrin concentration scalar is present and shear rate  is less than $\gamma^*:=\SI{100}{\per\second}$ relative to the fibrin threshold of $s_{I}^*:=\SI{600}{\nano\Molar}$.
    \begin{equation} \label{eq:Porosity}
    \phi = \begin{cases}
    \operatorname{max}\{\frac{3}{4}, 1-\frac{1}{4}\frac{s_{\text{I}}}{s_{\text{I}}^*}\} & \text{ if } \dot{\gamma}\leq \dot{\gamma}^*\\
    1 & \text{ else}
    \end{cases}
    \end{equation}
    In areas where a porosity of 0.75 is achieved, the viscous resistance  of \SI{1e12}{} changes to \SI{1e-12}{} as shown in equation \eqref{eq:ViscousResistance}.

\begin{equation} \label{eq:ViscousResistance}
\alpha^{-1} =
\begin{cases} 
10^{-12}, & \text{if } s_{\text{I}}\geq s_{\text{I}}* \text{ and } \dot{\gamma}\leq \dot{\gamma}^* \\ 
10^{12}, & \text{else}
\end{cases}
\end{equation}

Although \eqref{eq:Transport} is advected by the flow field from \eqref{eq:Momentum}, the system is two-way coupled through thrombus growth. Updates to the porosity $\phi$ and permeability $\alpha$ feed back into \eqref{eq:Momentum} most directly via the Darcy resistance term, and more generally through the $\phi$-weighted mass and momentum terms.


\subsection{Aneurysm wall thrombin release model}\label{subsec:Thrombin Release Model}

Determining the parameters for clot initiation in cerebral aneurysms presents a number of challenges. Under physiological conditions, clotting typically occurs following injury to the vessel wall. In cerebral aneurysms, there is no explicit injury site but the entire wall is weakened. Once clotting is initiated, a complex series of reactions take place to result in the eventual formation of a fibrin clot. While many descriptions of these reactions exist, the complexity of these systems makes it challenging to calibrate them on a per patient basis \cite{Ngoepe2016}. In this study, device placement is the trigger for clot formation in the aneurysmal sac, as it reduces the flow sufficiently to support adequate reaction of coagulation proteins for the formation of a thrombus.  Rather than modelling a complex network of reactions, the entire aneurysm wall is assumed to express thrombin, which enables the formation of fibrin as per equation \eqref{eq:MichaelisMenten} \cite{Jimoh-Taiwo2022}. As shown in \eqref{eq:Porosity} and \eqref{eq:ViscousResistance}, only areas which meet both biochemical and mechanical conditions can clot. 

\captionsetup[subfigure]{labelformat=parens, labelsep=space, font=large}
\begin{figure}[h]
    \centering
    \begin{subfigure}[b]{0.49\linewidth}
        \centering
        \includegraphics[width=\linewidth]{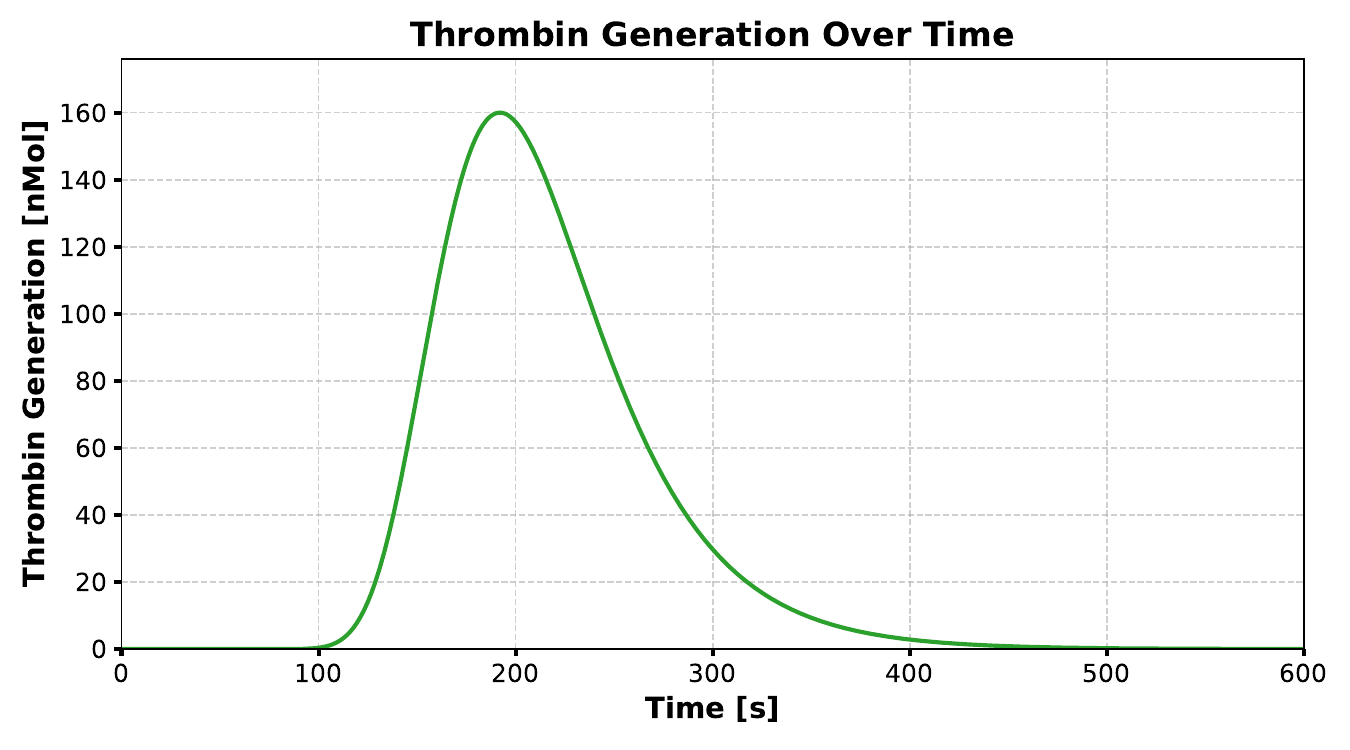}
        \caption{}
        \label{fig:releasefunction}
    \end{subfigure}
    \hfill
    \begin{subfigure}[b]{0.49\linewidth}
        \centering
        \renewcommand{\arraystretch}{1.2}
        \begin{minipage}[b]{\linewidth}
            \centering
            \begin{tabular}{ c c c c }
                \toprule
                Quantity & & & \\
                \midrule
                Peak Thrombin Value & $\frac{\beta e}{A}$ & $160$  & nMol \\
                Total Thrombin & $A$   & $300$ & nMolmin \\
                Time to peak & $t_p$   & $3.2$ & min \\
                \bottomrule
            \end{tabular}
            \vspace{0.5em}
        \end{minipage}
        \caption{}
        \label{tab:releasefunction_coeffs}
    \end{subfigure}
    \caption{
        \textbf{(Left)} Aneurysm wall thrombin release function $g_{\text{IIa}}(t)$ (see equation \eqref{eq:FunctionF}) over time, \textbf{(right)} the respective parameter values for the model-function.
    }
    \label{fig:Aneurysm wall thrombin release function vs time}
\end{figure}

To account for the expression of thrombin from the aneurysm wall, a thrombin release function, shown against time in Figure \ref{fig:Aneurysm wall thrombin release function vs time}, is applied at the aneurysm wall for all cases. The variable thrombin release model is derived from a thrombin generation curve fitting function developed by  \cite{Kremers2017} and applied in \cite{Hume2022}. The model is derived from thrombin generation curves derived from a static assay, which deviates from in vivo conditions. The main strength of this thrombin concentration function is the possibility of defining a patient-specific biochemical profile with just three measurable variables  \cite{Kremers2017}. While thrombin's interactions with other species are accounted for by the Michaelis-Menten model from equation \eqref{eq:MichaelisMenten}, its own concentration is defined by the function

\begin{align} \label{eq:FunctionF}
g_\text{IIa}(t) &= \frac{A}{\beta} \exp\left( \frac{t - t_p}{\beta} - \exp\left( \frac{t - t_p}{\beta} \right) \right),
\end{align}

where its coefficients are given in table \ref{tab:releasefunction_coeffs}, and while a resulting plot is shown in \ref{fig:Aneurysm wall thrombin release function vs time}.

Thrombin release is then modelled by a Neumann boundary condition at the aneurysm wall \cite{Kremers2017}:
\begin{equation} \label{eq:Flux}
\Gamma_\text{IIa}\nabla c_\text{IIa}\cdot \mathbf{n} = g_\text{IIa}(t),
\end{equation}
where $\mathbf{n}$ is the wall normal. \\

Subsections \ref{subsec:Navier-Stokes}-\ref{subsec:Thrombin Release Model} comprise the mathematical model components used in the simulation. While it is possible to verify aspects of the model, and significantly more complex models could be employed in a modular way switching out certain model components, the biochemical and haemodynamic complexities of in vivo and in vitro clot formation pose challenges for quantification of the entire system and therefore full validation remains an open challenge. Nevertheless, various approaches have been used to verify aspects of the computational thrombosis model presented in this work. Particle image velocimetry (PIV) was used to quantify the flow field in a 3D printed idealised aneurysm geometry and this data was used to validate the computational flow field \cite{Ho2019, Hume2022}. An experimental clot growth study, conducted using human-derived thrombin and fibrinogen, enabled quantification of occlusion outcome in an idealised 3D printed geometry \cite{Ngoepe2021}. This data was used to verify occlusion outcome in an idealised 2D aneurysm geometry \cite{Ngwenya2024}. This framework is now applied to the current study, which also considers occlusion outcome in realistic 3D aneurysm geometries.

%% file: OcclusionQuality.tex
\section{Occlusion quality rating}\label{sec:OcclusionQuality}

For endovascular treatment of aneurysms, the extent or quality of occlusion plays a crucial role in judging the success of the intervention. In endovascular coiling, for example, an aneurysm volume occlusion-ratio of 30-40\,\% is desirable for substantial reduction of flow-velocities and perfusion of the aneurysm, featuring an environment suitable for thrombus formation. Occlusion then further improves as the thrombus grows and (ideally) evenutally supports endothelial regrowth at the ostium and complete encapsulation of the coiled aneurysm \cite{marbacher2019recurrence}. Insufficient occlusion allows for residual flow into the aneurysmal sac, hindering the healing process. In situations with a gap at the aneurysm neck, insufficient coiling can even support an inflow jet of comparably high velocities and stresses on the aneurysm neck wall, triggering enlargement of the aneurysm sac and requiring reintervention. In the following sections, we introduce our numerical in-silico methods to assess the occlusion quality of device-treated aneurysms with and without thrombosis.


\subsection{Occlusion classification}\label{subsec:GeometricOcclusionAnalysis}
For the classification of coiling-occlusion and -success, the \ac{RRC} provides well established criteria \cite{Mascitelli496}. It defines four typical cases:\\

\begin{itemize}
    \item \textbf{Class I}: optimal packing
    \item \textbf{Class II}: insufficient packing at the \textit{neck} of the aneurysm
    \item \textbf{Class IIIa}: insufficient packing in the \textit{core-region} of the aneurysm 
    \item \textbf{Class IIIb}: insufficient packing close to the \textit{boundary} of the aneurysm 
\end{itemize}~\\
Class I is the most desirable outcome, while the other classes are potential \textit{candidates} for aneurysm regrowth due to insufficient occlusion and residual blood-flow into the weakened wall region. We emphasize the term ``\textit{candidates}'', as the actual outcome of an aneurysm occlusion procedure can depend on more factors than just the local packing density of the inserted device alone. These might include the angle of attack of blood-flow at the ostium or the general width of the aneurysm neck. In clinical practice, aneurysm occlusion analysis can be conducted by catheter-based injection of a tracer fluid into the parent vessel upstream of the sac, observing its subsequent distribution using medical imaging. The extent of occlusion is judged by the amount of tracer carried into the aneurysm by residual blood-flow, its residential volume and subsequent washout over a sequence of subsequent heart-beat cycles. Virtual simulation of such a tracer perfusion analysis is detailed in the upcoming section. This will allow to also compare occlusion outcomes purely based on the inserted device with such also taking into account the formed early thrombus.


\subsection{Virtual contrast agent analysis}\label{subsec:VirtualContrastAgent}
Our goal is to perform a virtual angiography that allows a \ac{RRC}. In clinical aneurysm angiography, iodinated contrast agents such as Iodixanol (Visipaque) are commonly used as tracers \cite{treweeke2016iodixanol, qureshi2017upright}. To this end, we (virtually) inject a contrast agent at the inflow boundary of the parent vessel's in-silico model and simulate the transport of the contrast agent's concentration $C:(0,T)\times \Omega \rightarrow [0,1]$ by means of an \ac{ADE}.

\begin{align}
    \label{eq:TracerTransportEquation1}
    \frac{\partial C(t,\vec{x})}{\partial t}+\nabla\cdot (C(t,\vec{x})\vec{u}(t,\vec{x})) - D_C\,\Delta C(t,\vec{x}) &= 0  && \qquad (t,\vec{x}) \in (0,T)\times\Omega
\end{align}

where the tracer concentration is switched on at $T_\text{inj}=1s$,  modelling injection of the tracer at the inlet of the flow domain. Note that the model assumes that the flow velocity $\vec{u}(t,\vec{x})$ is provided by the incompressible Navier--Stokes equation and $C(t,\vec{x})$ does not influence $\vec{u}(t,\vec{x})$. 
Initially, the domain is free of contrast agent, i.e.\ $C(0,\vec{x})=0$. At the inflow boundary $\Gamma_{\textup{in}}$, a Dirichlet boundary condition models the injection of contrast agent. After an initial delay of one heartbeat ($T_\text{inj}=1\,\text{s}$), the tracer concentration is set to $C=1$ for two seconds and to $C=0$ otherwise.
At the vessel walls $\Gamma_{\textup{wall}}$ and the outlet boundary $\Gamma_{\textup{out}}$, homogeneous Neumann boundary conditions are imposed. At the walls this reflects the no-penetration condition of the flow field, while at the outlet the tracer leaves the domain through advection with the velocity field.

\subsection{Numerical treatment of the ADE}
While fluid-flow is simulated by solving Navier--Stokes equations with a multi relaxation time Lattice Boltzmann method \cite{kruger_lattice_2017} first, for the subsequent simulation of tracer-transport in the blood vessel with \textit{given} flow-field $\vec{u}(t,\vec{x})$, we propose a finite-volume method for the advection- and diffusion-operators in equation (\ref{eq:TracerTransportEquation1}). Following the approach in \cite{hundsdorfer_numerical_nodate}, we construct a total variation diminishing (TVD) finite-volume scheme on a Cartesian grid.Time integration is performed using the second-order explicit trapezoidal rule. The diffusion term is discretized with a second-order finite-volume method, which reduces to the standard central difference scheme on a Cartesian grid. The advection term is approximated by a flux-limited upwind scheme. It is first-order accurate near sharp concentration fronts and achieves third-order accuracy in sufficiently smooth regions. As flux limiter, we employ the Koren limiter \cite{koren1993robust}.
Since time integration is explicit, the time step is restricted by the CFL condition.
The computational domain is represented directly on the Cartesian grid. Curved boundaries are approximated by a grid-aligned staircase representation. 
This is justified since the passive scalar is computed in postprocessing on a precomputed velocity field, and the washout assessment (in the spirit of the Raymond--Roy classification) is based on macroscopic contrast clearance within the aneurysm sac rather than on near-wall concentration gradients.

\subsection{Projection-based digital subtraction angiography}
\begin{figure}[h!]
    \begin{center}
        \includegraphics[height=0.3\linewidth, angle=0]{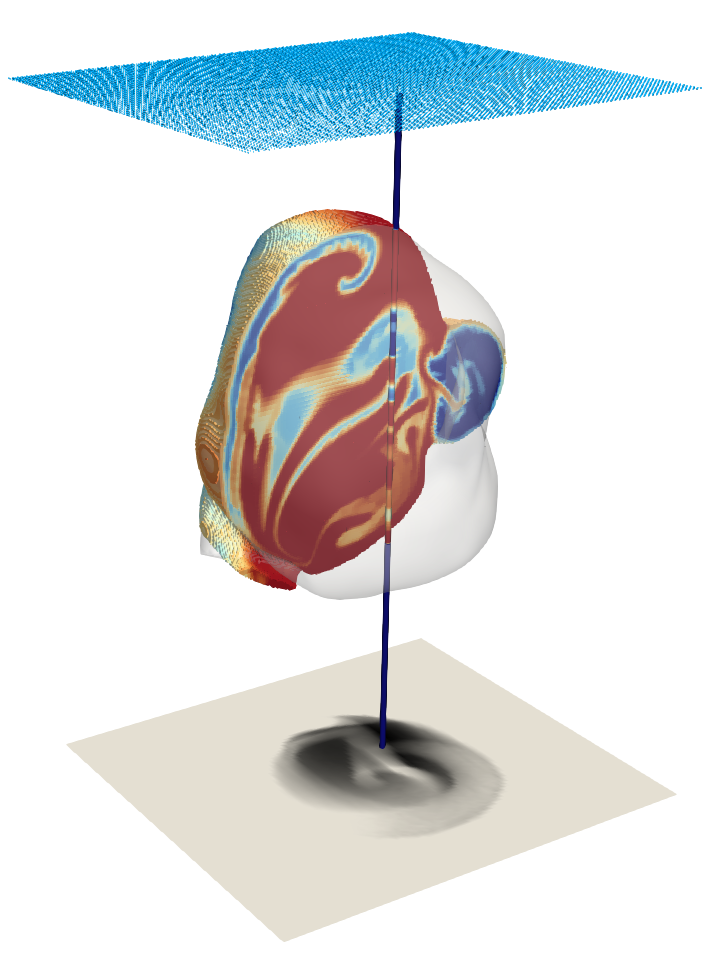}\hspace{1.5cm}\raisebox{2mm}{\includegraphics[height=0.275\linewidth]{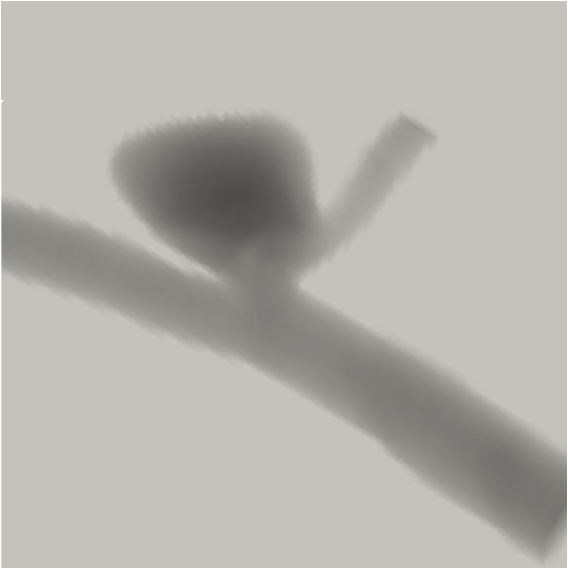}}
    \caption{\textbf{(Left)} Visualization of the virtual \ac{DSA} ``measurement'' method sending rays from the emitter plane (blue dots on the left) through the tracer concentration field, summing up concentration values via path-integrals and projecting the concentration onto the receiver plate (right). \textbf{(Right)} Virtual angiography obtained from the simulation. 
The geometry corresponds to aneurysm E of the MATCH case reported in \cite[Fig.~1]{berg2018multiple}. 
A qualitative comparison with the angiographic image shown therein indicates similar contrast filling behavior.  The proportionality factor $k$ (see equation \eqref{eq:DSAFactor}) was chosen to be 0.6 in order to scale the grey-level contrast to match the data.}
    \label{fig:VirtualDSA}
    \end{center}
\end{figure}

After the tracer flow-field is obtained, a \ac{DSA} is virtually emulated. In this imaging technique, X-rays are emitted by a source (emitter) and captured by a sensor (receiver). The emitter- and receiver-plane are parallel aligned allowing for the patient to be positioned in between them in order to be investigated. For our virtual angiography, we assume that the contrast agent acts as an absorbing medium, reducing the intensity of the emitted X-Ray beam. A simplified model for intensity reduction is given by the Beer-Lambert law
\[
I = I_0 e^{- \int_{\textup{ray}}\lambda(s)\,\textup{d}s},
\]
describing the reduction of the initial intensity $I_0$ down to $I$ when passing through a medium of optical density $\lambda(s)$, where the integral is carried out over the straight ray path. When the rays are captured by the receiver they cast a reduced intensity shadow visualizing the (residual) flow within the aneurysm and vasculature. In a \ac{DSA}, the intensity is evaluated twice; once before addition of contrast agent, $I_1$, and once after contrast agent injection, $I_2$. We assume that the \textit{tissue} has an optical density of $\lambda_t(s)$, which then changes to $\lambda_t(s) + kC(s)$ after adding the contrast agent, where $C(s)$ is the contrast agent concentration and $k$ is a proportionality constant. To improve visibility of the contrast agent enhanced regions, both intensities are subtracted, $\ln(I_2)-\ln(I_1)$, effectively removing the surrounding tissue from the image \cite{Haupert2007}. Mathematically this yields the expression
\begin{equation}
\label{eq:DSAFactor}
R = -(\ln(I_2)-\ln(I_1))= -\ln\left(\frac{I_2}{I_1}\right) = -\ln\left(\frac{I_0 e^{-\int_{\textup{ray}}\lambda_t(s) + kC(s)\, \textup{d}s}}{I_0 e^{-\int_{\textup{ray}}\lambda_t(s)\, \textup{d}s}} \right) = \int_{\textup{ray}}k C(s)\, \textup{d}s,
\end{equation}
indicating that to evaluate the signal $R$ \textit{virtually} from our simulation results, it is sufficient to evaluate the path-integral of the concentration $C$ over the length of the ray. Our algorithm, implemented as a simulation post-process step in the software ParaView\footnote{ParaView software webpage: \url{https://www.paraview.org/}} \cite{Ahrens2005ParaView, Ayachit2015ParaViewGuide}, for performing the virtual \ac{DSA} then proceeds in the following way, where a respective image visualizing the procedure is shown in Fig.\,\ref{fig:VirtualDSA} on the left.\\

To obtain qualitatively comparable contrast distributions, the proportionality factor $k$ (see \eqref{eq:DSAFactor}) was chosen based on a visual comparison with the angiographic appearance of \textit{case E} reported in \cite[Fig.~4]{berg2018multiple}. The calibrated value was $k = 0.6$.\\

\begin{enumerate}
    \item We construct two parallel planes in space that are separated by a distance $L$, such that the aneurysm is located between them.\\[2mm]
    \item We then define a set of rays starting orthogonal from one of the planes (the emitter) directed towards the other (the receiver).\\[2mm]
    \item We evaluate the path-integral along each of the rays. This then allows us to construct the projected field $R$ which is then defined on the sensor-plane.
\end{enumerate}~\\

%% file: NumericalExperiments.tex
\section{Numerical experiments}\label{sec:NumericalExperiments}

The aforementioned methods are applied to the three aneurysm cases introduced in Fig.\,\ref{fig:OurThreeCases}. For each case, an empty, pre-operative aneurysm is used as a reference, which is then compared to results for appropriate treatment methods (coiling, \ac{FD}, stent-assisted coiling). We also include a \textit{failed} coil placement for Case 1, where parts of the coil protrude into the parent vessel to assess the risk of intraluminal thrombus formation for the protruding wires. First, we present the thrombus formation results for each case by comparing occlusion outcomes for respective devices.  The virtual \ac{DSA} based tracer injection, aneurysm perfusion and washout analysis is then presented in subsection \ref{subsec:TracerResults} for all three cases.


\subsection{Case 1 - Small saccular aneurysm}\label{subsec:Case1}
\begin{figure}[h!]
    \centering
    \includegraphics[width=0.25\linewidth]{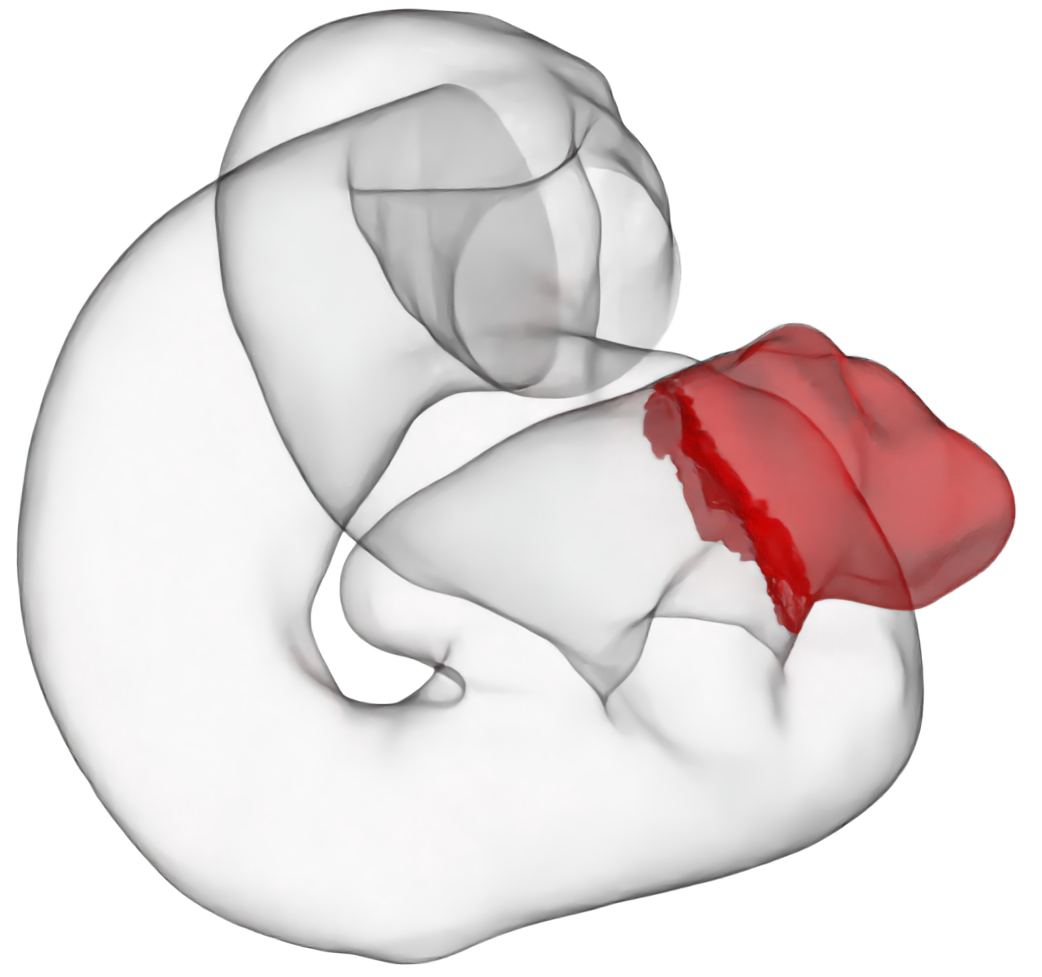}\includegraphics[width=0.25\linewidth]{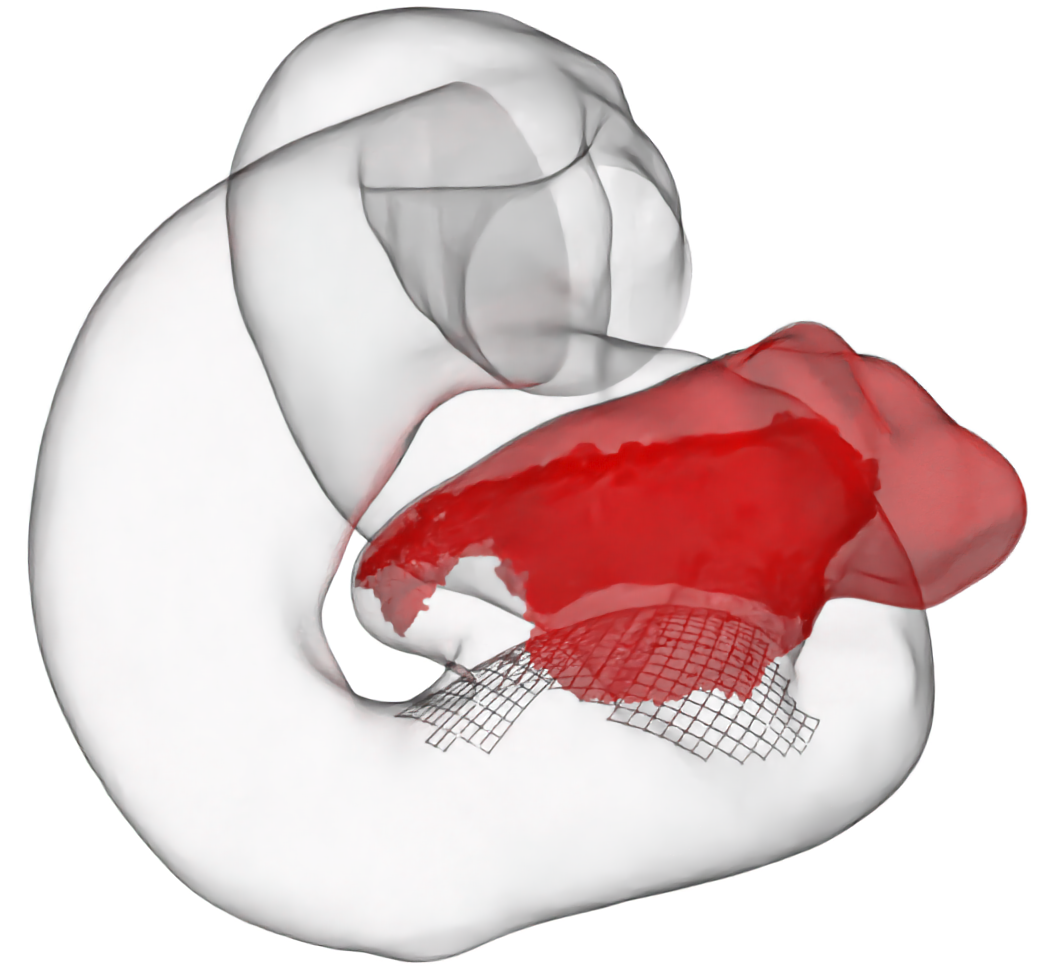}\includegraphics[width=0.25\linewidth]{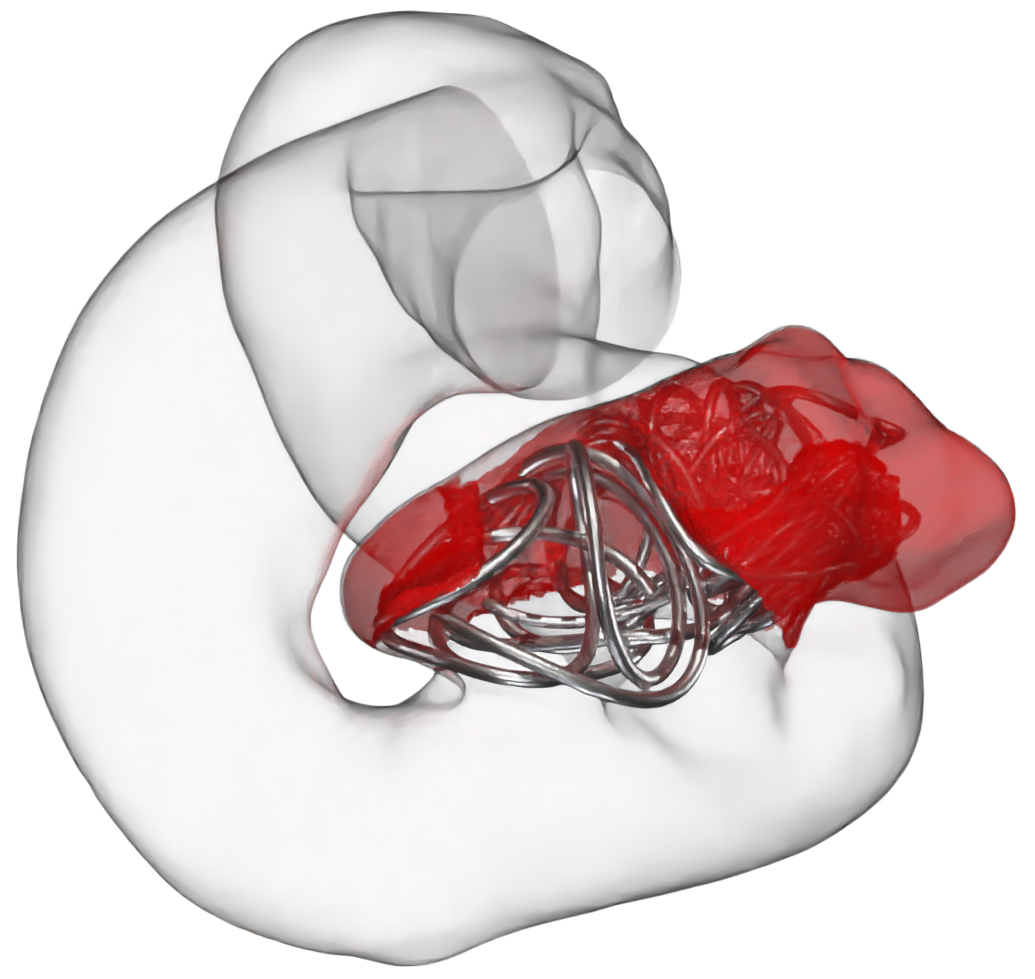}
    \caption{\footnotesize Simulated final thrombi (in red) in different treatment scenarios. \textbf{(Left)} empty, untreated aneurysm (for reference), \textbf{(Middle)} \ac{FD} treated aneurysm with the thrombus mainly attaching to the aneurysm walls, leaving a considerable void volume inside, \textbf{(Right)} Coiling-based treated aneurysm with thrombus growing ``along'' the coiling wire.}
    \label{fig:Case1_Thrombi}
\end{figure}

Case 1 represents a relatively small saccular \textit{non}-bifurcation aneurysm, where any of the three treatment approaches would be appropriate. We therefore use this case to compare pure coiling treatment with pure \ac{FD} placement. The coil-insertion procedure is simulated via the placement model from Sec.\,\ref{sec:EndovascularDevices},\ref{subsec:Coiling} (see again Fig.\,\ref{fig:OurThreeCases}, left, for the final coil geometry used here) while the \ac{FD} mesh is the one from Fig.\,\ref{fig:FlowDiverterImages}, right. For both cases, the resulting thrombi are depicted in Fig.\,\ref{fig:Case1_Thrombi} alongside the reference thrombus formed in the empty (untreated) aneurysm.

The thrombus results for this case enable direct \textbf{comparison of the treatment methods} with respect to the resulting thrombus occlusion volume, yielding the occlusion volumes and ratios reported in table \ref{tab:OcclusionVoluminaCase1}. From these metrics, it is clear that both treatments outperform the empty aneurysm reference case, as expected, and that coiling, in this case, presents an additional advantage over the \ac{FD} treatment. 

\begin{table}[h!]
	\small
    \centering
    \begin{tabular}{|c|c|c|c||c|}
    \cline{2-4}
         \multicolumn{1}{c|}{}& \multicolumn{3}{c||}{Treatment case} &  \multicolumn{1}{c}{}\\ \cline{5-5}
         \multicolumn{1}{c|}{}& \textbf{Empty} & \textbf{Stented}  & \textbf{Coiled} & \textbf{Coil volume} \\ \hline
         \textbf{Thrombus volume} & \SI{31.27}{mm^3} & \SI{35.69}{mm^3} & \SI{36.57}{mm^3} & \SI{4.36}{mm^3} \\\hline
         \textbf{Aneurysm volume} & \multicolumn{4}{c|}{\SI{71.05}{mm^3}} \\ \hline
         \multirow{2}{*}{\textbf{Percentage occlusion}} & 44.01\,\% & 50.23\,\% & 51.47\,\% & 6.14\,\% \\ \cline{2-5}
         & \multicolumn{2}{c|}{--} & \multicolumn{2}{c|}{\hspace{-0.7cm}$\Sigma$\hspace{0.7cm} 57.61\,\%} \\ \cline{1-5}
    \end{tabular}
    \caption{\textit{Purely geometric} occlusion ratio comparison of the different treatment approaches' resulting thrombi shown in Fig\,\ref{fig:Case1_Thrombi}. In the coiling case, both, the volume of the thrombus as well as the coil itself do contribute to occlusion, hence the additional computed sum.}
    \label{tab:OcclusionVoluminaCase1}
\end{table}

What \textit{cannot} be inferred from the occlusion metrics, which present a volumetric perspective, is a \textbf{thrombus shape / influence analysis}. The tracer simulation based on virtual \ac{DSA} from Sec.\,\ref{sec:OcclusionQuality} presents such an analysis in Fig.\,\ref{fig:DSA_Case1} in the latter part of this section.

\begin{figure}[h!]
    \centering
    \includegraphics[width=0.25\linewidth]{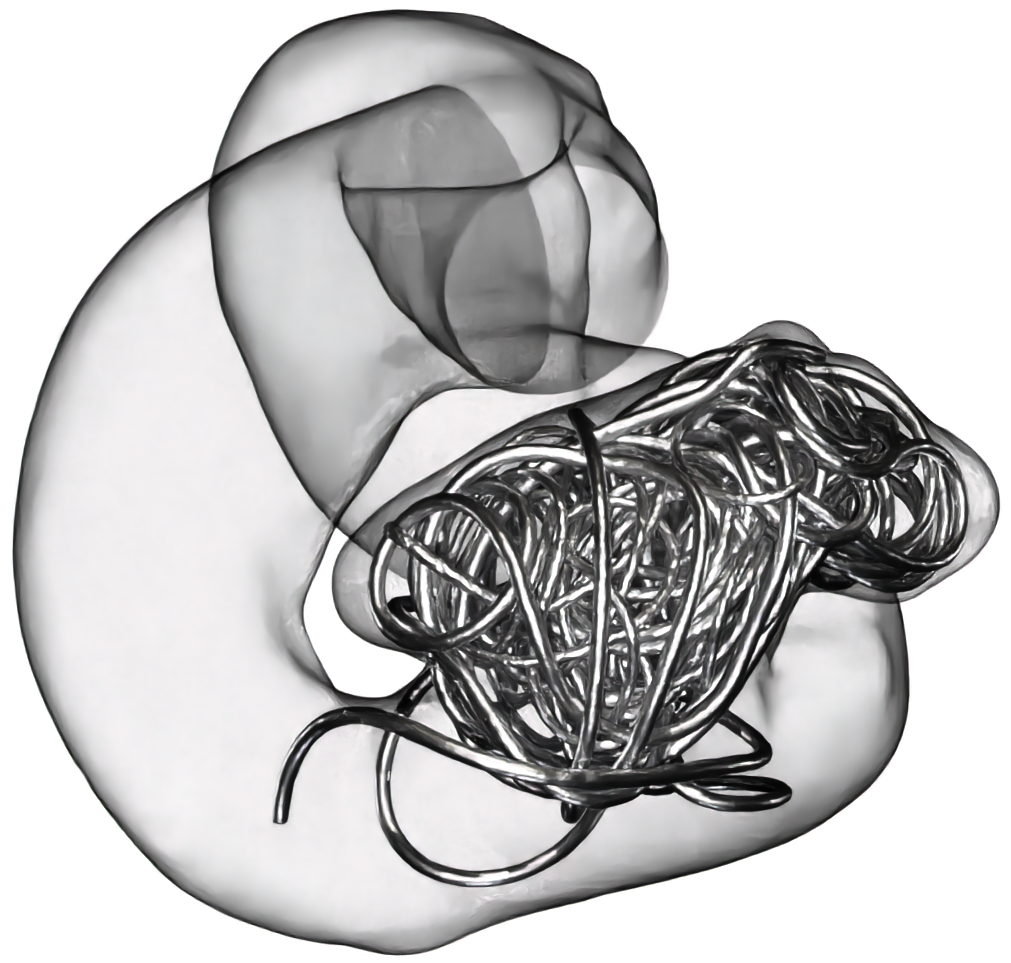}\includegraphics[width=0.25\linewidth]{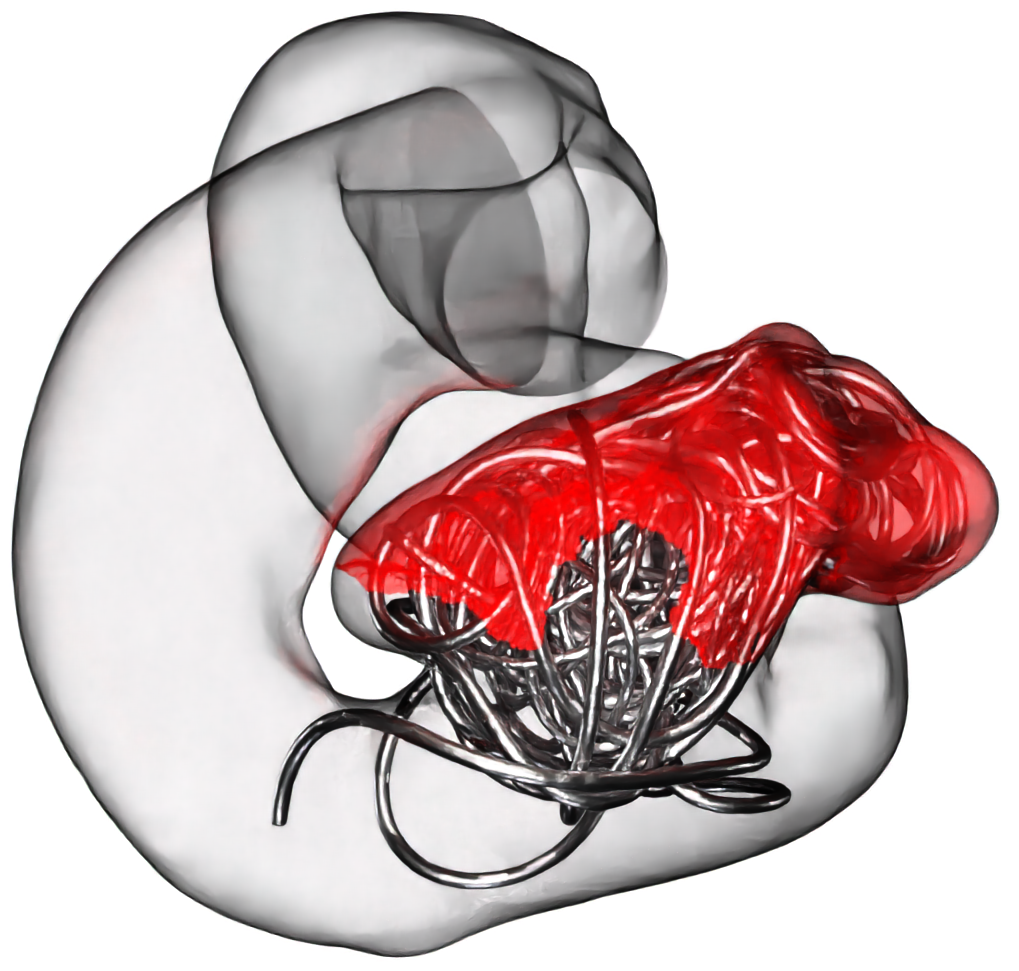}\includegraphics[width=0.25\linewidth]{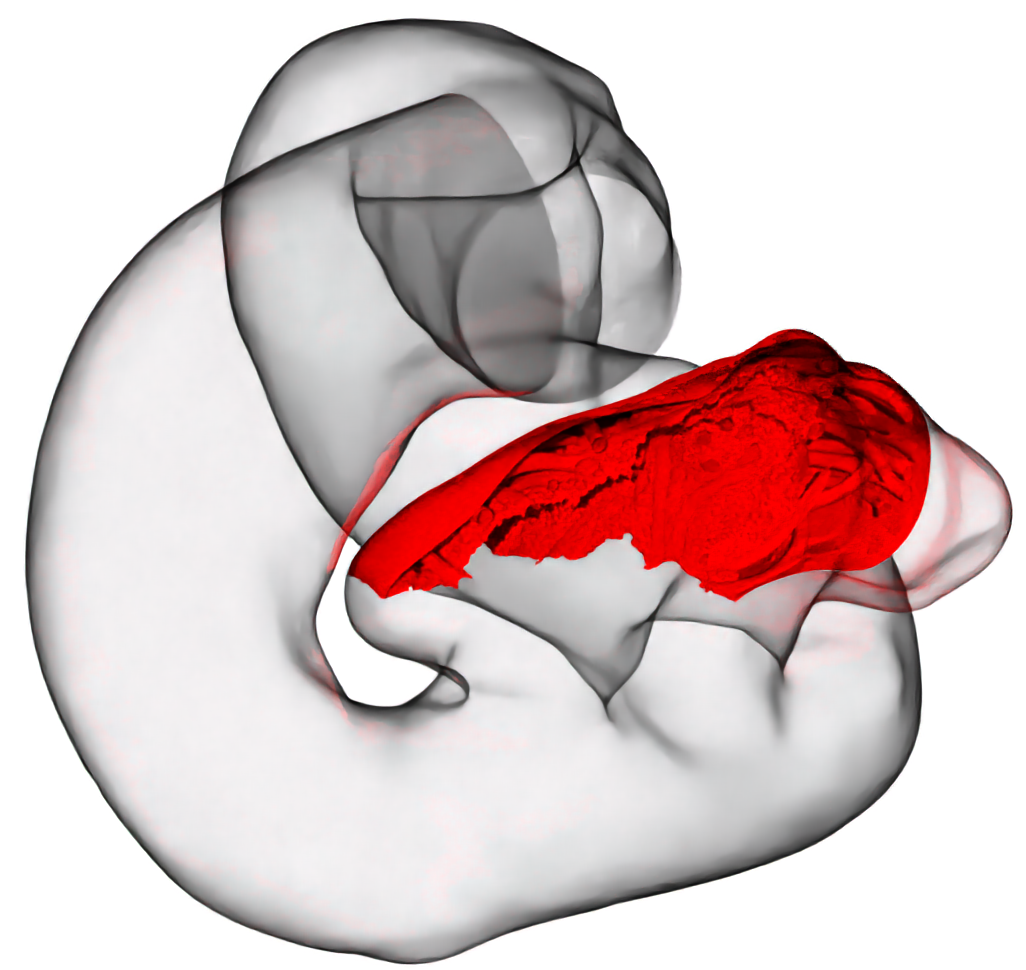}\includegraphics[width=0.25\linewidth]{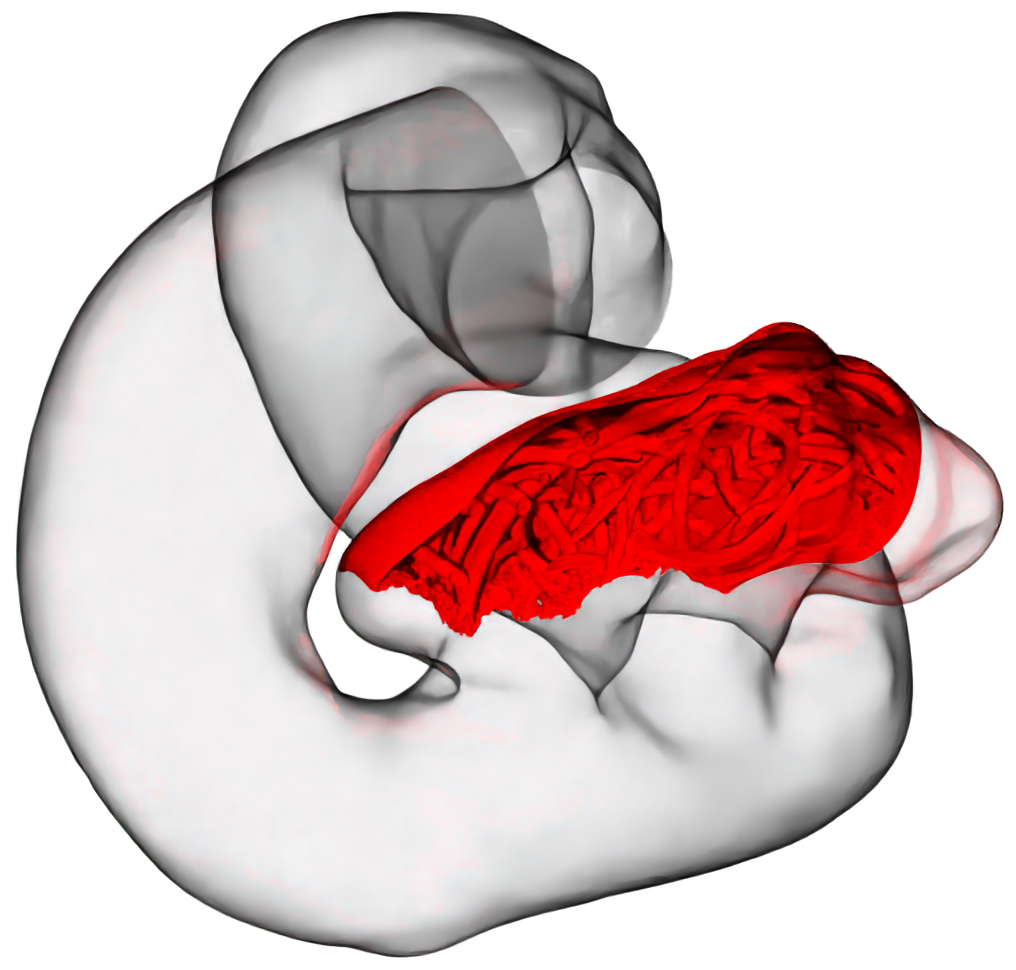}
    \caption{\footnotesize Failed coiling case with the coil protruding into the parenting vessel. \textbf{(First left)} Failed Coil geometry, \textbf{(Second left)} resulting thrombus comparable to the one of the successful coiling in Fig.\,\ref{fig:Case1_Thrombi}, on the right. The two images on the right compare \textbf{(Second from right)} the (cross section of) the thrombus of the failed coil with \textbf{(Very right)} the thrombus resulting from the same failed coil, but with an alternative clotting model where also the coil itself (not just the aneurysm dome) releases thrombin as a thrombogenic substance.}
    \label{fig:Case1_BadCoilThrombi}
\end{figure}

\paragraph{Successful vs. failed coiling comparison}
As previously mentioned, \textit{purposely failed} coil insertion is simulated, where the resulting coil extensively protrudes into the parent vessel, risking blockage and hence a potential stroke \cite{xiao2022embolic}. Fig.\,\ref{fig:Case1_BadCoilThrombi} shows the resulting thrombus of the failed coiling case.

Even though the coil protrudes into the parent vessel, early fibrin-thrombus formation is localised to the aneurysm sac, due to the shear-rate conditions for thrombus formation in our model being achieved there.  This finding persists even when thrombogenic proteins are released not only from the aneurysm walls, but also from the coils \cite{Kim1998ThrombogenicityMicrocoils}, as shown in Fig.\,\ref{fig:Case1_BadCoilThrombi} on the very right. Although the thrombus is larger in size, it is still confined to the aneurysm.


\subsection{Case 2 - Fusiform aneurysm}\label{subsec:Case2}

\begin{figure}[h!]
    \centering
    \includegraphics[width=0.25\linewidth]{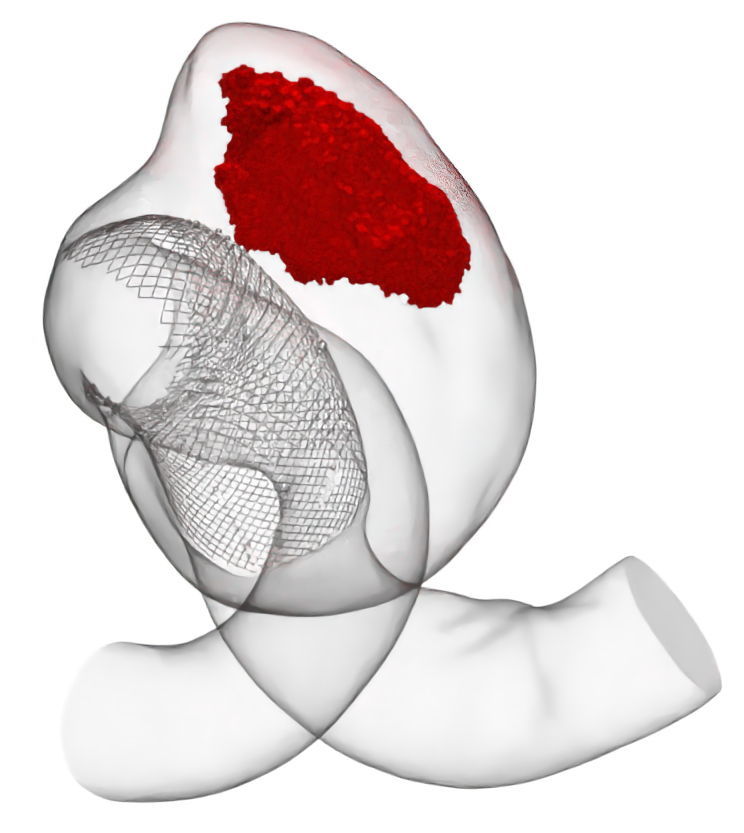}\includegraphics[width=0.25\linewidth]{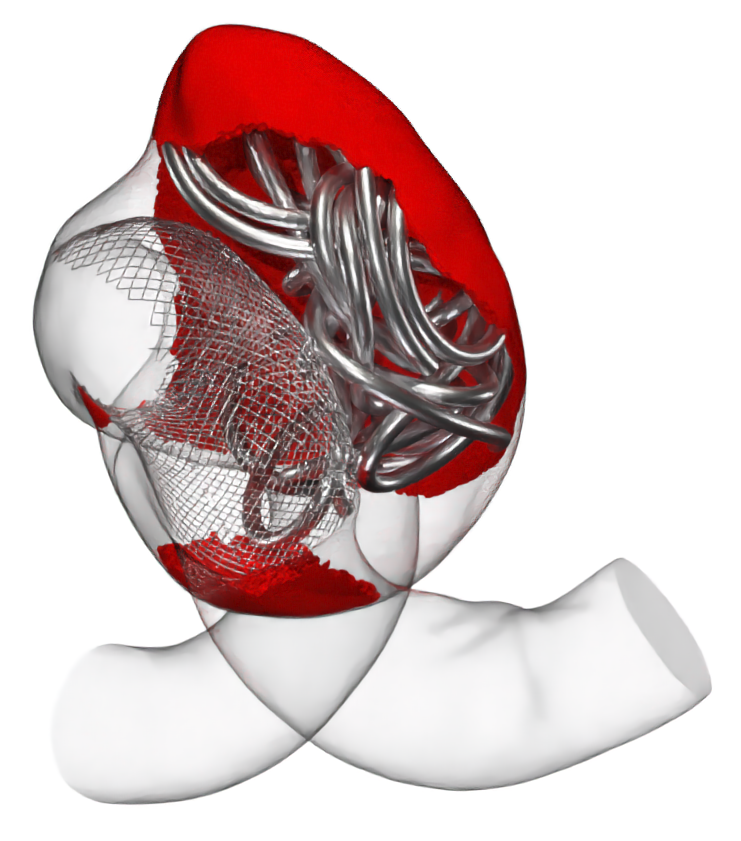}
    \caption{\footnotesize Simulated final thrombi (in red) comparing different treatment cases. Note that the case with an empty\,/\,untreated aneurysm did not produce \textit{any} thrombus and is hence left out here. \textbf{(Left)} \ac{FD}-only treatment with very scarce thrombus growth on the aneurysm dome only. \textbf{(Right)} Stent-assisted coiling treatment using three coils with the resulting thrombus covering larger parts of the aneurysm dome.}
    \label{fig:Case2_Thrombi}
\end{figure}

Case 2 represents a large fusiform aneurysm, for which a pure coiling treatment is not an option. Hence, only a \textit{stent-assisted} coiling procedure is simulated for this case. Due to the volumetric size of this aneurysm, we perform a multi-coil insertion procedure with three coils to reach a packing density of approximately 9.24\,\% of the whole aneurysm including the parent vessel. We also consider \ac{FD} treatment only. The \textit{empty} aneurysm case did not lead to \textit{any} clot formation, since the flow and recirculation were too rapid for the shear rate threshold for clotting to be met. Hence, Fig.\,\ref{fig:Case2_Thrombi} includes the two treated cases only.\\

Table \ref{tab:OcclusionVoluminaCase2} shows that stent-assisted coiling led to a larger occlusion ratio compared to \ac{FD}-treatment alone. This case, compared to Case 1, showed a greater difference between the two distinct treatment methods, with stent-assisted coiling resulting in a total occlusion ratio at least ten times that of flow diversion alone. Overall, the occlusion ratio numbers are smaller than in Case 1. This might arise from the aneurysm in Case 2 being 5.6 times larger, while our thrombosis model only accounts for the first few minutes following device placement.

\begin{table}[h!]
	\small
    \centering
    \begin{tabular}{|c|c|c|c||c|}
    \cline{2-4}
         \multicolumn{1}{c|}{}& \multicolumn{3}{c||}{Treatment case} &  \multicolumn{1}{c}{}\\ \cline{5-5}
         \multicolumn{1}{c|}{}& \textbf{Empty} & \textbf{Stented (no coils)}  & \textbf{Stent-assisted coiling} & \textbf{Coiling volume} \\ \hline
         \textbf{Thrombus volume} & \SI{0}{mm^3} & \SI{10.4}{mm^3} & \SI{78.75}{mm^3} & \SI{38.09}{mm^3} \\\hline
         \textbf{Aneurysm volume} & \multicolumn{4}{c|}{\SI{412.18}{mm^3}} \\ \hline
         \multirow{2}{*}{\textbf{Percentage occlusion}} & 0\,\% & 2.52\,\% & 19.10\,\% & 9.24\,\% \\ \cline{2-5}
         & \multicolumn{2}{c|}{--} & \multicolumn{2}{c|}{\hspace{-0.7cm}$\Sigma$\hspace{0.7cm} 28.34\,\%} \\ \cline{1-5}
    \end{tabular}
    \caption{Purely geometrically computed occlusion ratios of the thrombi (and coiling devices) shown in Fig.\,\ref{fig:Case2_Thrombi}. Again, in the stent-assisted coiling case, the thrombus volume and the volume of the coil(s) themselves has to be added for the total occlusion volume.}
    \label{tab:OcclusionVoluminaCase2}
\end{table}


\subsection{Case 3 - Giant saccular aneurysm}\label{subsec:Case3}

\begin{figure}[h!]
    \centering
\begin{minipage}{0.32\linewidth}
  \centering
  \includegraphics[width=\linewidth]{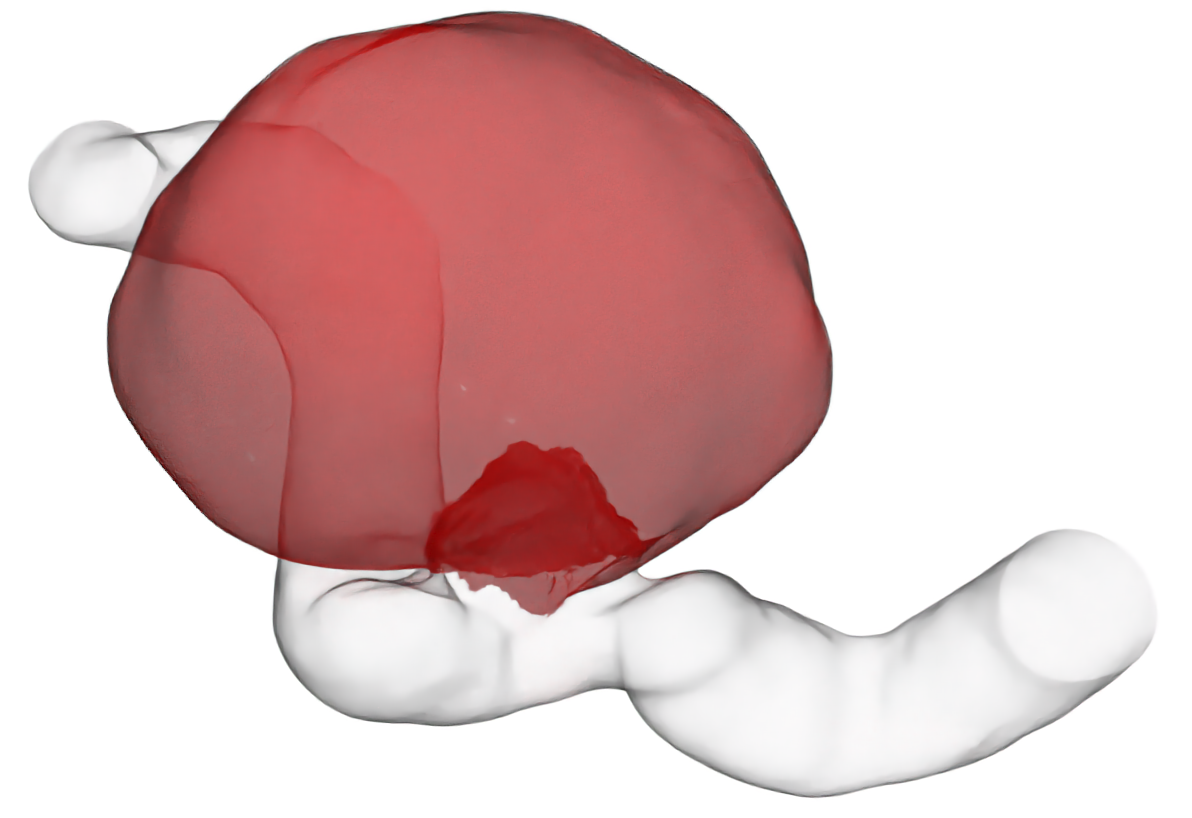}
\end{minipage}\hfill
\begin{minipage}{0.32\linewidth}
  \centering
  \includegraphics[width=\linewidth]{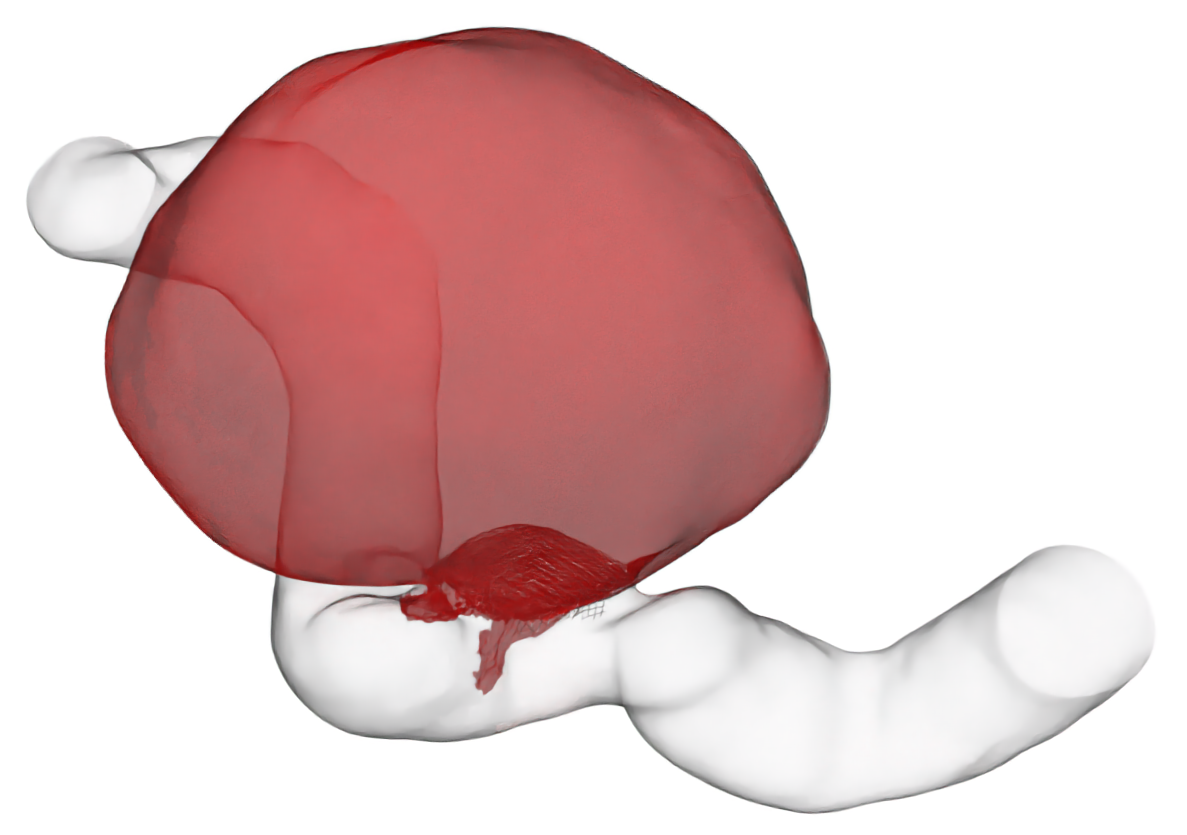}
\end{minipage}\hfill
\begin{minipage}{0.32\linewidth}
  \centering
  \includegraphics[width=\linewidth]{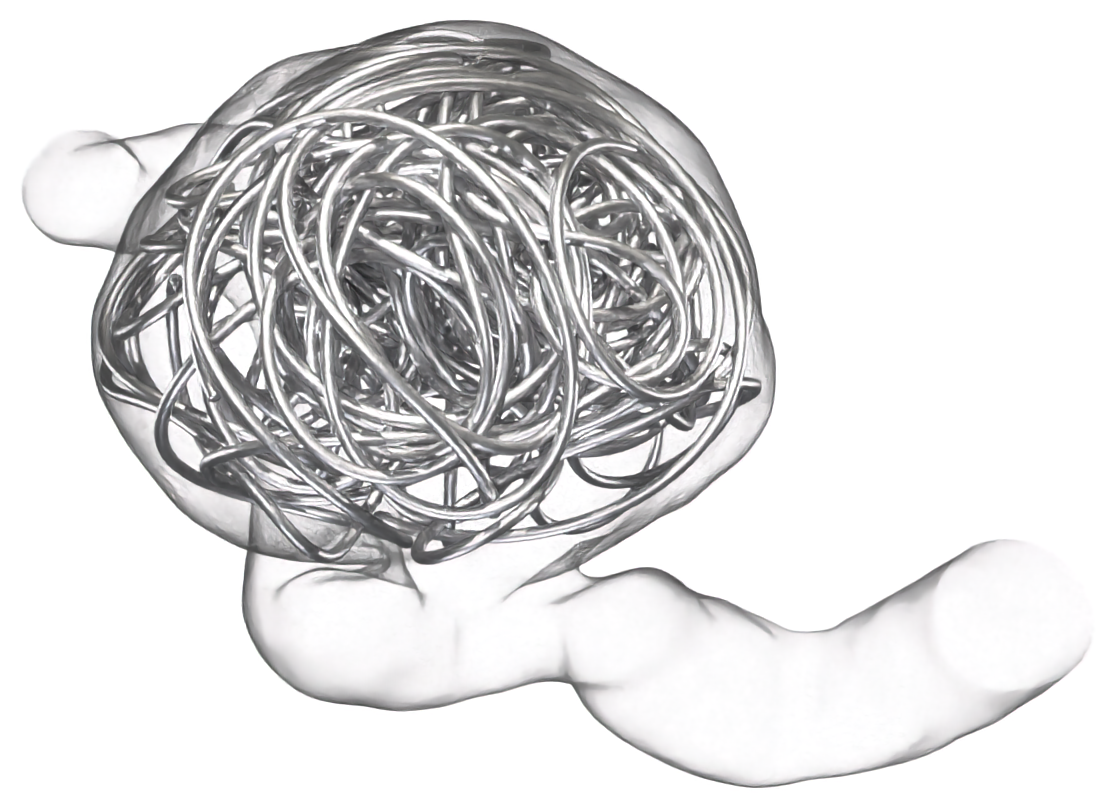}
\end{minipage}

\vspace{0.5em}

\begin{minipage}{0.32\linewidth}
  \centering
  \includegraphics[width=\linewidth]{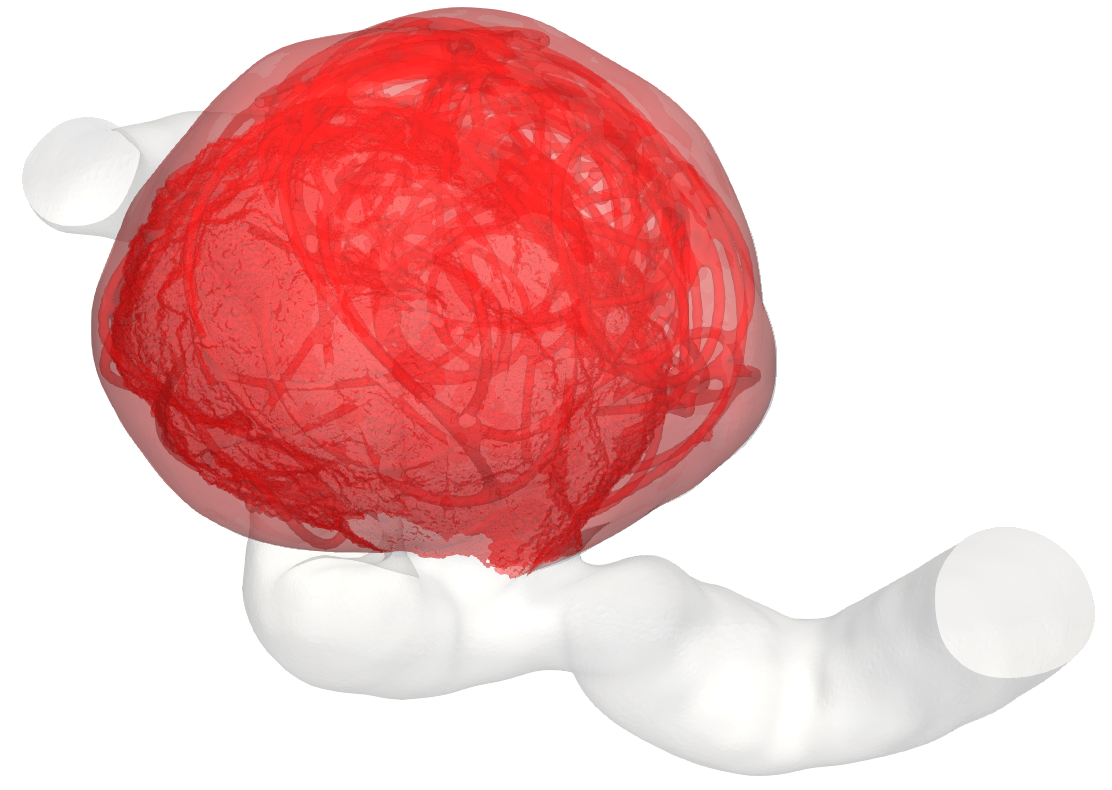}
\end{minipage}\hfill
\begin{minipage}{0.32\linewidth}
  \centering
  \includegraphics[width=\linewidth]{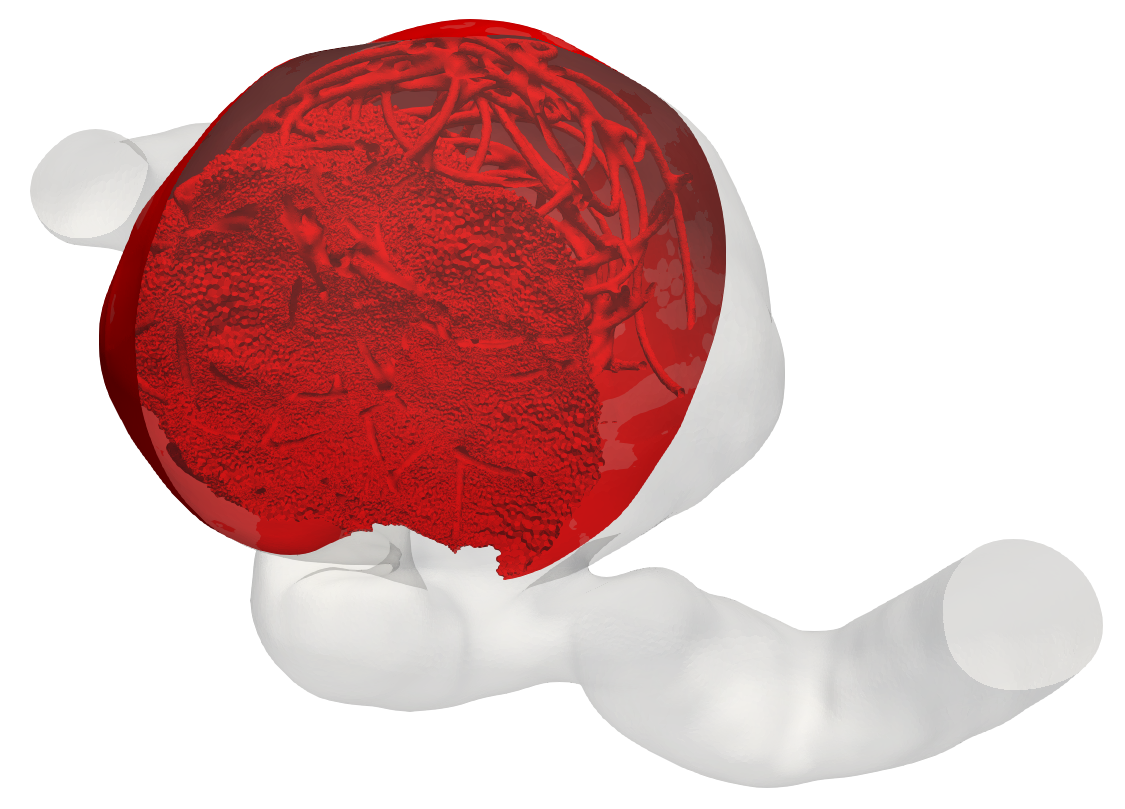}
\end{minipage}\hfill
\begin{minipage}{0.32\linewidth}
  \centering
  \includegraphics[width=\linewidth]{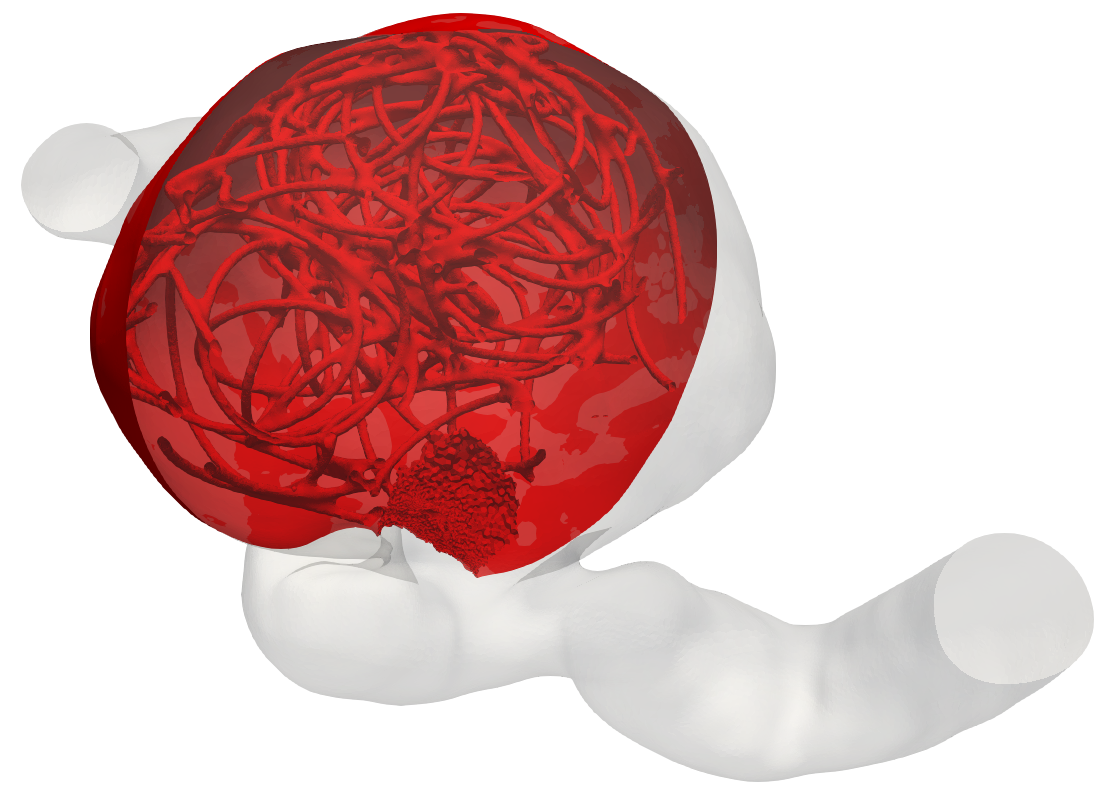}
\end{minipage}
    \caption{\footnotesize Simulated final thrombi (in red) in different treatment cases. \textbf{(Top left)} Empty, untreated aneurysm, \textbf{(Top middle)} \ac{FD} treated aneurysm, \textbf{(Top right)} Visualization of the coiling. \textbf{(Bottom left)} Thrombus with coils, \textbf{(Bottom middle)} Cross-section of the coil-induced thrombus, \textbf{(Bottom right)} Again a cross-section, but once more for the alternative model where not only the aneurysm dome but also the coil emits thrombin.}
    \label{fig:Case3_Thrombi}
\end{figure}

Case 3 is a giant saccular aneurysm, for which endovascular coiling is appropriate due to its narrow neck. However, to reach sensible values of packing density, the large volume of the aneurysm makes it necessary to also insert several (here up to five) coils with a combined length of approximately 2 m. The aneurysm size renders \ac{FD}-based treatment a viable alternative. Although the shape of the aneurysm does not require any additional position-confinement for the coils, we also include stent-assisted coiling in the virtual \ac{DSA} analysis.  For thrombosis, we consider coiling and flow diversion, with the thrombi shown in Fig.\,\ref{fig:Case3_Thrombi}. Similar to Fig.\,\ref{fig:Case1_Thrombi}, we also compare the coiling-induced thrombus in two different modelling situations. We present our standard thrombosis model, where only the aneurysm dome-surface emits thrombin and an alternative model, where the coiling wire is also thrombogenic. This additional comparison is included in the last column of table \ref{tab:OcclusionVoluminaCase3}.

\begin{table}[h!]
	\small
    \centering
    \hspace*{-5mm}\begin{tabular}{|c|c|c|c||c||c|}
    \cline{2-4}\cline{6-6}
        \multicolumn{1}{c|}{}& \multicolumn{3}{c||}{Treatment case} &   & \\ \cline{5-5}
        \multicolumn{1}{c|}{}& \textbf{Empty} & \textbf{Stented}  & \textbf{Coiling} & \textbf{Coil volume} & \textbf{Coiling + thr. em.}\\ \hline
         \textbf{Thrombus volume} & \SI{2146.46}{mm^3} & \SI{2189.88}{mm^3} & \SI{1129.59}{mm^3} & \SI{152}{mm^3} & \SI{2017.28}{mm^3} \\\hline
         \textbf{Aneurysm volume} & \multicolumn{5}{c|}{\SI{2193.09}{mm^3}} \\ \hline
         \multirow{3}{*}{\textbf{Percentage occlusion}} & 97.87\,\% & 99.85\,\% & 51.5\,\% & 6.93\,\% & 91.98\,\% \\ \cline{2-6}
         & \multicolumn{2}{c|}{--} & \multicolumn{2}{c||}{\hspace{-0.7cm}$\Sigma$\hspace{0.7cm} 58.43\,\%} & \multicolumn{1}{c}{} \\ \cline{4-6}
         & \multicolumn{3}{c||}{--} & \multicolumn{2}{c|}{\hspace{-0.7cm}$\Sigma$\hspace{0.7cm} 98.91\,\%} \\ \cline{1-6}
    \end{tabular}
    \caption{Occlusion volumes/ratios of the final thrombi (and coiling devices) shown in Fig.\,\ref{fig:Case3_Thrombi}. In the two coiling cases always the combined (summed up) volume of coiling(s) and their resulting thrombus has to be considered, hence the two additional sum computations in the last two rows of the table. The last column contains the values for the alternative modelling case, where also the coil's surfaces do emit thrombin therefore acting thrombogenic.}
    \label{tab:OcclusionVoluminaCase3}
\end{table}

Similar to Case 1, the additional thrombogenic activity of the coil in the alternative model leads to more voluminous thrombus filling up most of the void space that is still left inside the aneurysm in the regular coiling case. However, much more remarkable in this case is the difference between the empty, stented and coiled thrombus formation outcome. Even though the case is predestined for coiling from a purely mechanical perspective (low risk of coil protrusion or relocation into the parent vessel), a \textit{purely geometric} look at the resulting thrombus indicates something different when compared to the \ac{FD} treated scenario, making the coiling-treatment look less favorable from an occlusion point of view. The large thrombus fraction predicted, even in the \textit{untreated} case, results from the combination of strong thrombin activation at the aneurysm wall and the presence of large low-shear recirculation zones inside the sac. These conditions promote fibrin formation in the present model. In vivo, additional anticoagulant mechanisms and endothelial responses may limit such rapid thrombosis, which are not represented in the current framework. At this point, we already want to hint forward to the subsequent virtual tracer angiography conducted in section \ref{subsec:TracerResults} for this case, which again speaks more in favor of coil-based treatment taking into account the \textit{actual remaining inflow-characteristics} of the treated aneurysms as it is usually done in clinical practice, where no 3D models of the resulting thrombus are available.


\subsection{Device Clot Progression Analysis}\label{subsec:ClotResults}
\begin{figure}[ht]
\begin{center}
  \includegraphics[width=.9\linewidth]{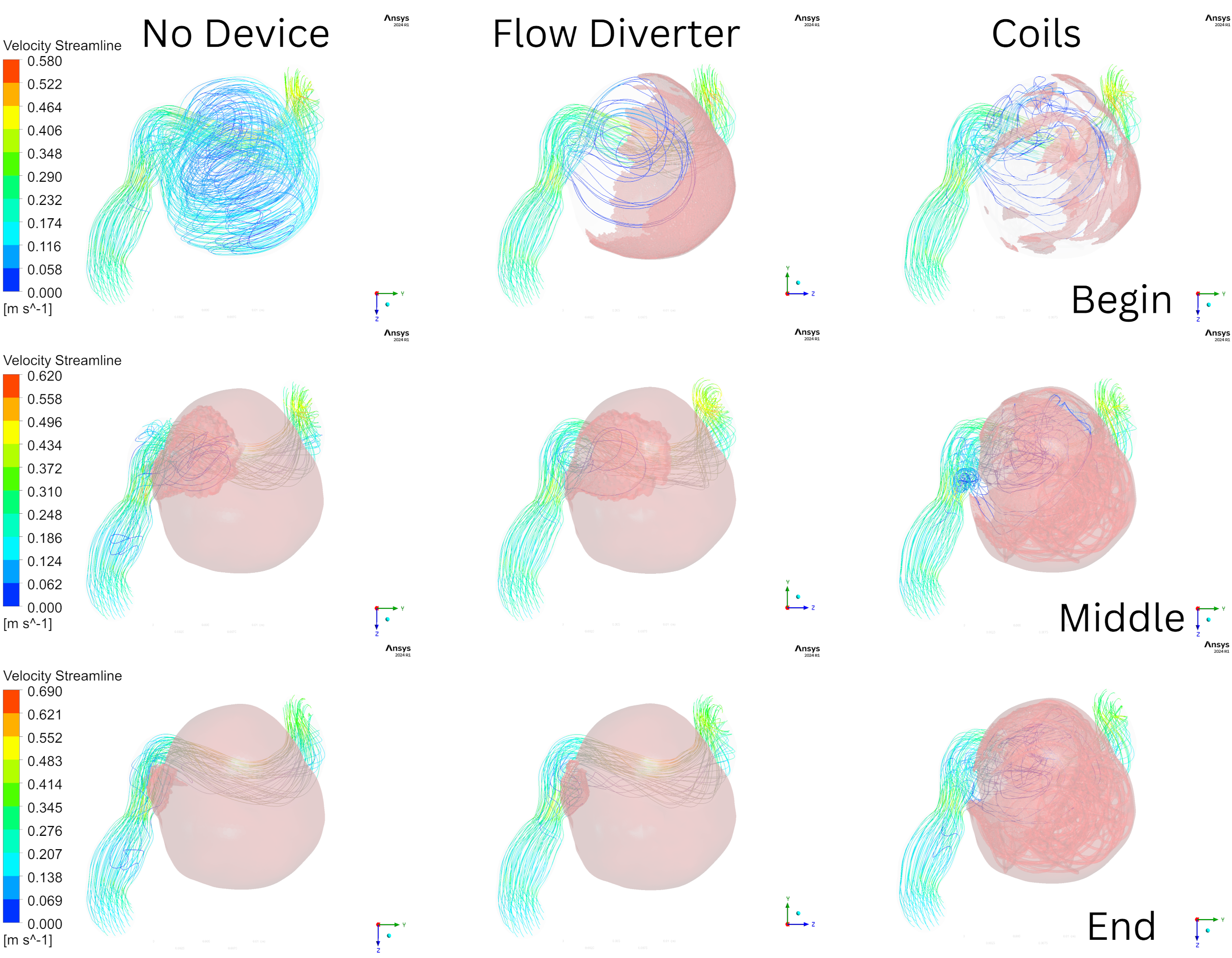}
\caption{Flow-pattern and early thrombus geometry comparison for different treatment cases: The \textbf{(rows)} stand for three characteristic points in time while the \textbf{(columns)} compare different treatment methods (first column shows the empty aneurysm for reference).}
\label{fig:BeginningClot}
\end{center}
\end{figure}

\begin{figure}[ht]
\begin{center}
  \includegraphics[clip=true,trim=10mm 10mm 3mm 10mm,width=.75\linewidth]{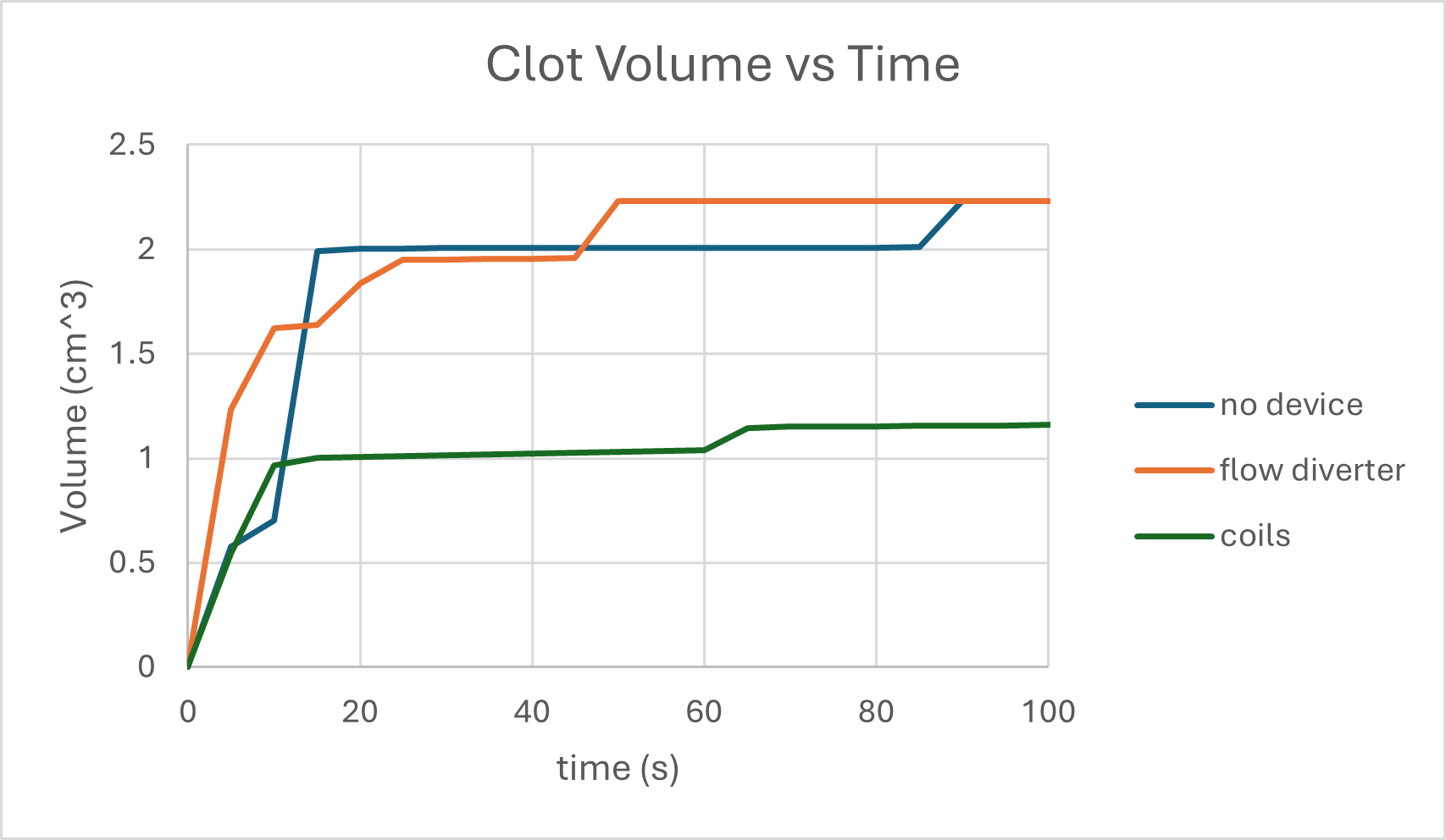}
\caption{Comparison of clot volume versus time for different treatment modes used in Case 3. The initial clot development rate are similar across all three treatment modes, however the volume stabilises at a lower value for coiling.}
\label{fig:clotvstime}
\end{center}
\end{figure}

Due to the different thrombus patterns observed during coiling and \ac{FD}-treatment of the giant aneurysm, shown in Fig.\,\ref{fig:Case3_Thrombi}, this subsection considers the influence of flow on early thrombus formation over time. Fig.\,\ref{fig:BeginningClot} clearly demonstrates the presence of a stable vortex in the no device and \ac{FD}-treated case. In the latter case, this favours shear- activated clot formation with relatively quick deposition of thrombus material at the aneurysm surface. In contrast, the presence of coils disrupts the vortex-flow pattern. The streamlines exhibit a rather diffuse course in between the wires of the coil, leading to a less favorable environment for shear-activated clotting. Comparison of clot volume over time, for these treatment scenarios, is illustrated in Fig. \ref{fig:clotvstime}. While the initial rate of clot development is similar for all three scenarios, the vorticial structures in the no device and flow diverter cases seem to support greater occlusion than the coil case, for the first 100 seconds. 

\clearpage


\subsection{Tracer-based virtual angiography results}\label{subsec:TracerResults}
This section evaluates the effect of thrombus formation on residual aneurysm perfusion using the virtual \ac{DSA} methodology described in Sec.~\ref{sec:OcclusionQuality}. This provides a flow-based assessment of occlusion quality that is more closely related to clinical angiographic evaluation.\\

\subsubsection{Case 3} 

\begin{figure}[hp]
    \centering
    \begin{subfigure}[t]{0.23\textwidth}
        \includegraphics[width=\linewidth]{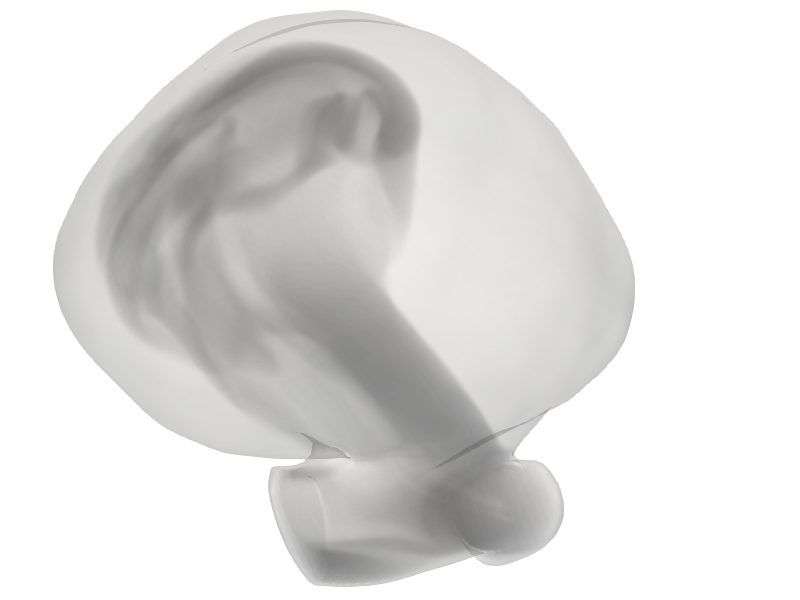}
        \caption*{(a)}
    \end{subfigure}
    \begin{subfigure}[t]{0.23\textwidth}
        \includegraphics[width=\linewidth]{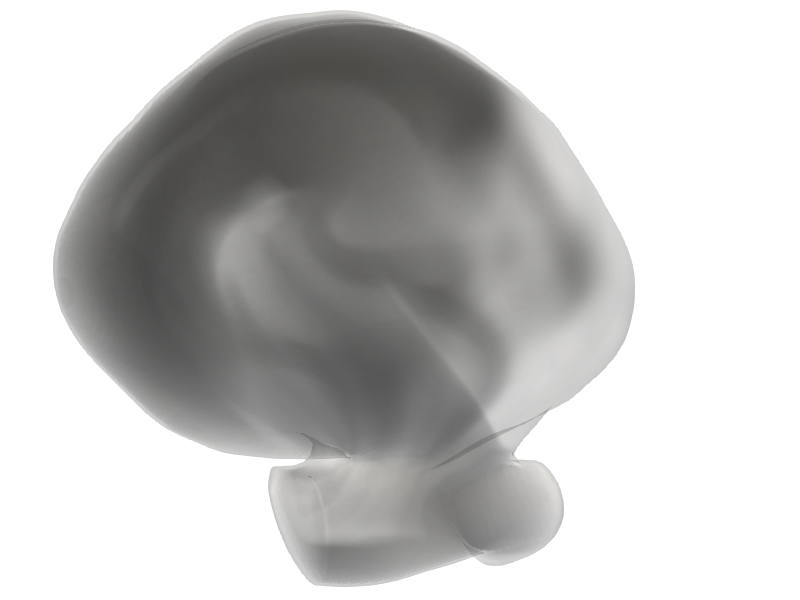}
        \caption*{(b)}
    \end{subfigure}
    \begin{subfigure}[t]{0.23\textwidth}
        \includegraphics[width=\linewidth]{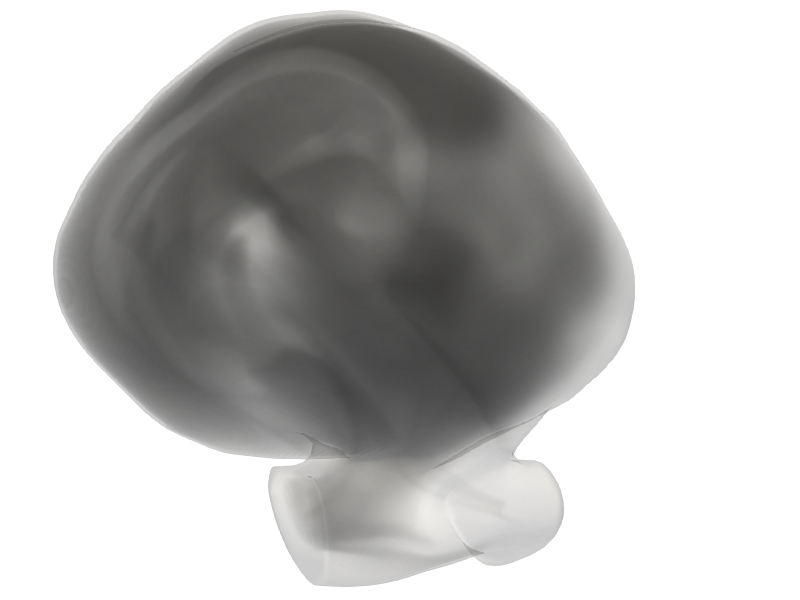}
        \caption*{(c)}
    \end{subfigure}
    \begin{subfigure}[t]{0.23\textwidth}
        \includegraphics[width=\linewidth]{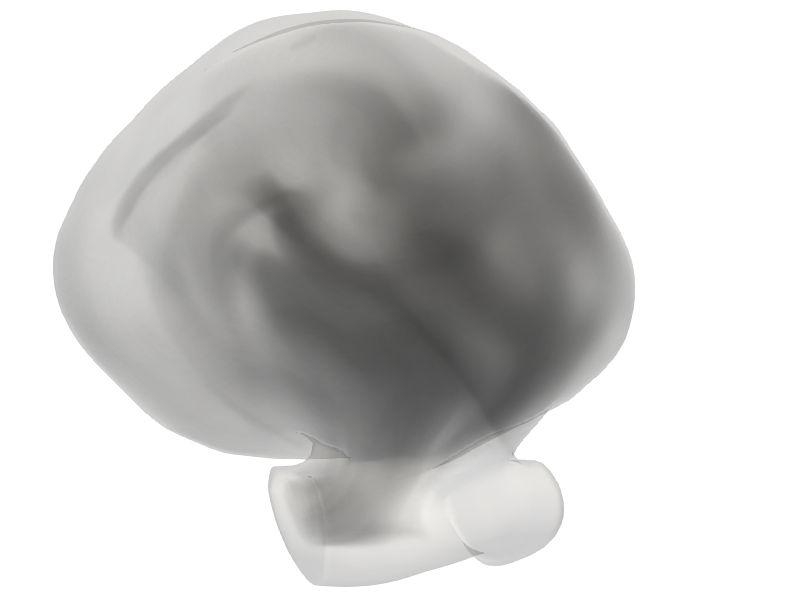}
        \caption*{(d)}
    \end{subfigure}
    
    \vspace{1mm}
    
    \begin{subfigure}[t]{0.23\textwidth}
        \includegraphics[width=\linewidth]{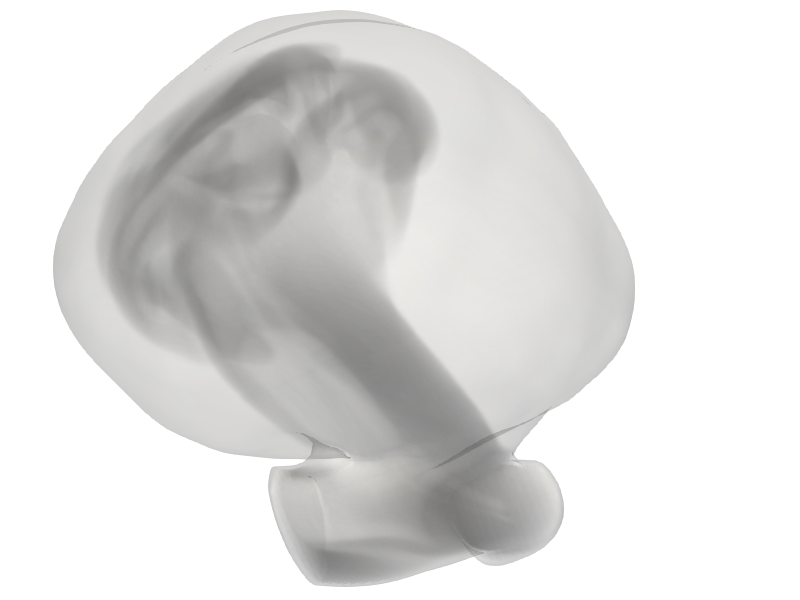}
        \caption*{(e)}
    \end{subfigure}
    \begin{subfigure}[t]{0.23\textwidth}
        \includegraphics[width=\linewidth]{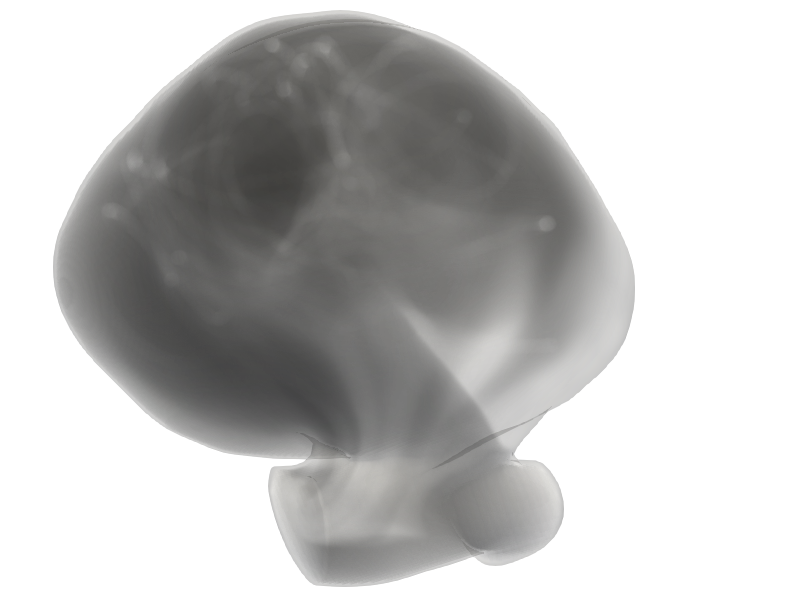}
        \caption*{(f)}
    \end{subfigure}
    \begin{subfigure}[t]{0.23\textwidth}
        \includegraphics[width=\linewidth]{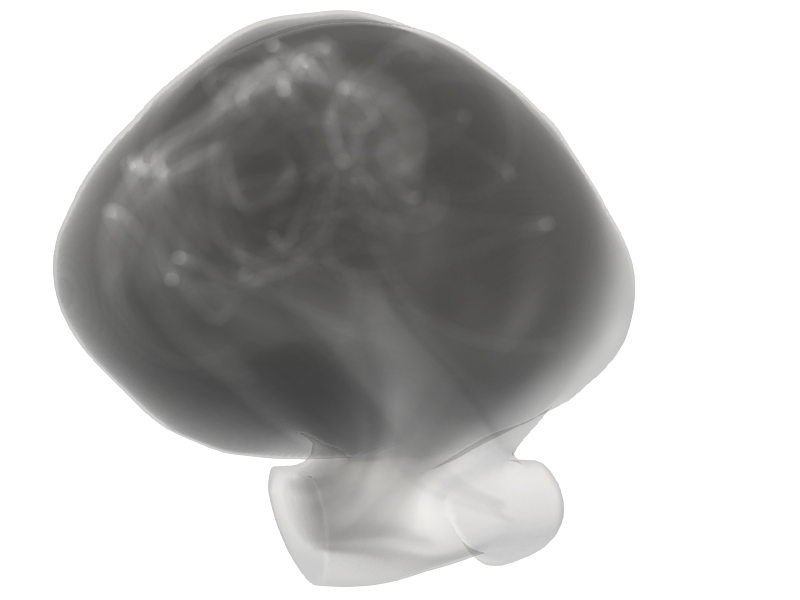}
        \caption*{(g)}
    \end{subfigure}
    \begin{subfigure}[t]{0.23\textwidth}
        \includegraphics[width=\linewidth]{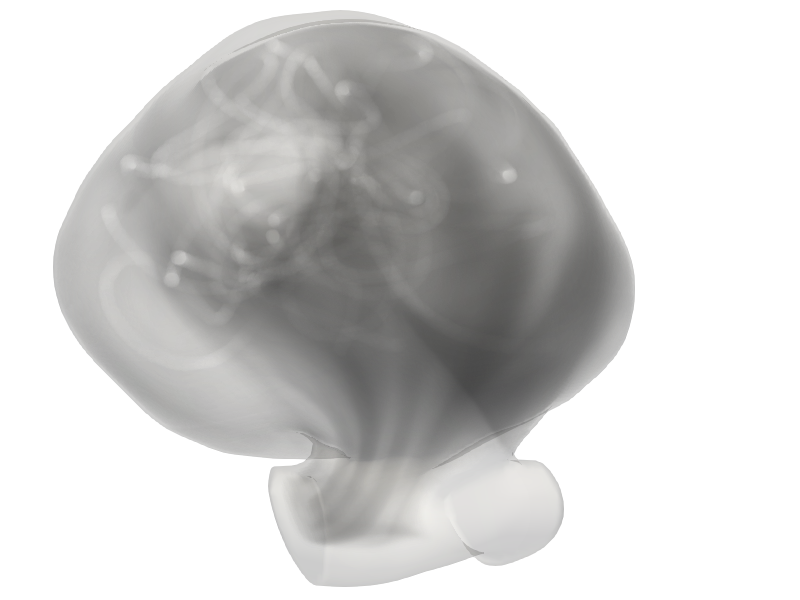}
        \caption*{(h)}
    \end{subfigure}
    
    \vspace{1mm}
    
    \begin{subfigure}[t]{0.23\textwidth}
        \includegraphics[width=\linewidth]{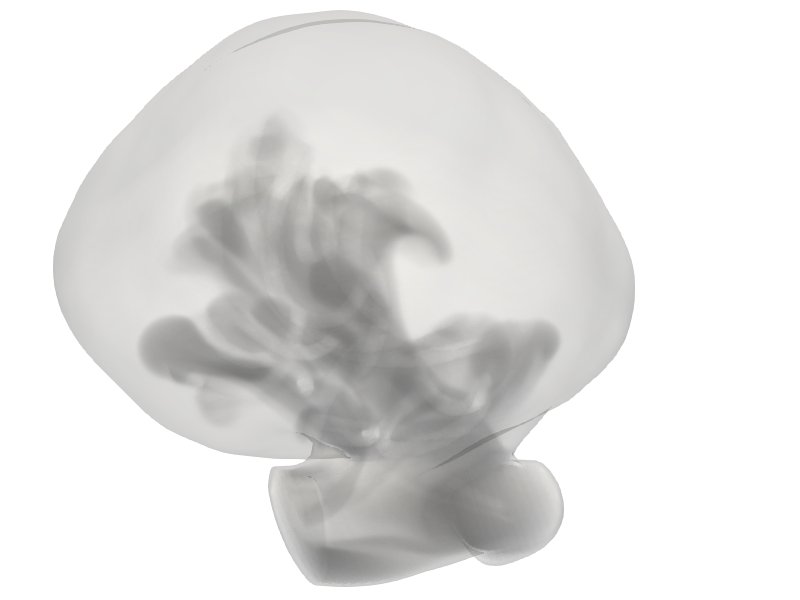}
        \caption*{(i)}
    \end{subfigure}
    \begin{subfigure}[t]{0.23\textwidth}
        \includegraphics[width=\linewidth]{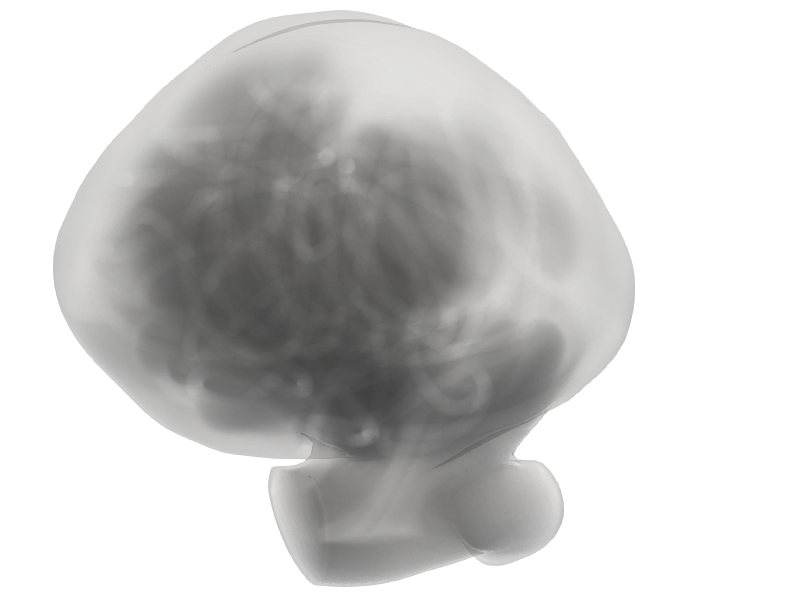}
        \caption*{(j)}
    \end{subfigure}
    \begin{subfigure}[t]{0.23\textwidth}
        \includegraphics[width=\linewidth]{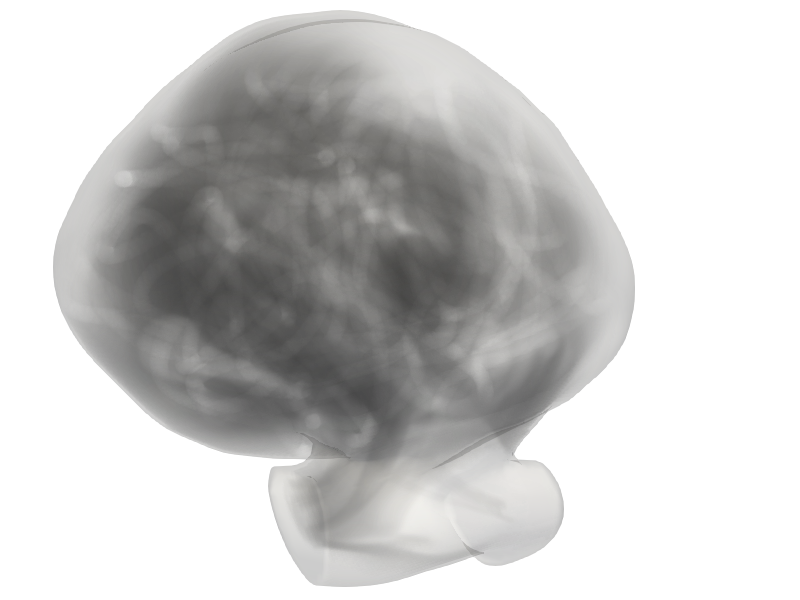}
        \caption*{(k)}
    \end{subfigure}
    \begin{subfigure}[t]{0.23\textwidth}
        \includegraphics[width=\linewidth]{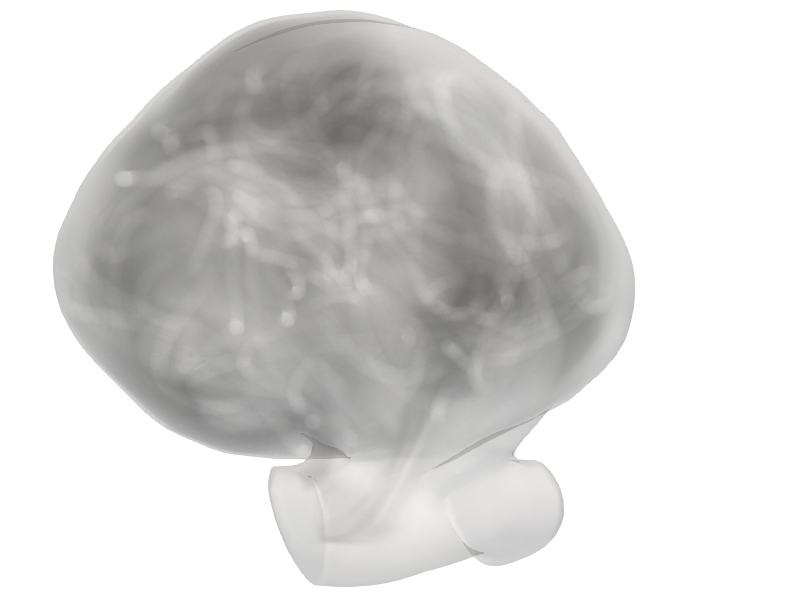}
        \caption*{(l)}
    \end{subfigure}
    
    \vspace{1mm}
    
    \begin{subfigure}[t]{0.23\textwidth}
        \includegraphics[width=\linewidth]{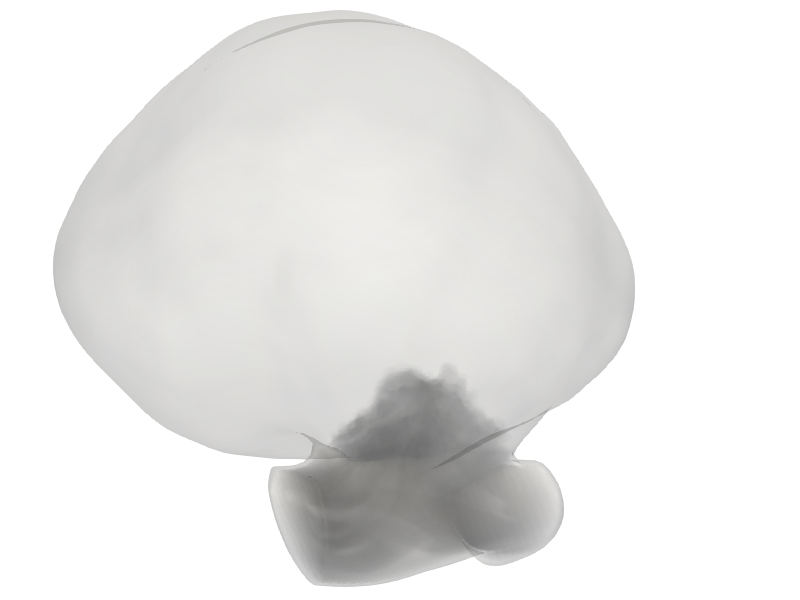}
        \caption*{(m)}
    \end{subfigure}
    \begin{subfigure}[t]{0.23\textwidth}
        \includegraphics[width=\linewidth]{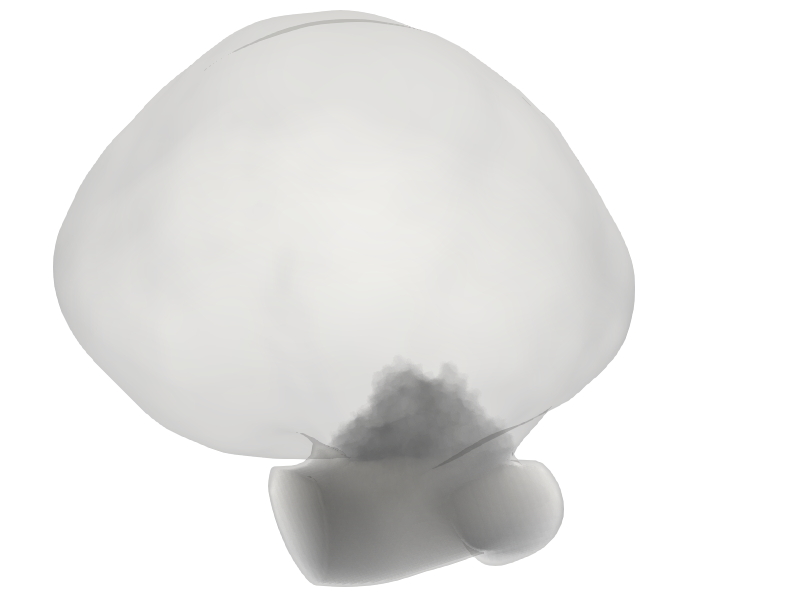}
        \caption*{(n)}
    \end{subfigure}
    \begin{subfigure}[t]{0.23\textwidth}
        \includegraphics[width=\linewidth]{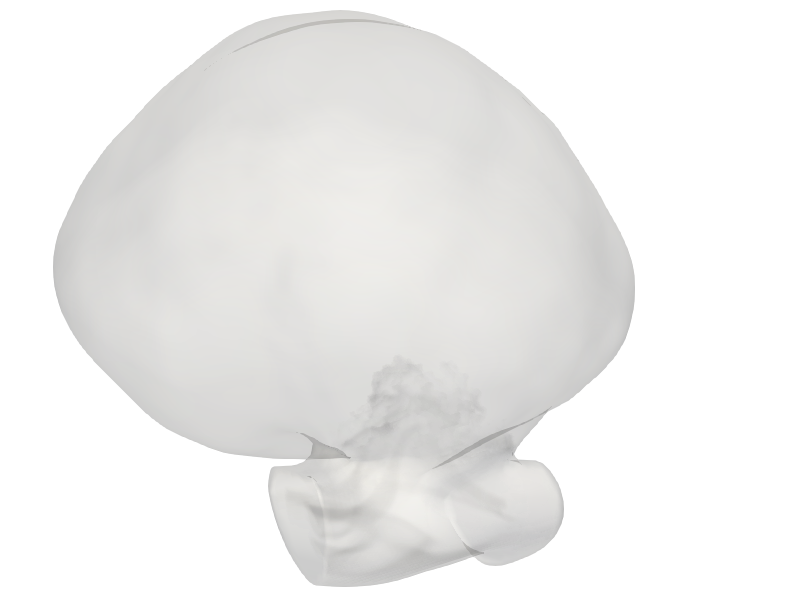}
        \caption*{(o)}
    \end{subfigure}
    \begin{subfigure}[t]{0.23\textwidth}
        \includegraphics[width=\linewidth]{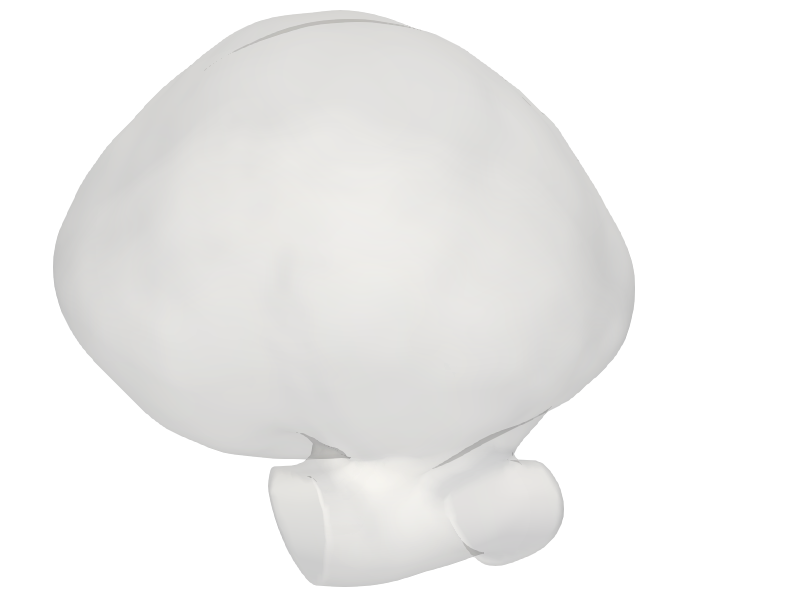}
        \caption*{(p)}
    \end{subfigure}

    \vspace{1mm}
    
    \begin{subfigure}[t]{0.23\textwidth}
        \includegraphics[width=\linewidth]{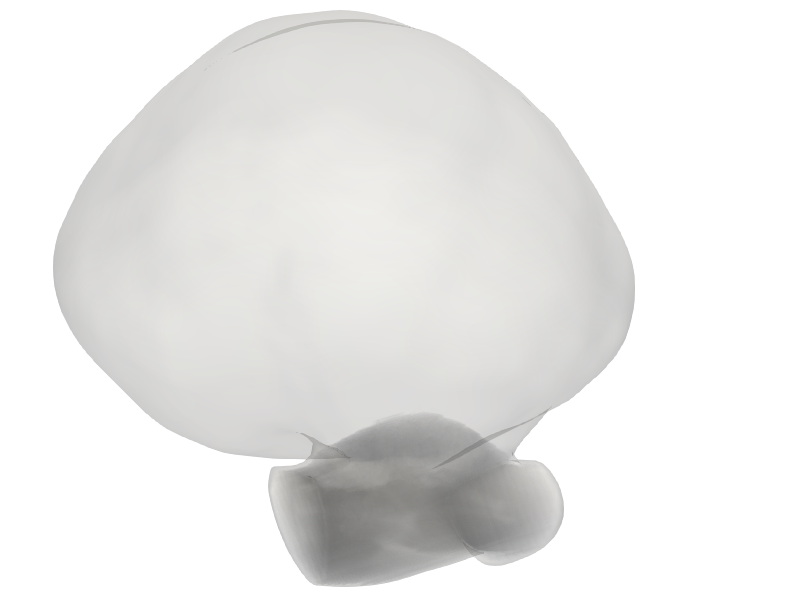}
        \caption*{(q)}
    \end{subfigure}
    \begin{subfigure}[t]{0.23\textwidth}
        \includegraphics[width=\linewidth]{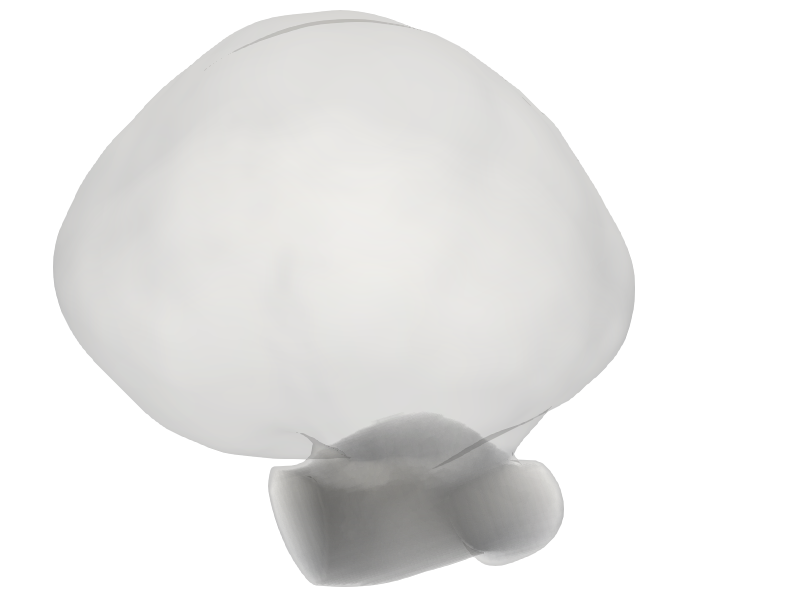}
        \caption*{(r)}
    \end{subfigure}
    \begin{subfigure}[t]{0.23\textwidth}
        \includegraphics[width=\linewidth]{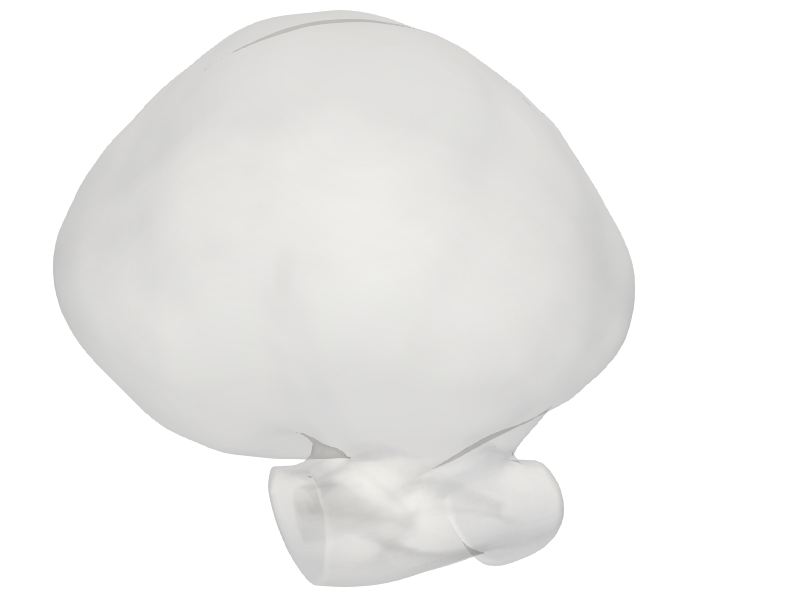}
        \caption*{(s)}
    \end{subfigure}
    \begin{subfigure}[t]{0.23\textwidth}
        \includegraphics[width=\linewidth]{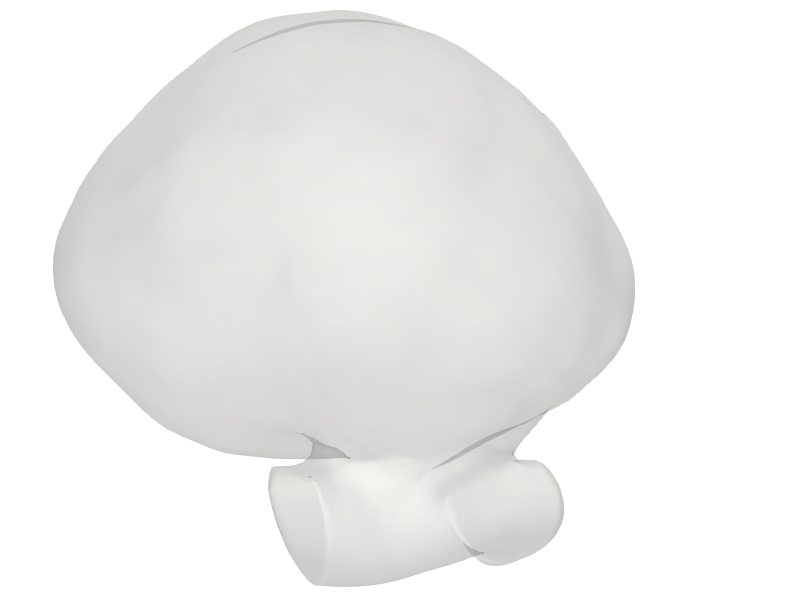}
        \caption*{(t)}
    \end{subfigure}

    \caption{3D tracer field visualization of Case 3 showing the aneurysm occlusion in different setups. Each \textbf{column} corresponds to the temporal snapshot at peak systoles resulting from a simulation over four heart-beat cycles. In the \textbf{first row} (a)-(d): No coil placed. In the \textbf{second row} (e)-(h): 2 coils placed. \textbf{Third row} (i)-(l): 5 coils placed. \textbf{Fourth row} (m)-(p): After coil induced thrombosis. \textbf{Last row} (q)-(t): After coil+stent induced thrombosis.}
    \label{fig:DSA_Case3}
\end{figure}

Case 3 is considered first. Figure~\ref{fig:DSA_Case3} shows three-dimensional visualisations of the calibrated tracer concentration field for the different treatment configurations. Each column corresponds to a temporal snapshot taken at peak systole during four consecutive cardiac cycles, while each row represents a different device or thrombus configuration.

The simulations capture tracer injection into the parent vessel, subsequent tracer transport into the aneurysm where present, and the following washout phase. The rows compare the untreated aneurysm, two coil-packing configurations, coil-induced thrombosis, and stent-assisted coiling with thrombosis. The latter configuration represents the most inflow-restrictive treatment case considered for this geometry.


The results show that increasing coil packing density reduces tracer penetration into the aneurysm; however, the reduction remains relatively modest when coils are considered without thrombus formation. A substantially larger reduction is observed once coil-induced thrombus formation is included, with tracer transport into the aneurysm almost completely suppressed. The stent-assisted coiling case produces a further reduction in tracer inflow, although the difference relative to the coil-induced thrombus case is small. This suggests that thrombus formation around the coil structure is a dominant contributor to aneurysm isolation in this case.

Comparison with the geometric thrombus distribution shown in Fig.~\ref{fig:Case3_Thrombi} supports this interpretation. Although the coil-induced thrombus contains a residual void region near the center of the aneurysm, the virtual \ac{DSA} results indicate that this thrombus morphology is sufficient to substantially reduce residual inflow. The resulting reduction in aneurysm perfusion therefore supports the effectiveness of the coiling treatment in the present case.

\begin{figure}[ht]
    \centering
    \begin{subfigure}[t]{0.23\textwidth}
        \includegraphics[width=\linewidth]{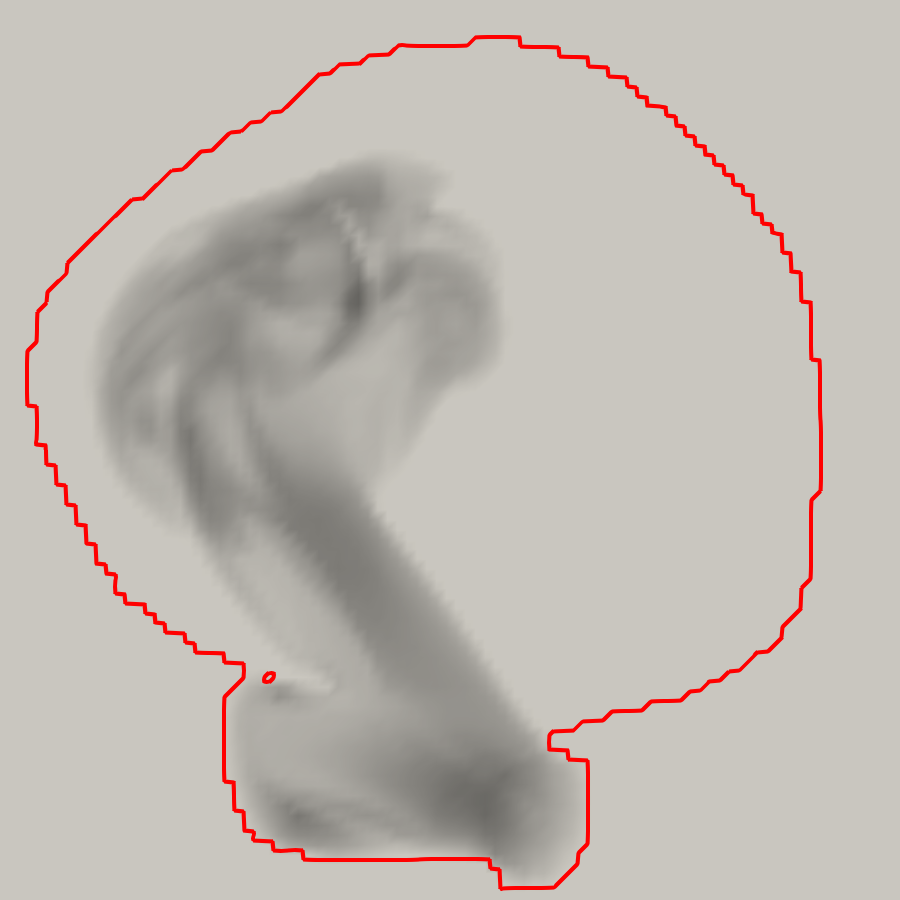}
        \caption*{(a)}
    \end{subfigure}
    \begin{subfigure}[t]{0.23\textwidth}
        \includegraphics[width=\linewidth]{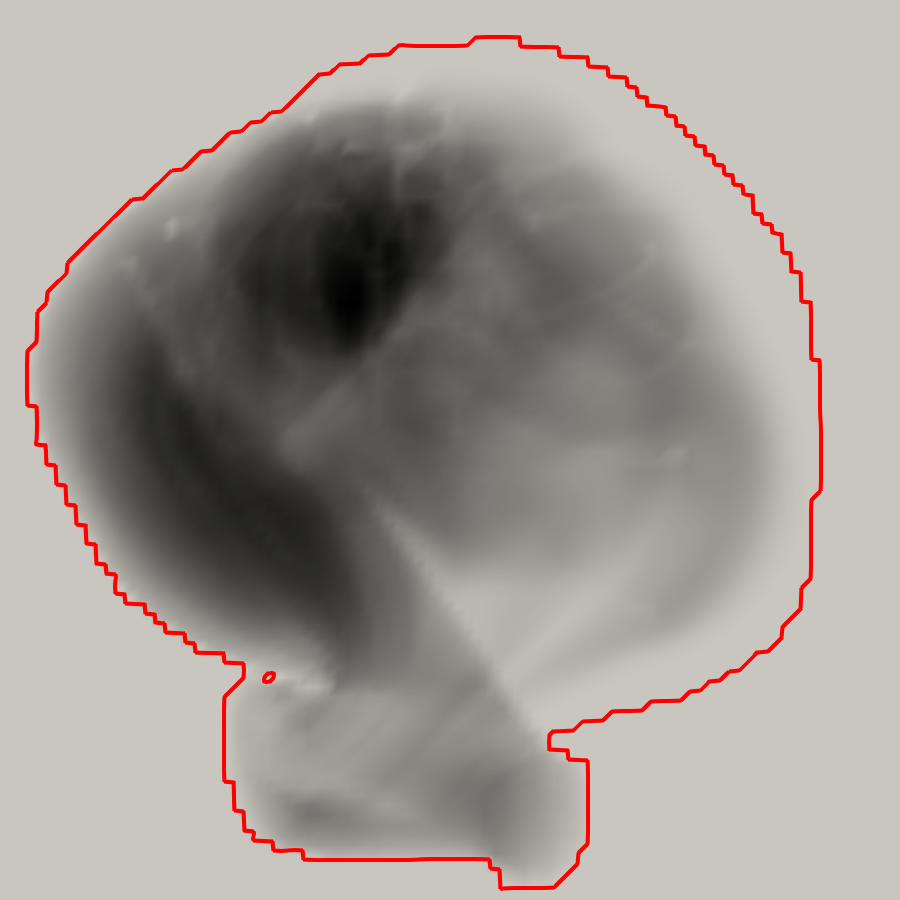}
        \caption*{(b)}
    \end{subfigure}
    \begin{subfigure}[t]{0.23\textwidth}
        \includegraphics[width=\linewidth]{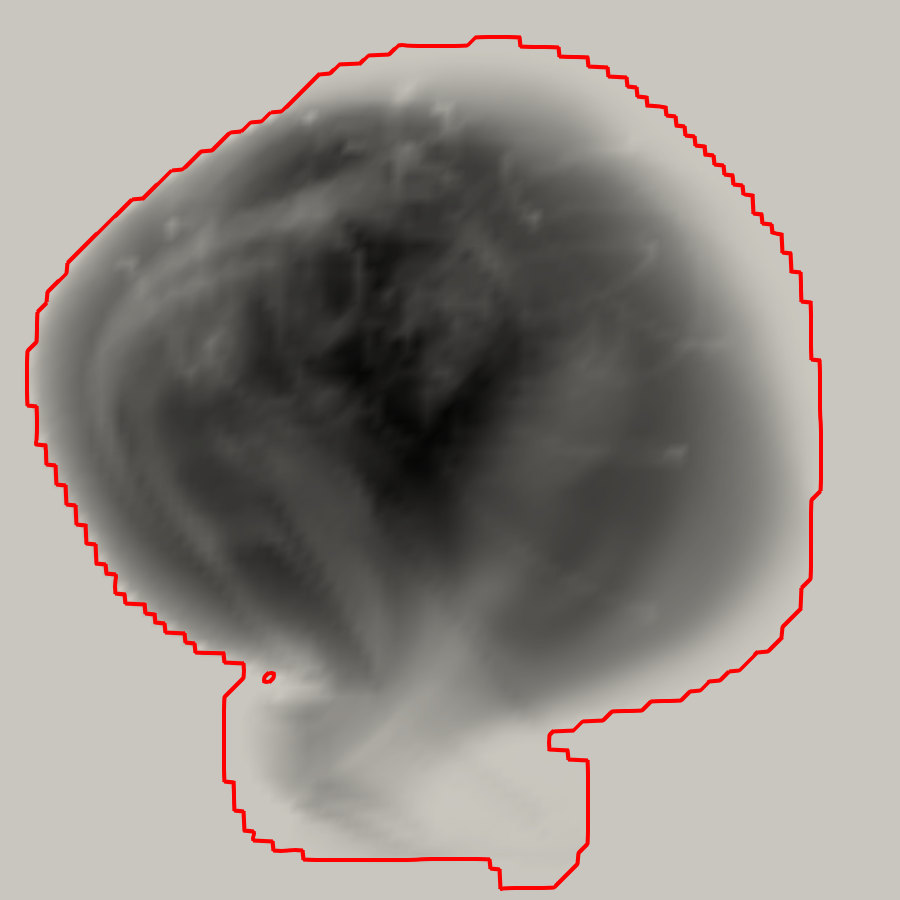}
        \caption*{(c)}
    \end{subfigure}
    \begin{subfigure}[t]{0.23\textwidth}
        \includegraphics[width=\linewidth]{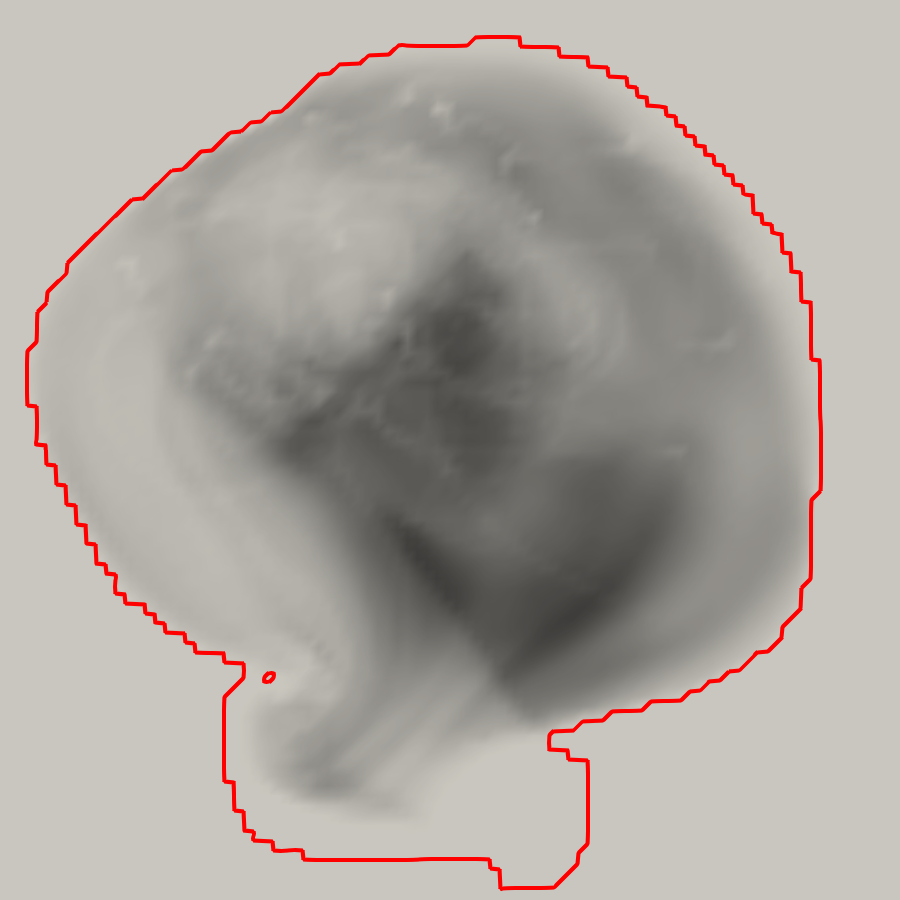}
        \caption*{(d)}
    \end{subfigure}
    
    \vspace{1mm}
    
    \begin{subfigure}[t]{0.23\textwidth}
        \includegraphics[width=\linewidth]{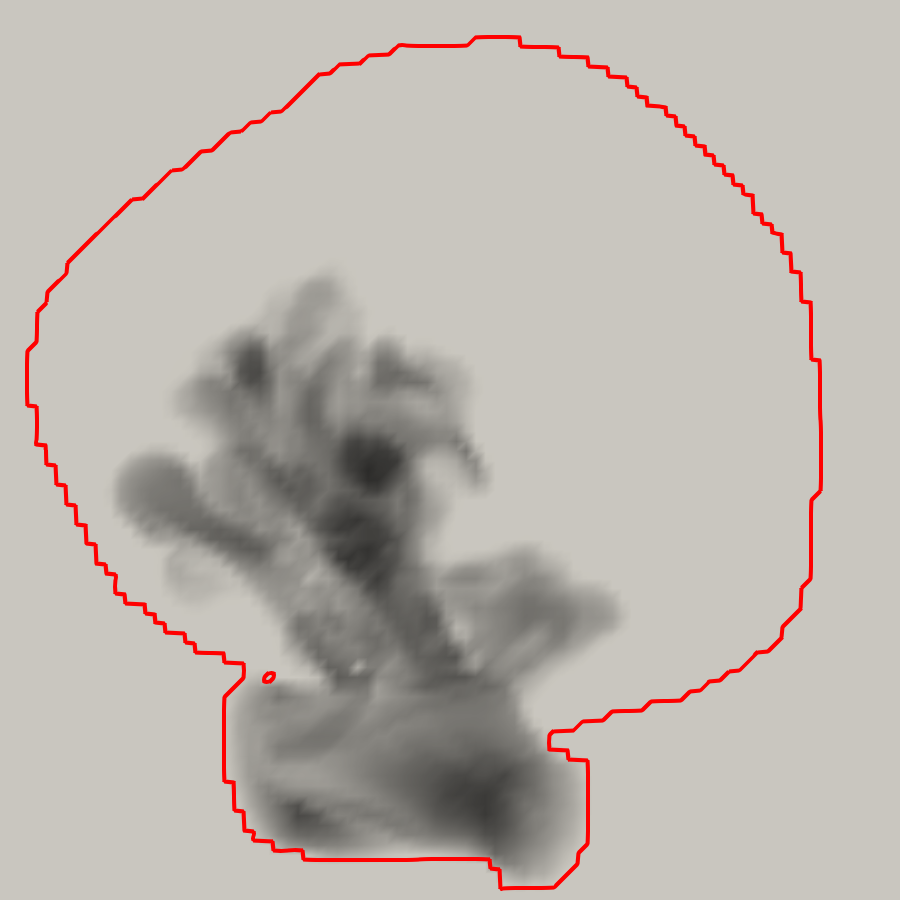}
        \caption*{(e)}
    \end{subfigure}
    \begin{subfigure}[t]{0.23\textwidth}
        \includegraphics[width=\linewidth]{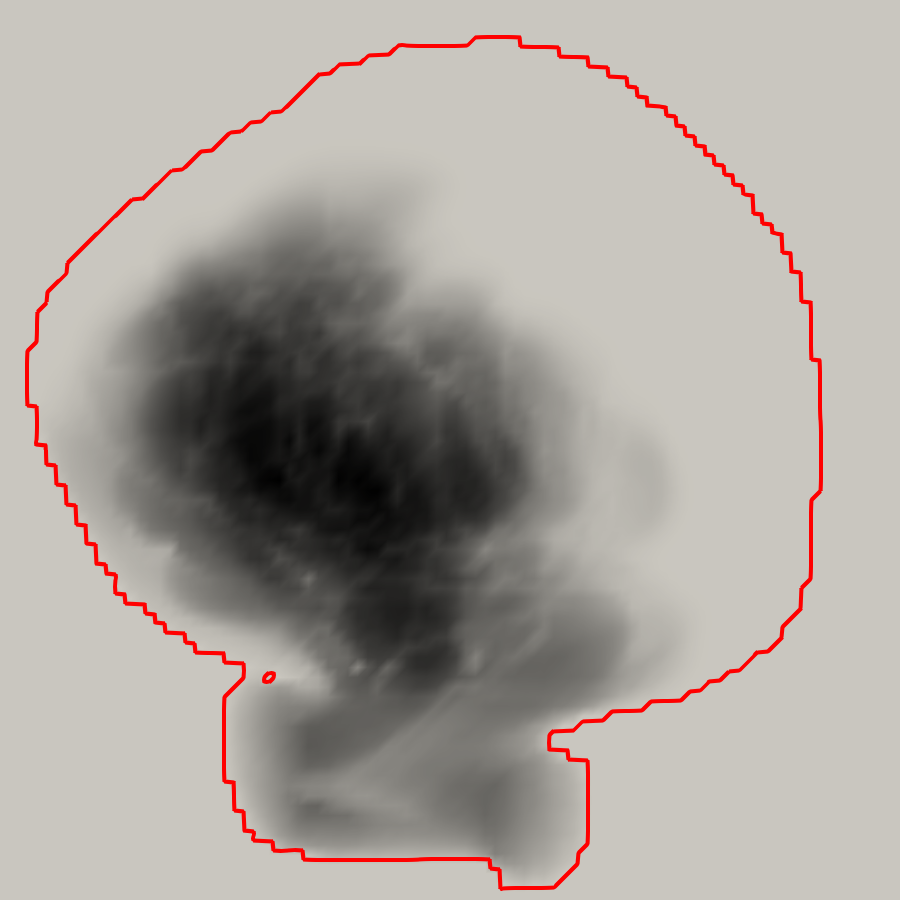}
        \caption*{(f)}
    \end{subfigure}
    \begin{subfigure}[t]{0.23\textwidth}
        \includegraphics[width=\linewidth]{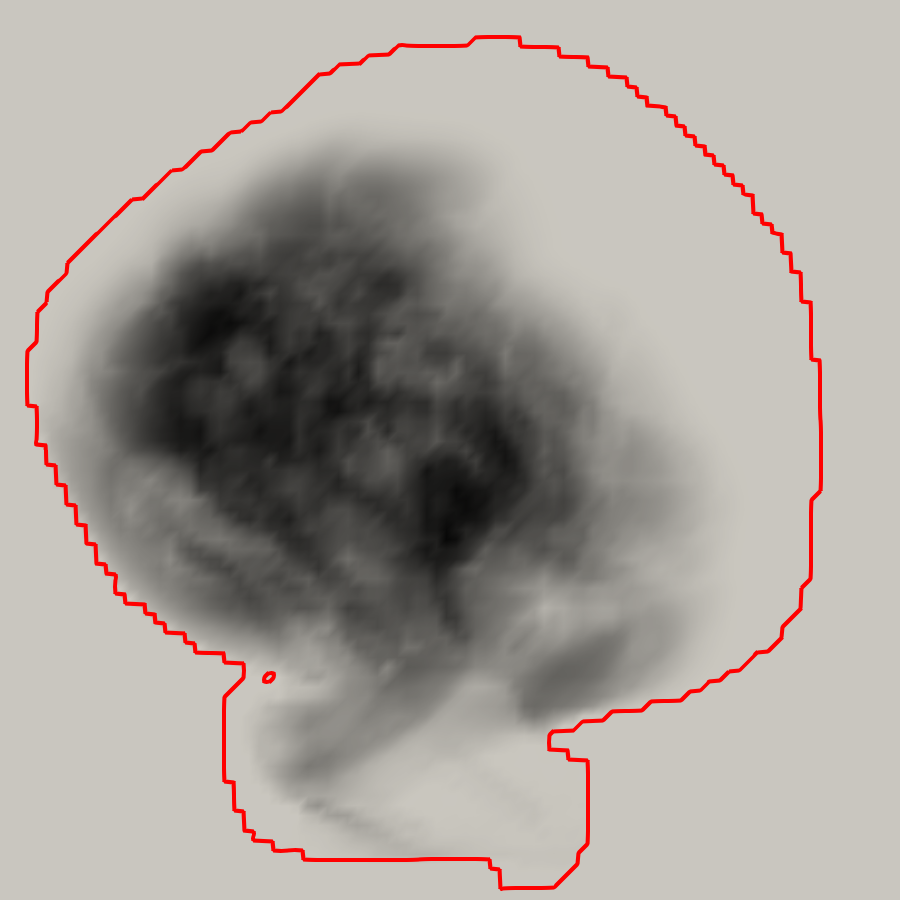}
        \caption*{(g)}
    \end{subfigure}
    \begin{subfigure}[t]{0.23\textwidth}
        \includegraphics[width=\linewidth]{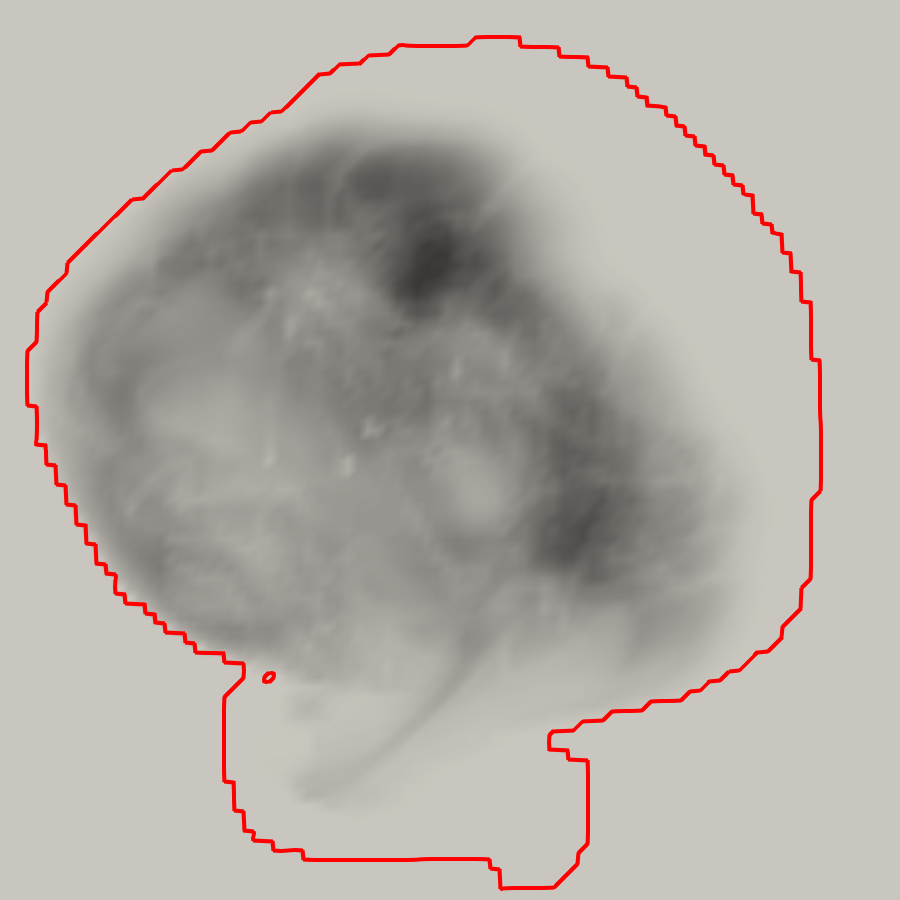}
        \caption*{(h)}
    \end{subfigure}

    \caption{\textit{2D} \ac{DSA} of Case 3 showing the \textit{projected} occlusion results for two setups similar to Fig.\,\ref{fig:DSA_Case3}. As within that figure, each \textbf{column} again corresponds to a temporal snapshot at peak systole. \textbf{The first row} (a)-(d) shows the two coils placed case (row 2 from Fig.\,\ref{fig:DSA_Case3}), \textbf{the second row} (e)-(h) shows a three coils placed situation.}
    \label{fig:DSA_Case3_Projected}
\end{figure}

Figure~\ref{fig:DSA_Case3_Projected} shows the corresponding two-dimensional virtual \ac{DSA} projections. These projections provide an image representation closer to clinical angiography and can be used for grey-level calibration against clinical data. However, the three-dimensional tracer fields in Fig.~\ref{fig:DSA_Case3} provide a more detailed basis for assessing tracer distribution and washout within the aneurysm

\subsubsection{Case 1}

\begin{figure}[h!]
    \centering
    \begin{subfigure}[t]{0.23\textwidth}
    \includegraphics[width=\linewidth]{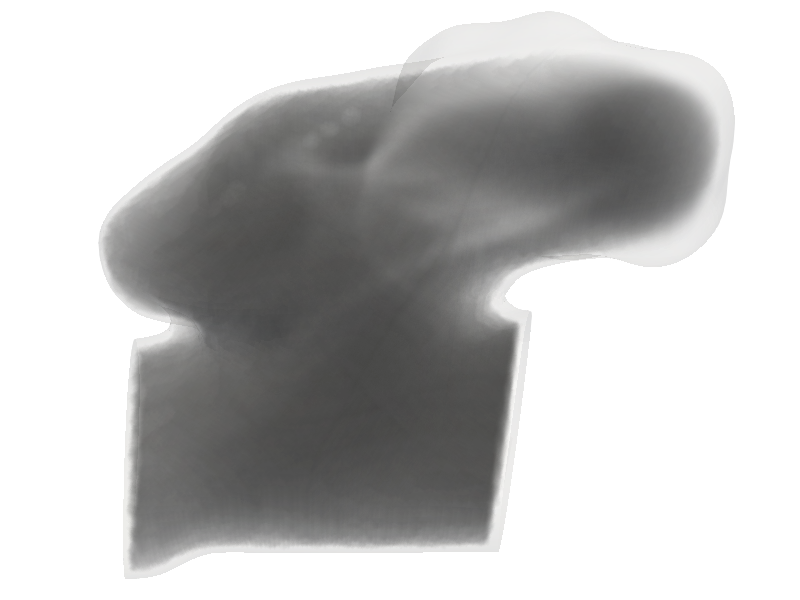}
        \caption*{(a)}
    \end{subfigure}
    \begin{subfigure}[t]{0.23\textwidth}
        \includegraphics[width=\linewidth]{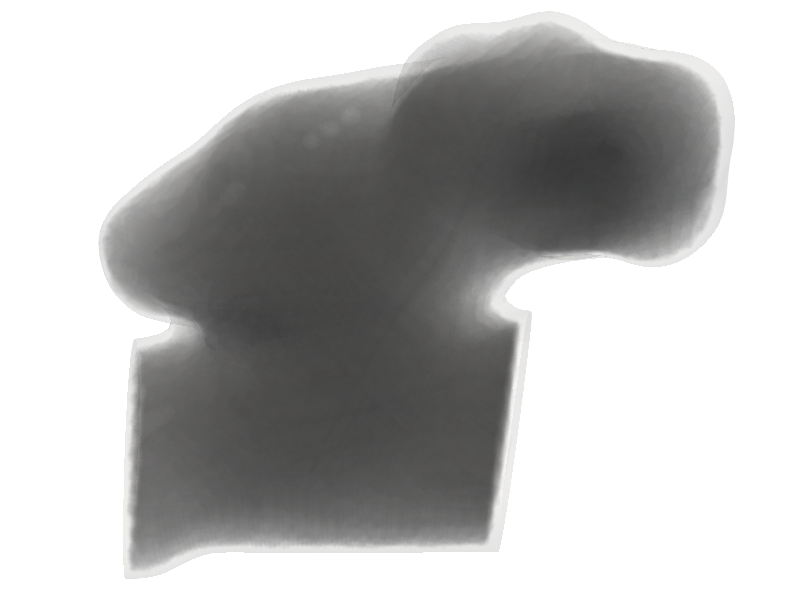}
        \caption*{(b)}
    \end{subfigure}
    \begin{subfigure}[t]{0.23\textwidth}
        \includegraphics[width=\linewidth]{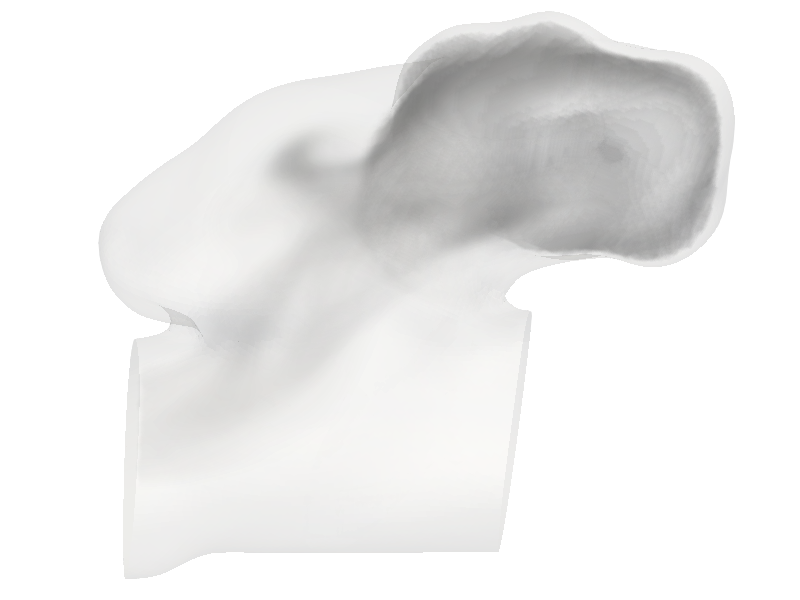}
        \caption*{(c)}
    \end{subfigure}
    \begin{subfigure}[t]{0.23\textwidth}
        \includegraphics[width=\linewidth]{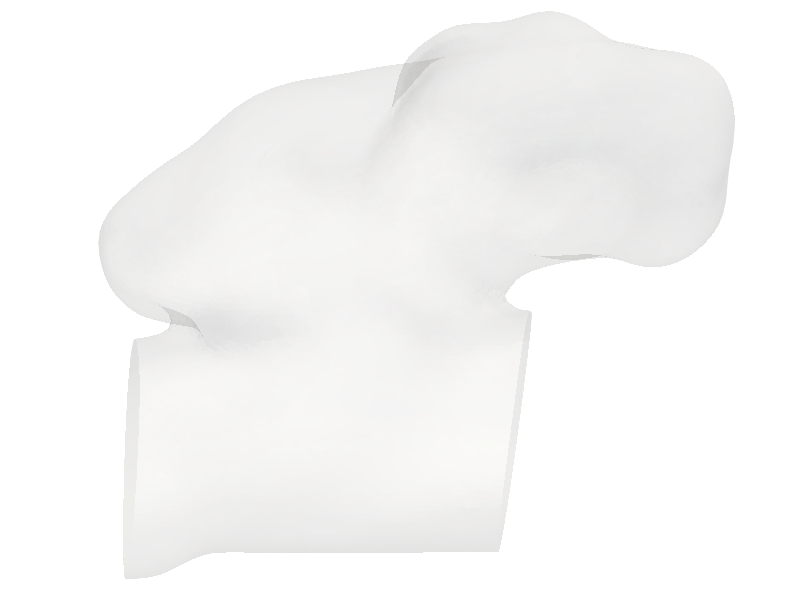}
        \caption*{(d)}
    \end{subfigure}
    
    \vspace{1mm}
    
    \begin{subfigure}[t]{0.23\textwidth}
        \includegraphics[width=\linewidth]{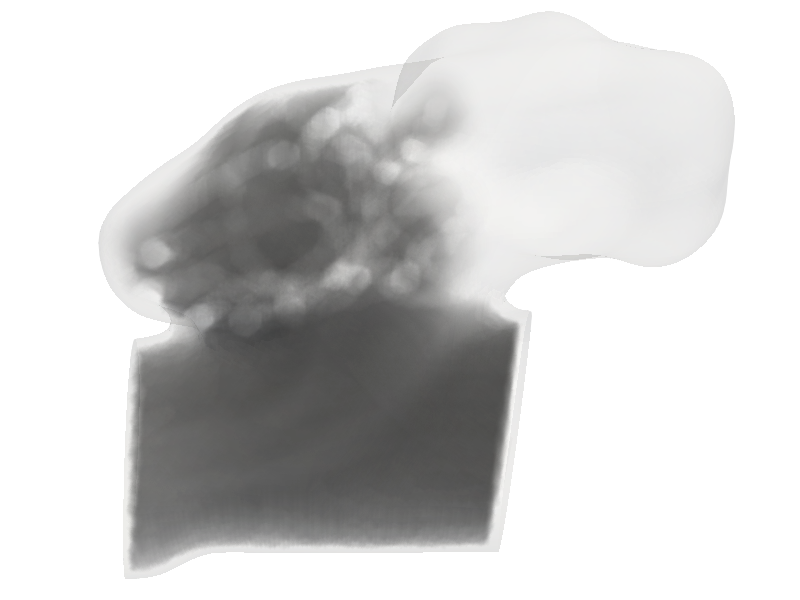}
        \caption*{(e)}
    \end{subfigure}
    \begin{subfigure}[t]{0.23\textwidth}
        \includegraphics[width=\linewidth]{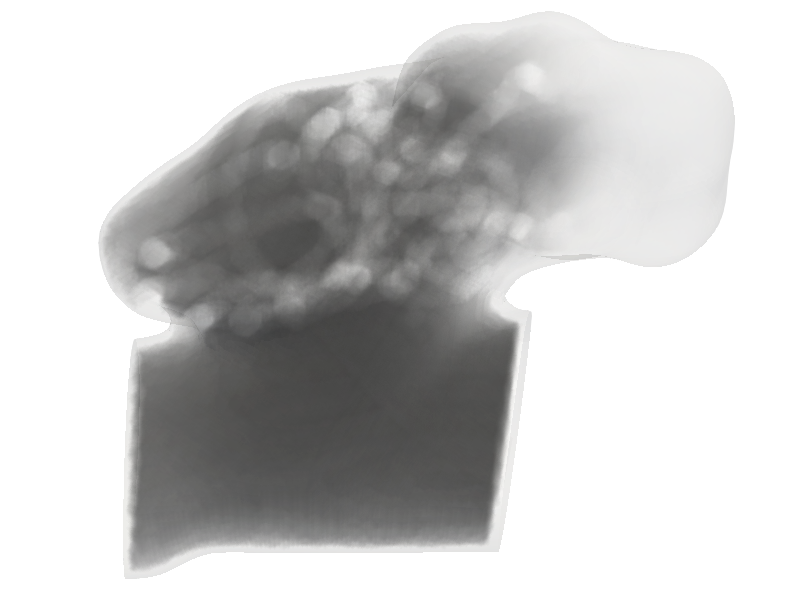}
        \caption*{(f)}
    \end{subfigure}
    \begin{subfigure}[t]{0.23\textwidth}
        \includegraphics[width=\linewidth]{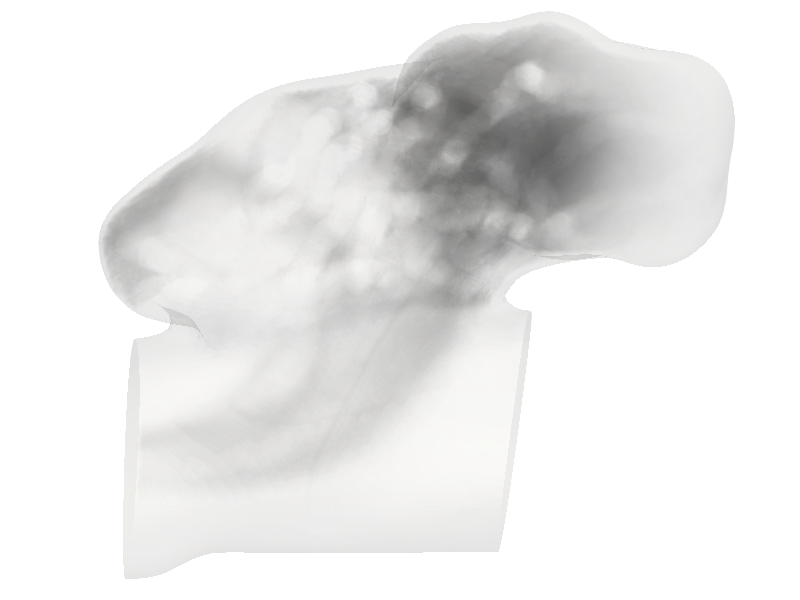}
        \caption*{(g)}
    \end{subfigure}
    \begin{subfigure}[t]{0.23\textwidth}
        \includegraphics[width=\linewidth]{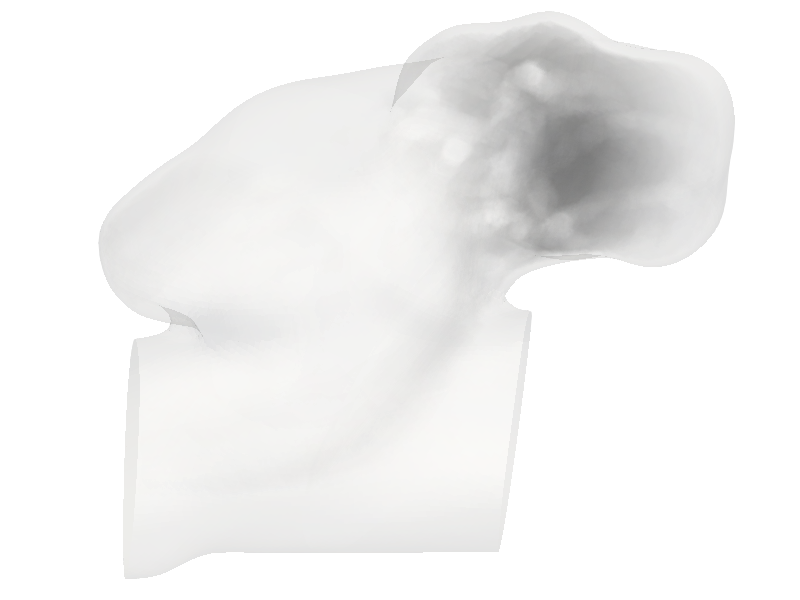}
        \caption*{(h)}
    \end{subfigure}

    \vspace{1mm}

    \begin{subfigure}[t]{0.23\textwidth}
        \includegraphics[width=\linewidth]{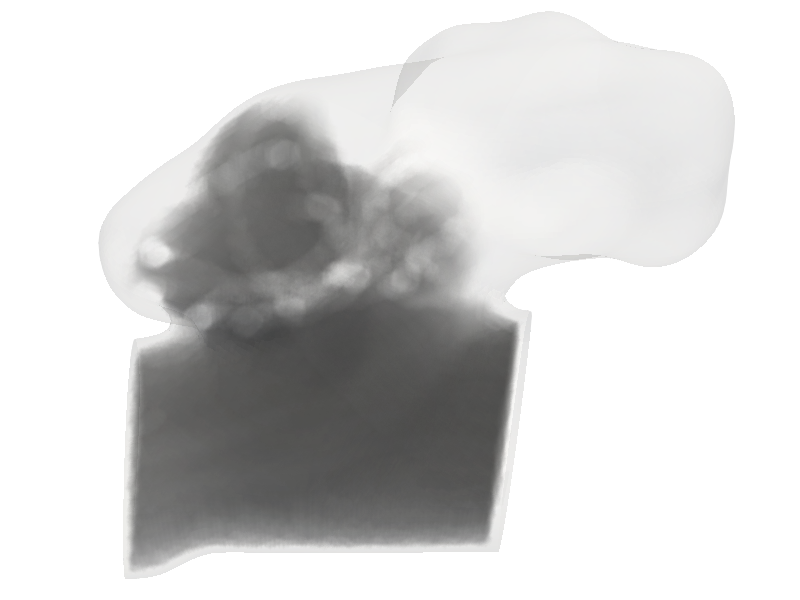}
        \caption*{(i)}
    \end{subfigure}
    \begin{subfigure}[t]{0.23\textwidth}
        \includegraphics[width=\linewidth]{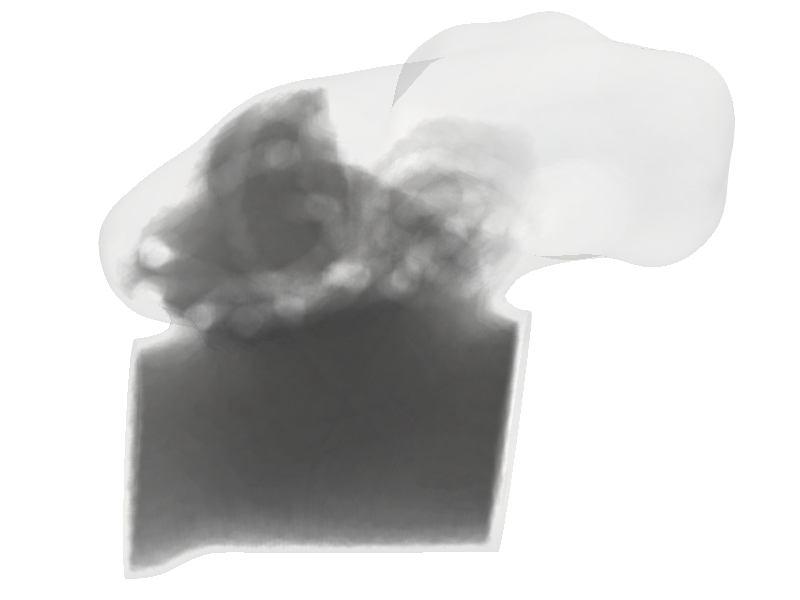}
        \caption*{(j)}
    \end{subfigure}
    \begin{subfigure}[t]{0.23\textwidth}
        \includegraphics[width=\linewidth]{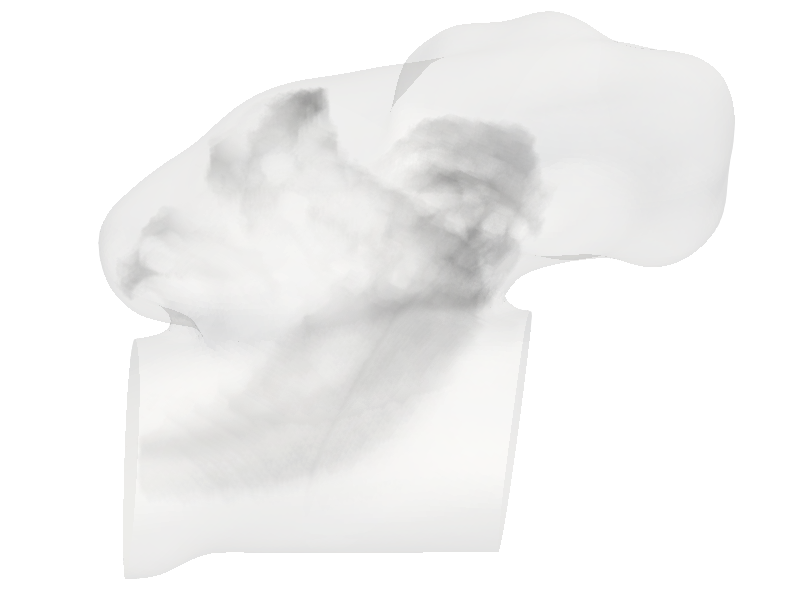}
        \caption*{(k)}
    \end{subfigure}
    \begin{subfigure}[t]{0.23\textwidth}
        \includegraphics[width=\linewidth]{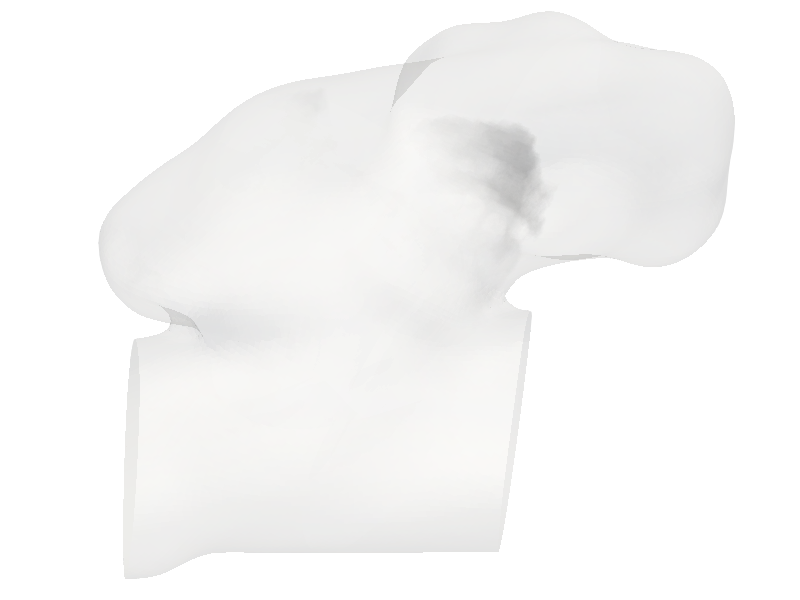}
        \caption*{(l)}
    \end{subfigure}

    \caption{3D tracer field visualization of Case 1 showing aneurysm occlusion in different setups. Each \textbf{column} corresponds to the temporal snapshot at peak systoles resulting from a simulation over four heart-beat cycles. The \textbf{first row} (a)-(d): No coil placed. The \textbf{second row} (e)-(h): coil placed, but no thrombosis. The \textbf{third row} (i)-(l): After coil-induced thrombosis.}
    \label{fig:DSA_Case1}
\end{figure}

Figure~\ref{fig:DSA_Case1} compares tracer inflow and washout in Case 1 for the untreated aneurysm, the coiled configuration without thrombosis, and the coiled configuration with thrombus formation. The results show that coiling alone reduces the initial tracer inflow into the aneurysm, as seen by comparing panels (e,f) with panels (a,b). However, tracer material remains trapped between the coil wires during the subsequent washout phase, resulting in greater residual tracer retention than in the untreated configuration, as shown by the comparison between panels (h) and (d).

In contrast, the inclusion of coil-induced thrombus substantially suppresses tracer penetration into the distal region of the aneurysm, as shown in panel (l). This indicates that, for Case 1, thrombus formation provides a major contribution to effective aneurysm isolation and substantially improves the occlusion quality achieved by the coiling treatment.

\subsubsection{Case 2}

Case 2 represents the most geometrically complex configuration considered here, owing to the fusiform morphology of the aneurysm. As a result, the differences between treatment configurations in the virtual \ac{DSA} results are less pronounced than in the previous cases, but remain discernible.

\enlargethispage*{5mm}
Similar to Case 1, the endovascular devices alone primarily reduce tracer inflow during the early injection phase. However, tracer material remains present near the distal region of the aneurysm during the subsequent washout phase. When thrombus formation is included, residual inflow is reduced more substantially. In this case, the combined stent-assisted coiling configuration provides greater inflow restriction than the \ac{FD}-only treatment.

\begin{figure}[h]
    \centering
    \begin{subfigure}[t]{0.23\textwidth}
        \includegraphics[width=\linewidth]{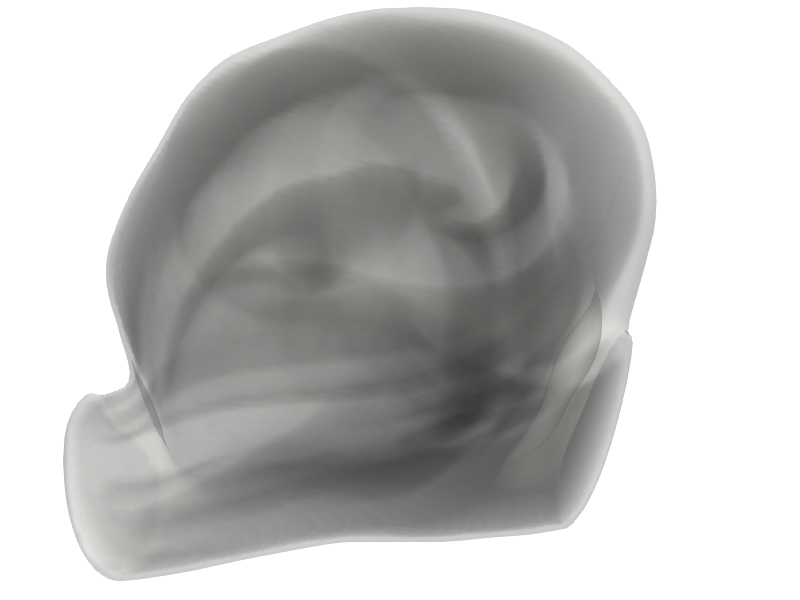}
        \caption*{(a)}
    \end{subfigure}
    \begin{subfigure}[t]{0.23\textwidth}
        \includegraphics[width=\linewidth]{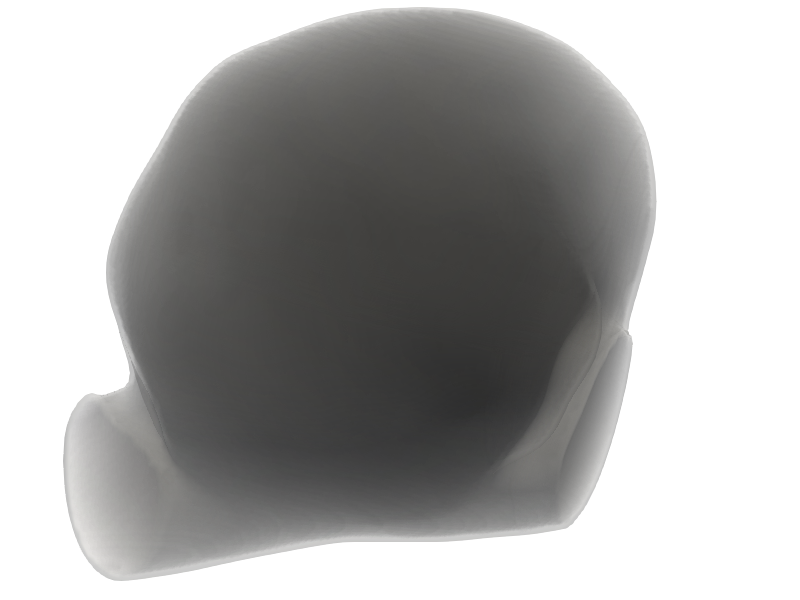}
        \caption*{(b)}
    \end{subfigure}
    \begin{subfigure}[t]{0.23\textwidth}
        \includegraphics[width=\linewidth]{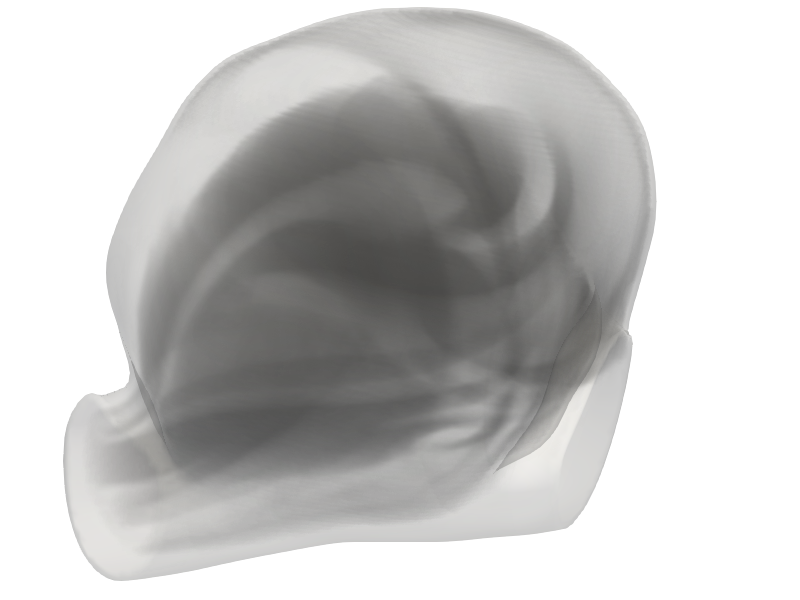}
        \caption*{(c)}
    \end{subfigure}
    \begin{subfigure}[t]{0.23\textwidth}
        \includegraphics[width=\linewidth]{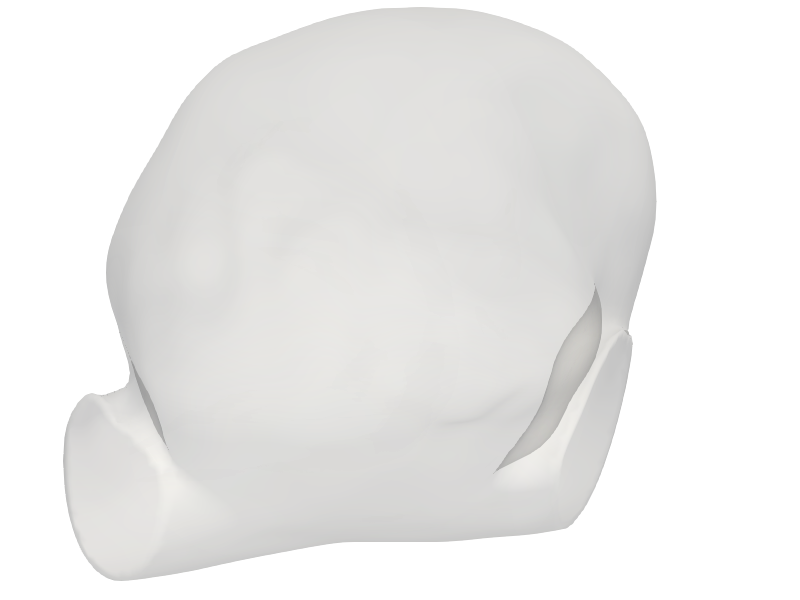}
        \caption*{(d)}
    \end{subfigure}
    
    \vspace{1mm}
    
    \begin{subfigure}[t]{0.23\textwidth}
        \includegraphics[width=\linewidth]{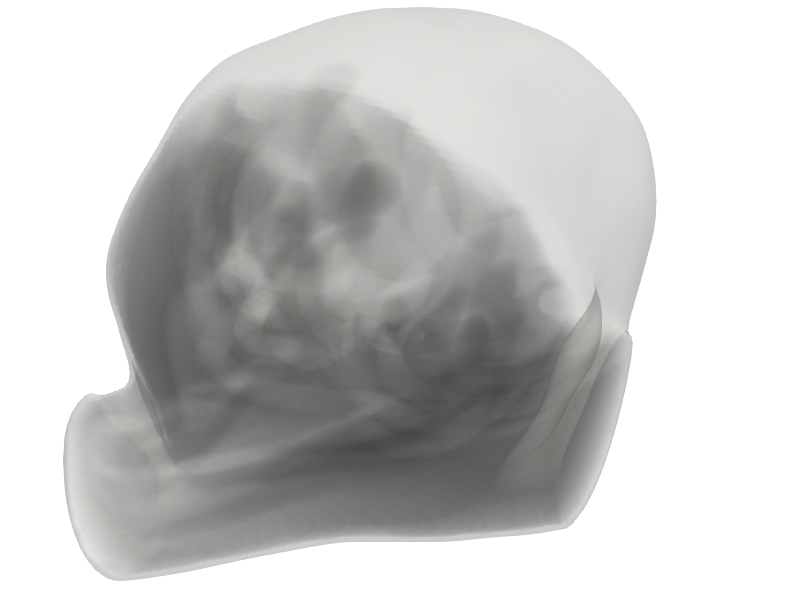}
        \caption*{(e)}
    \end{subfigure}
    \begin{subfigure}[t]{0.23\textwidth}
        \includegraphics[width=\linewidth]{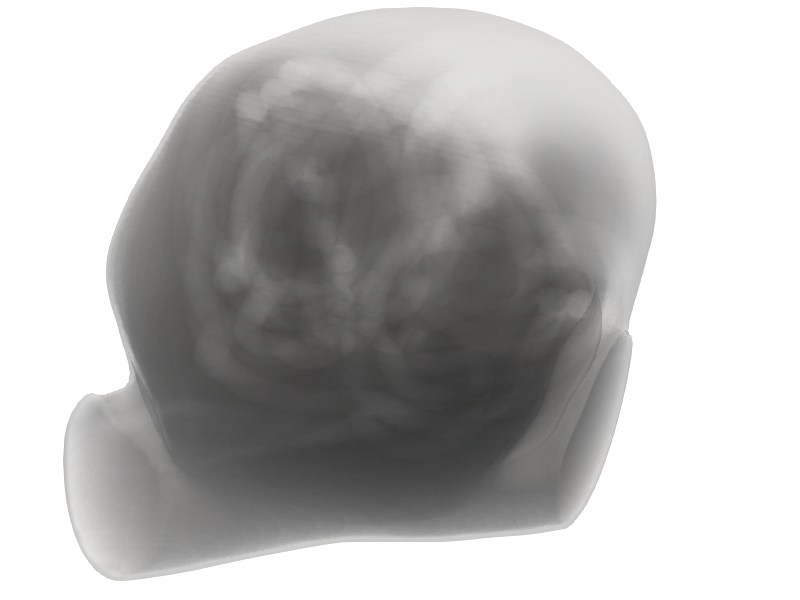}
        \caption*{(f)}
    \end{subfigure}
    \begin{subfigure}[t]{0.23\textwidth}
        \includegraphics[width=\linewidth]{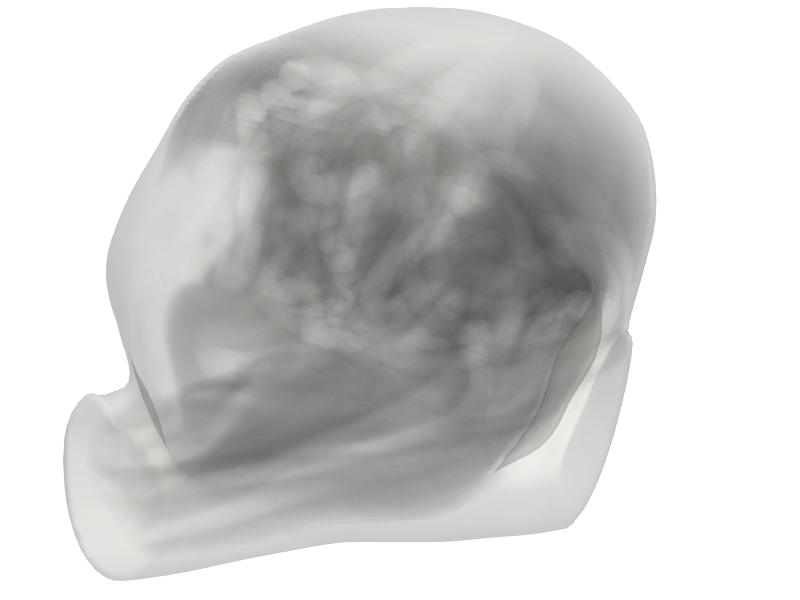}
        \caption*{(g)}
    \end{subfigure}
    \begin{subfigure}[t]{0.23\textwidth}
        \includegraphics[width=\linewidth]{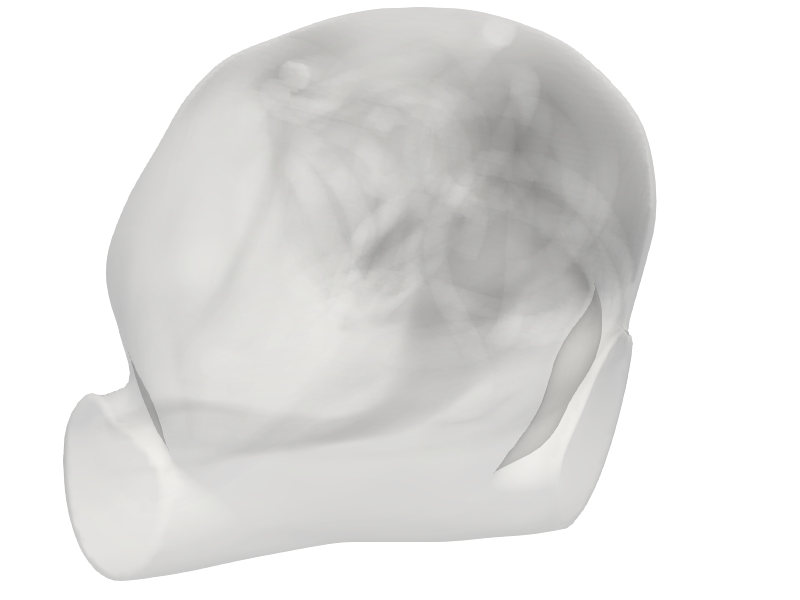}
        \caption*{(h)}
    \end{subfigure}

    \vspace{1mm}
    
    \begin{subfigure}[t]{0.23\textwidth}
        \includegraphics[width=\linewidth]{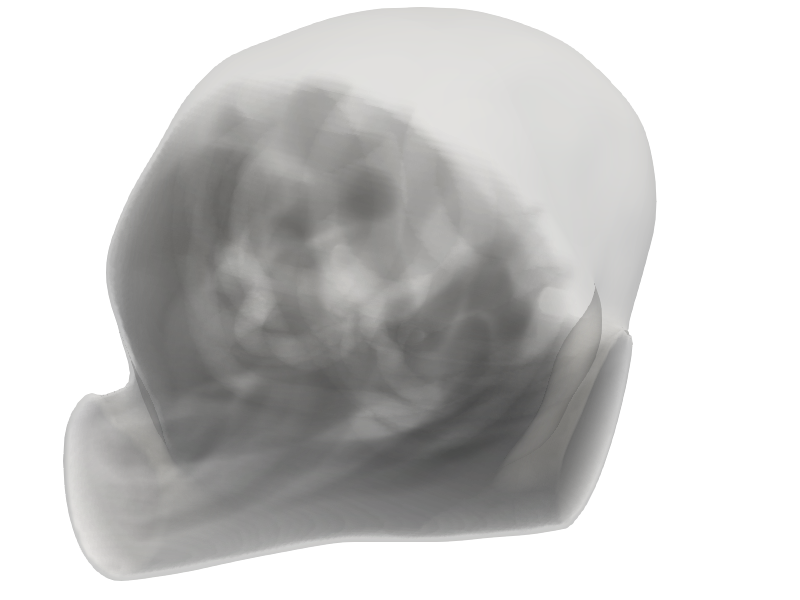}
        \caption*{(i)}
    \end{subfigure}
    \begin{subfigure}[t]{0.23\textwidth}
        \includegraphics[width=\linewidth]{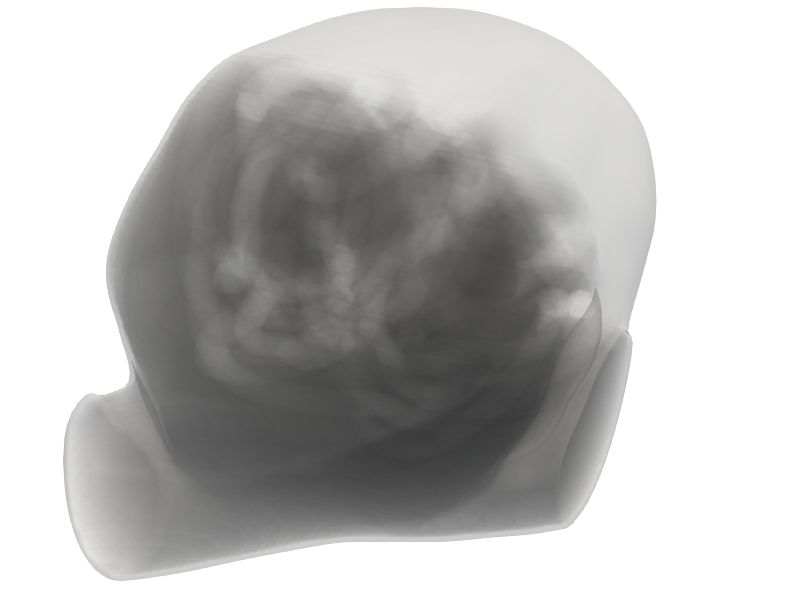}
        \caption*{(j)}
    \end{subfigure}
    \begin{subfigure}[t]{0.23\textwidth}
        \includegraphics[width=\linewidth]{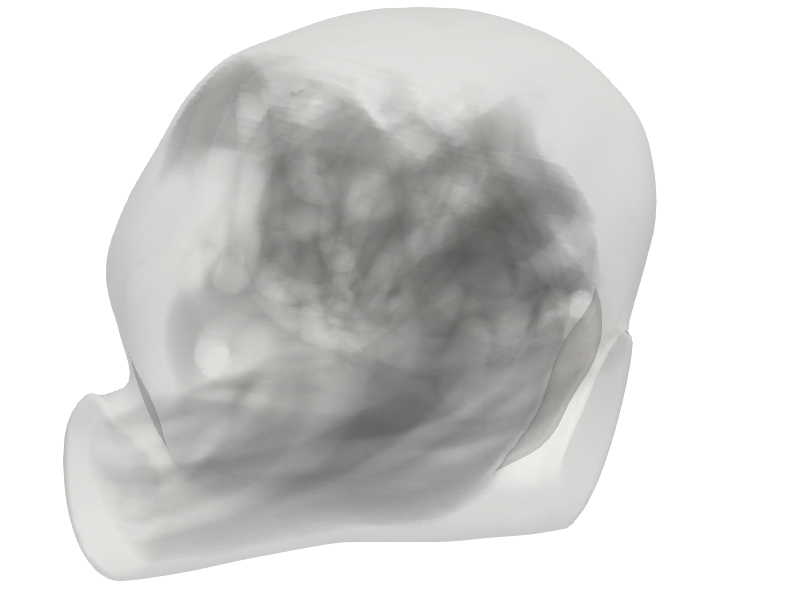}
        \caption*{(k)}
    \end{subfigure}
    \begin{subfigure}[t]{0.23\textwidth}
        \includegraphics[width=\linewidth]{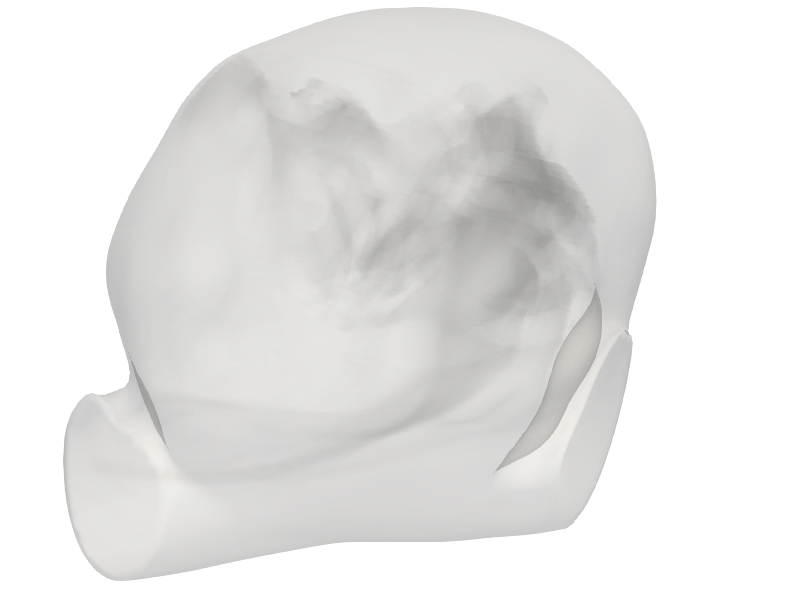}
        \caption*{(l)}
    \end{subfigure}

    \vspace{1mm}
    
    \begin{subfigure}[t]{0.23\textwidth}
        \includegraphics[width=\linewidth]{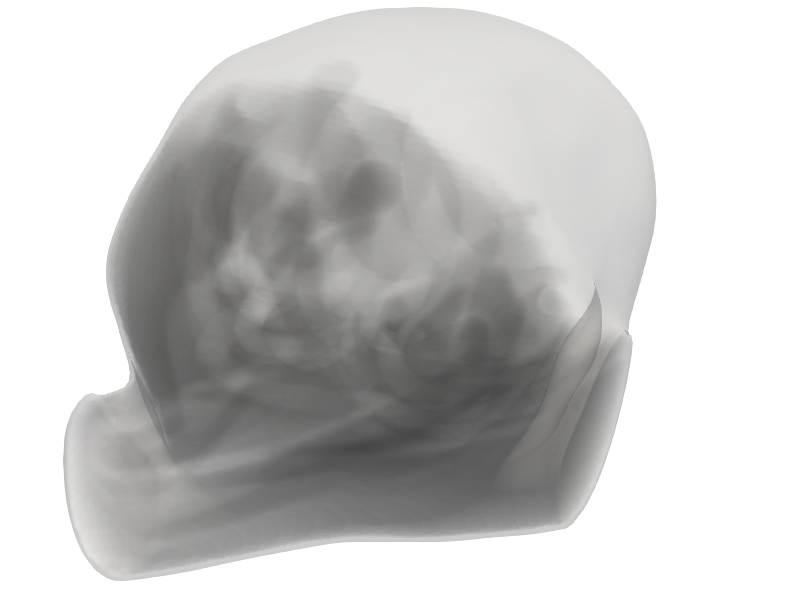}
        \caption*{(m)}
    \end{subfigure}
    \begin{subfigure}[t]{0.23\textwidth}
        \includegraphics[width=\linewidth]{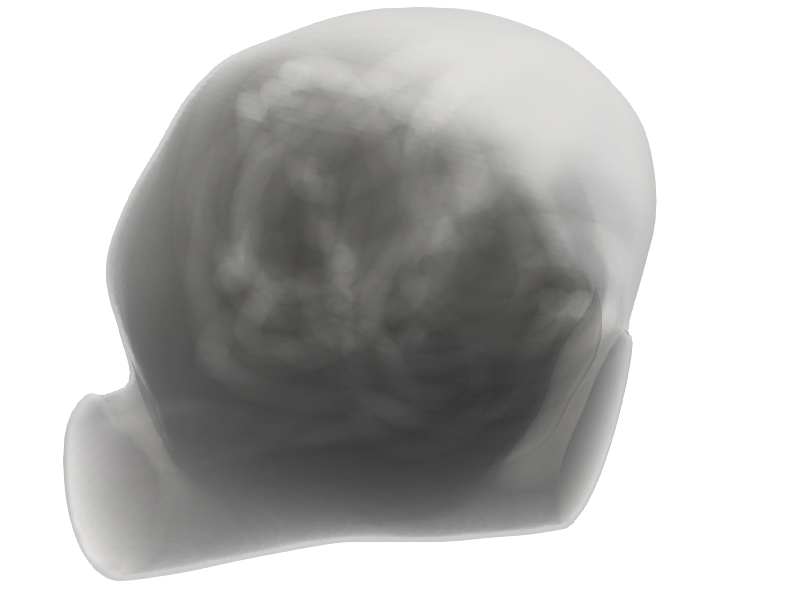}
        \caption*{(n)}
    \end{subfigure}
    \begin{subfigure}[t]{0.23\textwidth}
        \includegraphics[width=\linewidth]{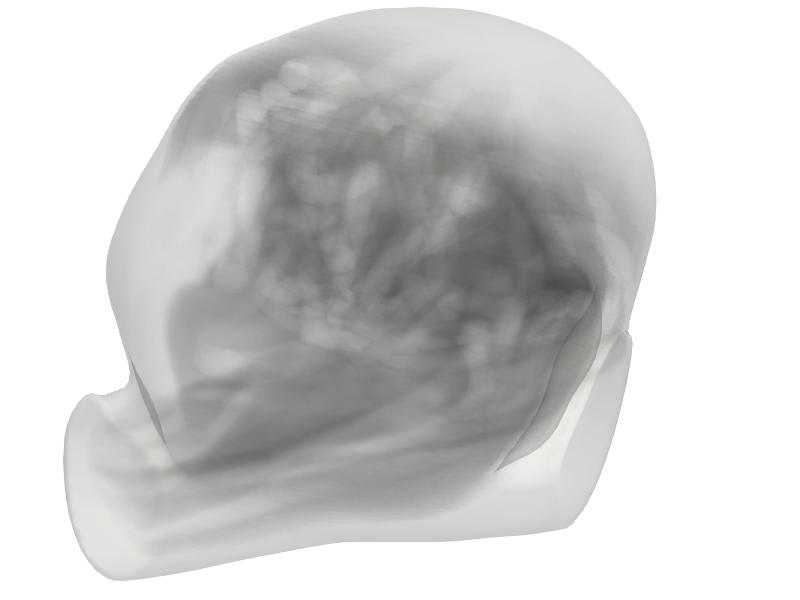}
        \caption*{(o)}
    \end{subfigure}
    \begin{subfigure}[t]{0.23\textwidth}
        \includegraphics[width=\linewidth]{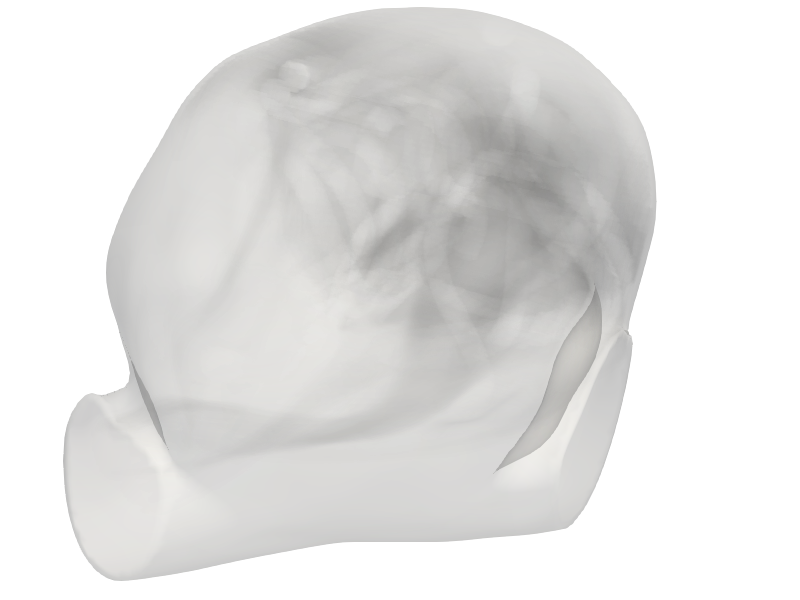}
        \caption*{(p)}
    \end{subfigure}

    \caption{3D tracer field visualization of Case 2 showing aneurysm occlusion in different setups. Each \textbf{column} corresponds to the temporal snapshot at peak systoles resulting from a simulation over four heart-beat cycles. \textbf{In the first row} (a)-(d): No coil placed. \textbf{In the second row} (e)-(h): stent-assisted coiling in place. \textbf{In the third row} (i)-(l): After stent-assisted coiling induced thrombosis. \textbf{In the last row} (m)-(p): After \ac{FD}-only induced thrombosis.}
    \label{fig:DSA_Case2}
\end{figure}


%% file: Discussion.tex
\section{Discussion}\label{sec:Discussion}
Based on simulations of contrast agent flow and thrombus occlusion, the study presented in this paper demonstrates the potential utility of virtual interventional planning tools for three different aneurysms, based on different treatment methods. Coiling, stent-assisted coiling and \ac{FD} treatment were considered in different combinations for a saccular, fusiform and giant aneurysm, depending on the most appropriate approach that would likely be employed in a clinical context. An important contribution of the study is the presentation of computational results that resemble approaches that are currently used in clinical practice. While the use of new quantitative metrics is often desirable, particularly in cases where simulation tools are able to provide metrics that are otherwise unavailable in existing clinical platforms, validating their exact impact on aneurysm evolution remains a challenge \cite{staarmann2019shear, villa2011neurist, meng2014high}. As such, replicating metrics that are known to be clinically reliable can be a good starting point for bridging the gap between applied clinical experience and mathematical modeling and simulation. 

In clinical practice, long-term and immediate occlusion outcomes are used as a metric for determining the success of neuroradiological intervention for the treatment of unruptured cerebral aneurysms \cite{Thompson2015, Turjman2014, AlSaiegh2022, Bender2018}. This is influenced by the presence of endovascular devices, which either reduce the amount of flow entering the aneurysmal sac or act as a physical barrier within the aneurysm. Occlusion is also the result of early thrombus formation, following intervention. Occlusion is reported as full or partial, depending on the extent to which flow into the sac has been reduced. In the case of full occlusion, there is an absence of flow in the aneurysm sac while a remnant of flow is observed for partial occlusion. These observations are typically based on contrast agent angiography. As such, the virtual \ac{DSA} (2D-projection) and general 3D contrast agent simulation method and results in this study align with those used in clinic, but also include additional detail gained from the thrombosis simulations. 

The virtual contrast agent results showed a clear visual distinction between the untreated and treated cases for Cases 1 and 2. The distinction between different treatments is less obvious for these two cases. Case 3 showed very little visual difference between the untreated case and the 2-coil treated case. The distinction between the treated cases was, however, marked. For the thrombosis modelling results, Cases 1 and 2 were partially occluded for all treatment modes. This was true even for the failed coiling case in Case 2. For Case 3,  the full spectrum of occlusion outcomes, ranging from partial to full occlusion, was observed across the different treatments. On closer inspection of the giant aneurysm in Case 3, it became evident that thrombus development in the presence of endovascular devices is closely tied to the presence of  stable vortical structures, thereby confirming our hypothesis. The no device and flow diverter cases, which had a stable vortical structures, achieved greater thrombus occlusion than the coiled case, which had diffuse streamlines. A computational study which examined thrombus development without devices also found that stable vortex modes supported thrombus development in cerebral aneurysms \cite{Ngwenya2024}. Clinical studies have also reported lower occlusion in giant aneurysms treated with coils, immediately after intervention \cite{Scalise2026, chalouhi2013comparison, abo2026comparison,LI2025111548}.   

Based on this very limited sample set, it would seem that the modelling platform is able to make distinctions between no treatment, treatment-induced partial occlusion and treatment-induced full occlusion. In cases where occlusion is partial for all treatment modes in the short term, the potential benefits of a specific treatment mode are hard to predict. The results in this study predict short-term occlusion outcome, on the scale of minutes, following device placement. When considering the success of endovascular intervention, clinical studies report on results immediately after intervention, and also on a longer term occlusive process, which often takes place over months. It is important to make a distinction between the different processes under consideration \cite{cekirge2016new}. As discussed, immediately after device placement, there is typically a reduction of flow into the aneurysmal sac. This change in local flow patterns gives rise to conditions that support coagulation, leading to the formation of a clot with a fibrin mesh. This is a relatively short-term process, taking place over the scale of minutes, which our model is able to capture. The longer term process, which is reported as long term occlusion in clinical studies, is not accounted for in this model. Although wound healing is related to clotting, the process requires different modelling tools to capture the very different timescales and physical processes at play \cite{monroe2012clotting}. A study examining thrombus organization in an aneurysm swine model showed that the healing process recruited a wider range of different cells and took place over a longer time period than the initial clot that formed in a matter of minutes \cite{lee2007thrombus}. The initial, acute thrombus was replaced by leukocyte and macrophage infiltration in the organization stage. The proliferative stage that follows is marked by myofibroblast and fibroblast proliferation, and also includes the initiation of collagen and extracellular matrix deposition. In the contraction stage, further collagen deposition was observed alongside aneurysm contraction. Interestingly, in this model, which was devoid of devices, healing occurred from the aneurysm walls inwards, i.e. in a centripetal direction. Although observed in swine, pig fibrinolytic and wound healing pathways have been shown to have similarity with those of humans, therefore histological and molecular analysis of wound healing results might be partially extendable to humans \cite{siller2008interspecies, sullivan2001pig}.

%% file: Conclusion.tex
\section{Conclusion}\label{sec:Conclusion}
Pairing the thrombosis CFD model with a residual-contrast transport model links mechanistic occlusion dynamics to an angiography-like observable and enables device strategies to be assessed in terms that align with clinical imaging interpretation. In our considered cases, regions of delayed washout observed prior to clot formation tend to spatially coincide with the regions in which thrombus subsequently develops. This suggests that haemodynamic patterns visible in contrast transport simulations may serve as spatial indicators for susceptibility to clot initiation. Further investigation into these patterns revealed that vortical structures play a key role in device-induced occlusion outcomes. 

The post-clot dye simulations provide a functional readout of the state after early thrombus formation by quantifying how much residual perfusion and exchange remain through persisting communicating pathways once the modeled clot has increased hydraulic resistance within the aneurysm. When interpreted alongside thrombosis-driven occlusion progression, post-clot dye behavior helps distinguish endovascular device configurations that achieve effective isolation, shown by minimal dye ingress or rapid clearance within the remaining lumen, from configurations that retain persistent dye-accessible regions despite partial clotting. Overall, the combined framework supports medium-term prognosis by evaluating whether device deployment produces stasis patterns that promote thrombus growth and whether the resulting thrombus sufficiently suppresses post-clot transport to indicate durable functional occlusion beyond short-term deployment criteria alone.

Looking ahead, the availability of suitable clinical angiographic data would allow calibration of patient-specific parameters in the mathematical model and further validation of the proposed framework. Comparing simulated and clinical image based spatiotemporal normalized contrast distribution by using the Wasserstein distance metric could then not only provide an objective quality measure for calibration and validation but also opens the door for a systematic data-driven model enhancement.

%% file: acronyms.tex
\begin{acronym}[WEB-Device] 
\acro{AVM}{Arteriovenous Malformation}
\acro{FD}{Flow Diverter}
\acro{RRC}{Raymond Roy Classification}
\acro{ADE}{Advection Diffusion Equation}
\acro{DSA}{Digital Subtraction Angiography}
\end{acronym}